\documentclass[runningheads]{llncs}
\usepackage{orcidlink}

\usepackage[T1]{fontenc}
\usepackage{graphicx}
\usepackage{subcaption}
\usepackage{enumitem}
\usepackage{listings}

\usepackage{mathrsfs}
\usepackage{amsmath}
\usepackage{amssymb}
\usepackage{tcolorbox}
\usepackage{multirow}
\usepackage{diagbox}
\usepackage{longtable}
\usepackage{float}
\usepackage{pdflscape}
\usepackage{lscape}

\DeclareUnicodeCharacter{03B5}{$\varepsilon$}

\usepackage[linesnumbered,ruled,vlined]{algorithm2e}

\usepackage{tikz}

\newcommand\R{\ensuremath{\mathbb{R}}}

\providecommand{\abs}[1]{\ensuremath{\left\lvert#1\right\rvert}}

\definecolor{verde}{rgb}{0.0,0.6,0.0}
\definecolor{jpurple}{rgb}{0.58,0,0.82}
\definecolor{darkgreen}{rgb}{0.0, 0.2, 0.13}
\definecolor{backcolor}{rgb}{0.96,0.96,0.93}
\definecolor{oldmauve}{rgb}{0.4, 0.19, 0.28}
\definecolor{markedcode}{rgb}{0.99, 0.7, 0.7}
\definecolor{extracted}{rgb}{0.7, 0.95, 0.7}
\definecolor{burlywood}{RGB}{222,184,135}
\definecolor{antiquewhite}{rgb}{0.98, 0.92, 0.84}
\definecolor{floralwhite}{rgb}{1.0, 0.98, 0.94}
\definecolor{snow}{rgb}{1.0, 0.98, 0.98}
\definecolor{electricultramarine}{rgb}{0.25, 0.0, 1.0}
\definecolor{electricpurple}{rgb}{0.75, 0.0, 1.0}
\definecolor{awesome}{rgb}{1.0, 0.13, 0.32}
\definecolor{emerald}{rgb}{0.31, 0.78, 0.47}
\definecolor{green(colorwheel)(x11green)}{rgb}{0.0, 1.0, 0.0}
\definecolor{green(munsell)}{rgb}{0.0, 0.66, 0.47}
\definecolor{amber}{rgb}{1.0, 0.75, 0.0}
\definecolor{airforceblue}{rgb}{0.36, 0.54, 0.66}
\definecolor{ballblue}{rgb}{0.13, 0.67, 0.8}
\definecolor{mygreen}{rgb}{0,0.6,0}
\definecolor{mygray}{rgb}{0.5,0.5,0.5}
\definecolor{mymauve}{rgb}{0.58,0,0.82}
\definecolor{codebg}{HTML}{fafafa}
\definecolor{numbergrey}{HTML}{7f7f7f}
\definecolor{coverbackground}{HTML}{002c5c}
\definecolor{bubblegum}{rgb}{0.99, 0.76, 0.8}
\definecolor{brown(web)}{rgb}{0.65, 0.16, 0.16}
\definecolor{burgundy}{rgb}{0.5, 0.0, 0.13}
\definecolor{burlywood}{rgb}{0.87, 0.72, 0.53}
\definecolor{caputmortuum}{rgb}{0.35, 0.15, 0.13}
\definecolor{candypink}{rgb}{0.89, 0.44, 0.48}
\definecolor{citrine}{rgb}{0.89, 0.82, 0.04}
\definecolor{corn}{rgb}{0.98, 0.93, 0.36}
\definecolor{deepcarrotorange}{rgb}{0.91, 0.41, 0.17}
\definecolor{pblue}{rgb}{0.13,0.13,1}
\definecolor{pgreen}{rgb}{0,0.5,0}
\definecolor{pred}{rgb}{0.9,0,0}
\definecolor{pgrey}{rgb}{0.46,0.45,0.48}

\begin{document}
\title{Multi-objective Integer Linear Programming approach for Automatic Software Cognitive Complexity Reduction}
\titlerunning{MO-ILP approach for Automatic Software CC Reduction}
%
\author{Adriana Novoa-Hurtado\inst{1}\orcidlink{0009-0006-6523-5427} \and
Rubén Saborido\inst{1}\orcidlink{0000-0002-0944-5941} \and
Francisco Chicano\inst{1}\orcidlink{0000-0003-1259-2990} \and
Manuel Giménez-Medina\inst{2}\orcidlink{0000-0001-9686-8127}}
\authorrunning{Novoa-Hurtado, Saborido, Chicano and Giménez-Medina}
%
\institute{ITIS Software, University of Málaga \\
\email{\{novoa,rsain,chicano\}@uma.es} \and
Ayesa Advanced Digital Services Limited (ASDA) \\
\email{mgimenezm@ayesa.com}}
\maketitle              
\begin{abstract}
Clear and concise code is necessary to ensure maintainability, so it is crucial that the software is as simple as possible to understand, to avoid bugs and, above all, vulnerabilities. There are many ways to enhance software without changing its functionality, considering the extract method refactoring the primary process to reduce the effort required for code comprehension. The cognitive complexity measure employed in this work is the one defined by SonarSource, which is a company that develops well-known applications for static code analysis. This extraction problem can be modeled as a combinatorial optimization problem. The main difficulty arises from the existence of different criteria for evaluating the solutions obtained, requiring the formulation of the code extraction problem as a multi-objective optimization problem using alternative methods. We propose a multi-objective integer linear programming model to obtain a set of solutions that reduce the cognitive complexity of a given piece of code, such as balancing the number of lines of code and its cognitive complexity. In addition, several algorithms have been developed to validate the model. These algorithms have been integrated into a tool that enables the parameterised resolution of the problem of reducing software cognitive complexity.

\end{abstract}
\section{Introduction}
Current software systems are increasingly larger as they evolve, making it difficult for a developer working on a large project to fully comprehend all the details. That is why it is essential to develop clear code, not only to understand it, but also because other developers will probably use or modify it. That is why there is much effort in the literature focused on ensuring code maintainability.

Cyclomatic Complexity was initially formulated as a measure of the ``testability and maintainability'' of a module's control flow. While it excels at measuring the former, its underlying mathematical model is unsatisfactory at producing a value that measures the latter. Recently, a novel software cognitive complexity metric has been proposed and integrated in the well-known static code tools SonarCloud\footnote{\url{https://sonarcloud.io} (last access, November 2025)} and SonarQube\footnote{\url{https://www.sonarqube.org} (last access, November 2025)}, an open-source service and platform, respectively, for continuous inspection of code quality. This cognitive complexity metric, which we refer to as \textbf{CC} (from Cognitive Complexity), reflects the relative difficulty of understanding and, therefore, maintaining methods, classes, and applications. The CC is given by a positive number that is increased every time a control flow sentence appears. Their nesting levels also contribute to the CC of the code. Although SonarSource\footnote{\url{https://www.sonarsource.com} (last access, November 2025)} suggests keeping code's cognitive complexity no greater than a threshold, software developers lack support to reduce the CC of their code.

One way to reduce the CC associated with a method is by extracting certain pieces of code into new methods without altering any existing functionality. This strategy is known as the Extract Method refactoring. Refactoring is a transformation that preserves the original behavior of the code. Software CC reduction involves searching for sequences of extract method refactoring operations to reduce the CC below a given threshold~\cite{ASCCR}.

In previous work, we modeled the reduction of the CC to a given threshold as an Integer Linear Programming (ILP) problem~\cite{ASCCRthroughtILP}, minimizing the number of code extractions to reduce the CC of Java methods to or below a given threshold, allowing the problem to be solved by exact solvers. This last approach is beneficial, as it does not result in too many new methods distorting the initial semantics of the code. However, preliminary experiments in a software factory revealed that just minimizing the number of extractions resulted in highly unbalanced methods, both in terms of CC and lines of code (hereafter referred to as \textbf{LOC}). We now introduce a multi-objective ILP (MO-ILP) optimization problem. We then propose and evaluate two main algorithms to efficiently solve this problem and reduce the CC of a method by obtaining sequences of code extractions while balancing the CC and LOC of the extracted methods. Thus, the contributions of this work are three:
\begin{itemize}[noitemsep]
    \item Formulating a multi-objective ILP variant of the software CC reduction problem. The three ILP objectives with which we work are to minimize the number of extractions performed to a given method and equilibrate the CC and LOC of all code extractions.
    
    \item Designing a tool, called \emph{ILP CC reducer tool}, that integrates several multi-objective ILP problem-solving algorithms that obtain a set of solutions aimed at reducing the CC of a given method.
    
    \item Carrying out experimental validation of the proposed multi-objective ILP model and the ILP CC reducer tool using a benchmark of methods from open-source and private industrial software projects in Java.
\end{itemize}

CC reduction will be achieved by working at method level and ensuring the threshold is below 15 for each of them, as suggested by SonarQube~\cite{ASCCR}.

The remainder of this paper is organized as follows. Section 2 motivates the challenge of reducing the CC of code and introduces the concepts and notation used in the rest of the document. Section 3 presents the formulation of the proposed multi-objective ILP model. The main algorithms available for two- and three-objective models are presented in this section, including the \emph{augmented $\epsilon$-constraint} algorithm and the \emph{hybrid method} algorithm. Section 4 shows the implementation of the ILP CC reducer tool at method level. It describes the functionalities of the tool's module that uses the ILP model, its architecture, and its implementation. Section 5 presents the experimental analysis, as well as the results obtained for two- and three-objective ILP models with the different algorithms developed. Section 6 revises the work related to refactoring and CC, as well as the contributions of this paper. Section 6 presents the conclusions derived from the experimental results and proposes future work, including potential extensions and improvements.

\section{Motivation}
To ensure clarity when illustrating the difficulties developers face when reducing the CC of code, we use the \texttt{routeAPacketTo} method in the \texttt{Host} class of the \texttt{cybercaptor-server}\footnote{\url{https://github.com/fiware-cybercaptor/cybercaptor-server.git} (last accessed June 2025)} open-source Java project. It is shown in Figure~\ref{fig:initial-running-example}, and it has CC = $20$, which is above the maximum recommended by SonarQube (CC~=~$15$).

\begin{figure}[!ht]
\begin{lstlisting}[columns=fullflexible, numbers=left,xleftmargin=0.5cm,xrightmargin=0cm,frame=single,framexleftmargin=0em,escapechar=ñ,stepnumber=1,
    showstringspaces=false,
    tabsize=1,
    breaklines=true,
    breakatwhitespace=false,
    basicstyle=\scriptsize,
    language=Java,
    mathescape=true]{}
public void routeAPacketTo(IPAddress ip, int ttl, List<Host> usedHosts) throws Exception {
    usedHosts.add(this);
    if (ttl == 0) {//Problem in routing
        throw new Exception("Routing problem...");
    }
    if (!hasIP(ip)) { //Packet not arrived
        Host nextHost = null;
        List<Host> directlyAccessibleHosts = getDirectlyAccessibleHosts();
        for (Host directlyAccessibleHost: directlyAccessibleHosts) {
            if (...) // If the packet is for a neighbour, we send it to him
                nextHost = directlyAccessibleHost;
        }
        if (nextHost != null) {
            nextHost.routeAPacketTo(ip, ttl - 1, usedHosts);
        } else {//We have to look in the routing table
            List<Host> directlyAccessible = getDirectlyAccessibleHosts();
            IPAddress nextIP = this.getRoutingTable().getNextHop(ip);
            boolean nextHostFound = false;
            for (Host aDirectlyAccessible: directlyAccessible) {
                if (...) { //Search the nextHop host object
                    aDirectlyAccessible.routeAPacketTo(ip, ttl - 1, usedHosts);
                    nextHostFound = true;
                }
            }
            if (!nextHostFound) { //Routing problem
                throw new Exception("Routing problem...");
            }
        }
    }
}
\end{lstlisting}
\caption[Method \texttt{routeAPacketTo} in the \texttt{Host} class of the project \texttt{cybercaptor-server}.]{Method \texttt{routeAPacketTo} in the \texttt{Host} class of the open-source project \texttt{cybercaptor-server}, used as running example in the project.}
\label{fig:initial-running-example}
\end{figure}

The number of different extract method refactoring opportunities of a method with $n$ sentences is bounded by $CR_{2}^{n} = \binom{n+2-1}{2} = \binom{n+1}{2}$, given that each extraction is delimited by its first and last sentences in the code. The running example in Figure~\ref{fig:initial-running-example} has $28$ lines of code, so an upper bound for the number of feasible code extractions is $\binom{29}{2} = 406$. Thus, analyzing all possible refactoring opportunities to find those that reduce the CC of a method under a given threshold becomes unmanageable for developers~\cite{ASCCR}. Furthermore, not all extractions are applicable, as they could break a control flow sentence (for instance, Lines 11 and 12 in Figure~\ref{fig:initial-running-example} cannot be extracted as a new method) or they do not contribute to the CC of the method (see Lines 14 and 16 in Figure~\ref{fig:initial-running-example}). For the running example, we computationally checked that there are $23$ applicable code extractions that reduce the initial CC of the method. However, if one needs to evaluate all the possible sequences of extract method refactoring operations, an upper bound is $2^{23}-1 = 8,388,607$ alternatives. 

As explained in the introduction, we previously modeled CC reduction as a single-objective integer linear programming problem. After surveying $36$ professional developers, they themselves confirmed that this type of solution was not adequate for improving code comprehension in some specific cases: developers preferred that the code also be balanced in terms of CC and LOC. It seemed natural then to incorporate these objectives into the problem while keeping the number of extractions in mind. Solving this problem would lead to solutions that maintain both the minimum possible number of code extractions while balancing the CC and LOC of extracted methods, thus allowing for clean, readable, and maintainable code.

\section{Background}
\label{sec:concepts}
The concepts used throughout the rest of the document are outlined in this section. There is a description and example of an Abstract Syntax Tree (AST) for code, which is useful for representing source code. Multi-objective integer linear programming is also introduced.

\subsection{Abstract Syntax Tree (AST)}
\label{subsec:AST}
The Abstract Syntax Tree (AST) is a data structure used in computer science to reflect the syntactic structure of a piece of code with a tree-like form. Each node in the AST represents an element that is present in the code, such as a variable declaration, a method call, a conditional sentence, a \texttt{for} or \texttt{while} loop, among others.

\begin{figure}[!ht]
\centering
\includegraphics[width=\linewidth]{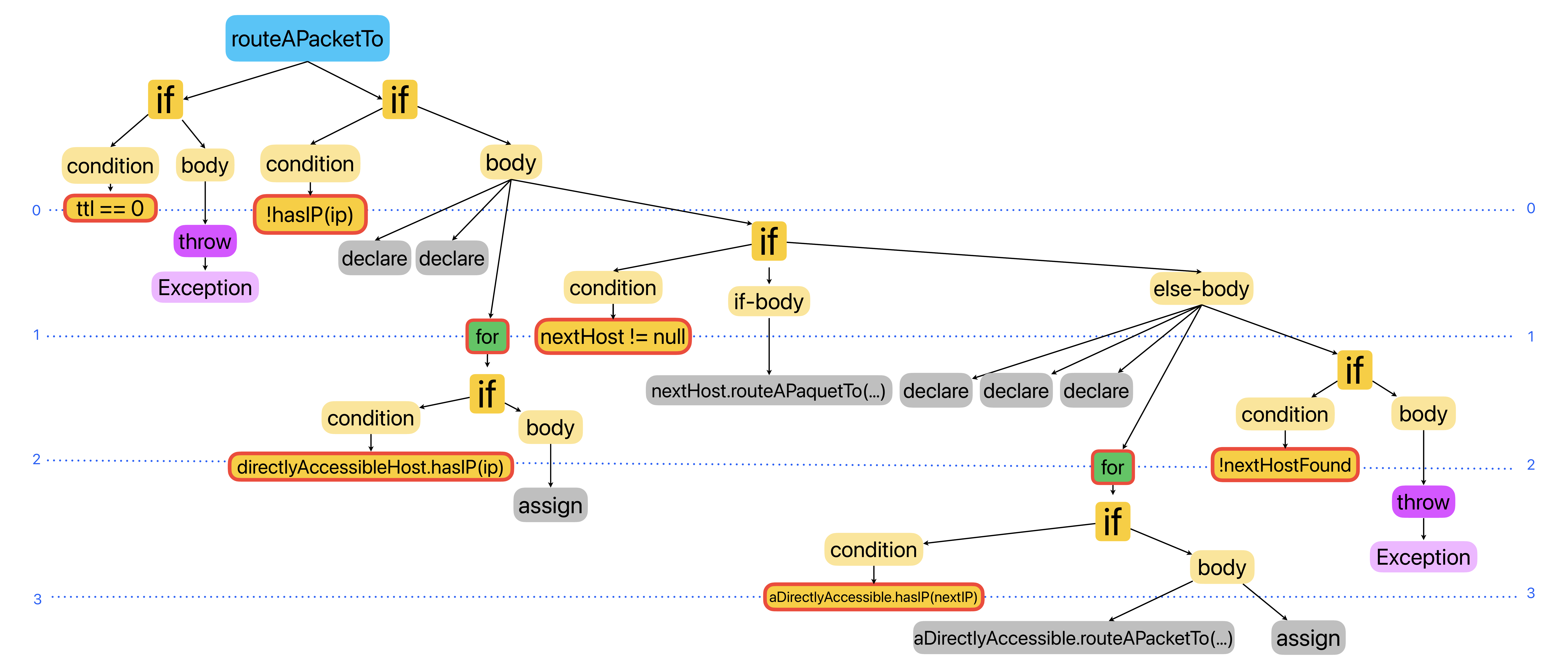}
\caption[AST for code in Figure~\ref{fig:initial-running-example}]{AST for code in Figure~\ref{fig:initial-running-example}.}
\label{fig:AST-tree}
\end{figure}

The AST for the running example code in Figure~\ref{fig:initial-running-example} can be found in Figure~\ref{fig:AST-tree}. Conditional sentences are shown in yellow, and \texttt{for} loops are shown in green in the example. Both conditional sentences and \texttt{for} loops contribute to the CC of a code, so they are delineated with a red border to reflect their role in program comprehension. The rest of the nodes in the AST do not contribute to the CC of the code. There are four nesting levels in this example, represented with horizontal blue dashed lines, which are also numbered from $0$ to $3$.

\subsection{Cognitive complexity (CC) of a code}
\label{subsec:CCreduction}
Cognitive Complexity is a relatively novel measure of how difficult a piece of code is to understand and handle intuitively. Unlike the more traditional Cyclomatic Complexity, which focuses on the number of possible execution paths in a program, CC shifts the attention toward human factors. This new concept focuses on understandability, i.e., how long it takes a developer to understand the code. Different metrics have been defined over the years to reduce it as much as possible. By minimizing CC, teams can improve collaboration, reduce onboarding time for new developers, and decrease the likelihood of introducing errors during future modifications.

\subsubsection{Measures for code complexity}
The software development community has long sought to predict the effort required for software development or quality assessment, leading to the proposal of various metrics. In 1983, Basili et al. conducted an empirical study to investigate the correlation between different measures and actual data~\cite{Basili1983}. Later in 1988, Weyuker presented abstract properties to formally compare software complexity models~\cite{Weyuker1988}. These properties allow developers to evaluate a large number of existing software metrics. Most metrics focus on method complexity, and then the complexities of each method are added. Weyuker states that the complexity of a program, as measured by the data flow, depends directly on the placement of statements and how the components interact via the potential flow of data. Kaner and Bond conducted an evaluation to validate proposed metrics until 2004, during which they criticized the way metrics programs are used. They attempted to evaluate measurements by answering 10 questions, such as their purpose, scope, and scale, among others~\cite{Kaner2004SoftwareEM}. Misra et al. proposed a suite of cognitive metrics for evaluating complexity for object-oriented code~\cite{Misra2012}. The progression observed over the years is that researchers have sought to define a formal metric to measure something subjective. Finally, it was possible to validate a measure that can be used to determine the effort spent by a developer to understand code.

\subsubsection{CC reduction}
Once the CC metric is defined, one would like to reduce it to a given threshold, resulting in a clean, easy-to-understand, and easy-to-manage code. CC reduction can be explained both qualitatively and quantitatively. Qualitatively, it is easier to understand a compact code by substituting a piece of code with a line that explains its functionality, i.e., the extracted method name. Recio Abad et al. found that the names offered by LLMs are highly acceptable~\cite{RSAINnamingMethods}. On the other hand, CC reduction can be explained quantitatively. If one subtracts the metrics corresponding to a target method, not only is the CC of that method reduced, but the sum of the CC of the refactored and its extractions is usually lower than the initial CC.

\subsection{Linear programming}
\label{sec:lienarProgramming}
Linear Programming aims to optimize a linear function subject to a set of inequalities~\cite{cormen2009}. Given a set of real numbers $a_1, a_2, ..., a_n$ and a set of variables $x_1, x_2, ..., x_n$, we define a linear function $f$ on those variables by 
\begin{equation}
    f(x_1, x_2, ..., x_n) = a_1 x_1 + a_2 x_2 + ... + a_n x_n = \sum_{j=1}^{n} a_j x_j .
\end{equation}

If $b$ is a real number and $f$ is a linear function, the equation $f(x_1, x_2, ..., x_n) = b$ is a linear equality, while $f(x_1, x_2, ..., x_n) \leq b$ and $f(x_1, x_2, ..., x_n) \geq b$ are linear inequalities. We use the general term linear constraints to denote either linear equalities or inequalities. In linear programming, we do not allow strict inequalities, but they are usually easy to transform into non-strict inequalities. A linear programming problem consists in minimizing or maximizing a linear function subject to a finite set of linear constraints. If we are to minimize, then we refer to the linear program as a minimization linear program. If we are to maximize, then we call the linear program a maximization linear program.

An integer linear programming problem (ILP) is a linear programming problem with the additional constraint that the variables $x$ must take integral values. Determining whether an integer linear program has a feasible solution is NP-hard, which means there is no known polynomial-time algorithm to solve this problem.
A mixed integer linear programming problem (MILP) is a linear programming problem that incorporates both integer and continuous variables.

\subsection{Multi-objective optimization}
\label{subsec:background_ILP}
It is common to find problems in the real world where we need to optimize multiple objectives simultaneously. In these cases, the objectives often conflict, as optimizing one typically results in the rest deteriorating. The goal then is to find a set of solutions that satisfy every constraint, where no objective can be improved without worsening at least one of the others. A multi-objective optimization problem can be formulated as follows~\cite{backgroud_rsain_FP_MO-ILP}:

\begin{align}
    \text{minimize}& \qquad \mathbf{f} (\mathbf{x}) = \left( f_1(\mathbf{x}), f_2(\mathbf{x}), ... ,f_k(\mathbf{x}) \right)^T \\
    \text{s.t}& \qquad \mathbf{x} \in S,
\end{align}

\noindent where:
\begin{itemize}
    \item $\mathbf{x} = \left( x_1, x_2, ..., x_n \right)^T \in S$ is the vector of decision variables.
    \item $S \subset \R^n$ is the decision space of feasible solutions.
    \item $f_i: S \rightarrow \R, \quad i = 1, ...,k,$ with $k \geq 2$ are the objective functions to be optimized.
    \item The image of $S$ in $\R^k$, i.e. $Z = \mathbf{f}(S) \subset \R^k$, is the objective space.
    \begin{itemize}
        \item The objective space is composed of the objective vectors as $\mathbf{f}(\mathbf{x})$, for any $\mathbf{x} \in S$.
    \end{itemize}
\end{itemize}

\subsubsection{Pareto front}
Since the objectives are often conflicting, finding a single decision vector (solution) that optimizes all the objectives simultaneously is usually impossible. Thus, much attention is given to the so-called Pareto optimal solutions for which none of the objectives can be improved without deteriorating at least one of the others.
The set of Pareto optimal solutions in the decision space is called the Pareto optimal set, and its image in the objective space is the Pareto front (PF)~\cite{backgroud_rsain_FP_MO-ILP}. Each element of the Pareto front is a point, but it is usually referred to as a ``solution'' due to a common abuse of language.

\begin{figure}[!ht]
    \centering
    \includegraphics[width=0.7\linewidth]{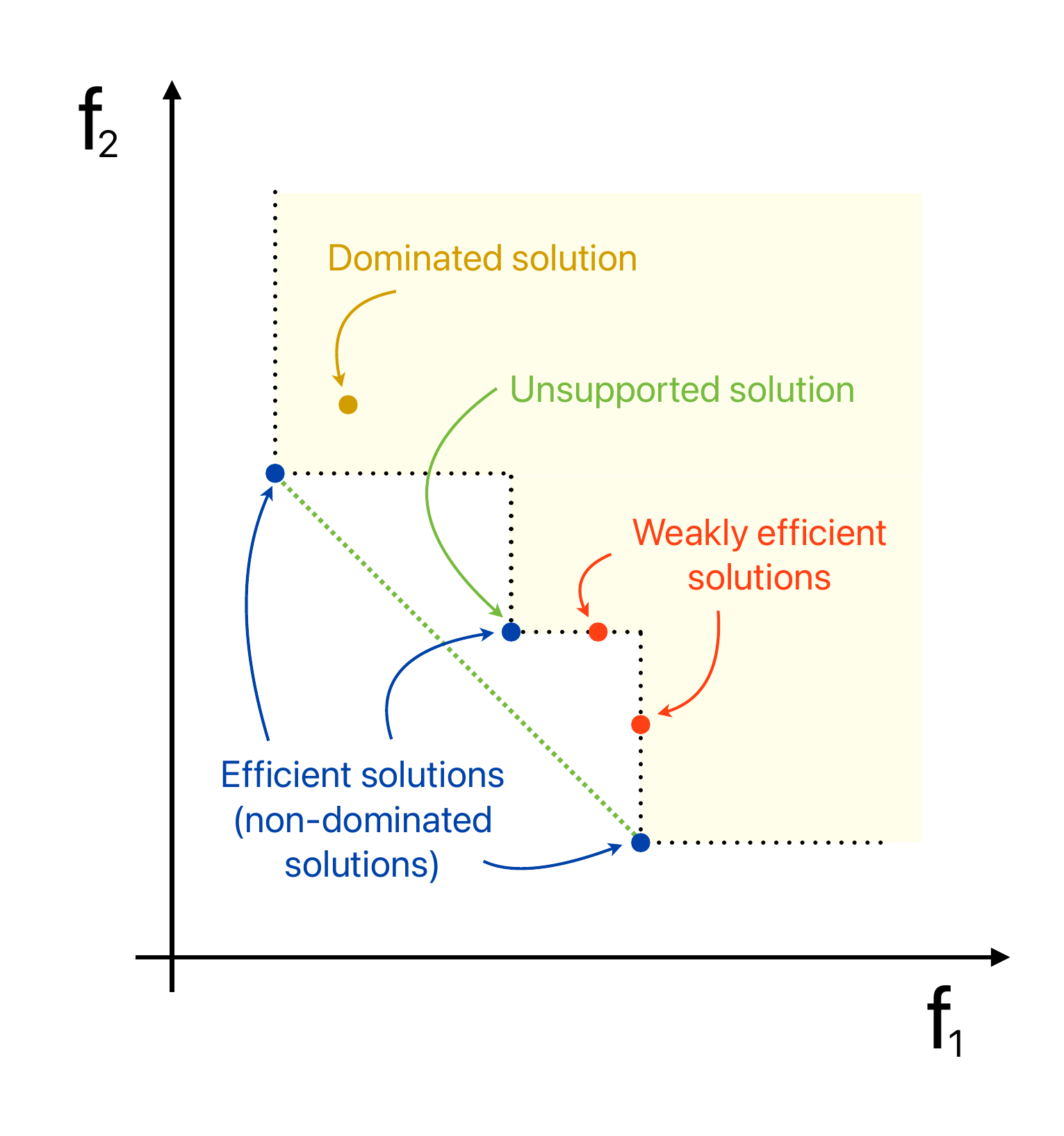}
\caption[Objective space example and types of solutions in PF]{Objective space example and types of solutions that may appear in the objective space. Dominated solutions do not belong to the PF.}
\label{fig:example_pareto_front}
\end{figure}

Now we define some important multi-objective concepts using Figure~\ref{fig:example_pareto_front}, which is an example of a PF in the objective space. For $\mathbf{z}, \mathbf{\overline{z}} \in \R^k$:
\begin{itemize}
    \item \textbf{Domination relation}: $\mathbf{z}$ \emph{dominates} $\mathbf{\overline{z}}$ if and only if $z_i \leq \overline{z}_i$ for all $i=1,...,k$, and $z_j < \overline{z}_j$ for, at least, one index $j$. The left-most blue solution in Figure~\ref{fig:example_pareto_front} dominates the brown solution (the brown solution is a \emph{dominated} solution).
    \item \textbf{Efficient solution} (see the blue solutions in Figure~\ref{fig:example_pareto_front}): one solution is efficient if there is no other solution in the search space that dominates it.
    \item \textbf{Unsupported efficient solution} (see the green solution in Figure~\ref{fig:example_pareto_front}): $\mathbf{z}$ is an unsupported efficient solution if it cannot be obtained by minimizing a linear combination of the objectives. Therefore, unsupported solutions, although they are non-dominated solutions, are not found in the convex envelope of the PF.
    \item \textbf{Weak domination relation}: $\mathbf{z}$ \emph{weakly dominates} $\mathbf{\overline{z}}$ if $z_i \leq \overline{z}_i$ for all $i=1,...,k$.
    \item \textbf{Strict domination relation}: $\mathbf{z}$ \emph{strictly dominates} $\mathbf{\overline{z}}$ if $z_i < \overline{z}_i$ for all $i=1,...,k$.
    \item \textbf{Weakly efficient solution}: A solution $z$ is \emph{weakly efficient} (red and blue solutions in Figure~\ref{fig:example_pareto_front}) if it is not strictly dominated by any other solution in the search space. All the efficient solutions are also weakly efficient.
\end{itemize}

A non-dominated solution set is a set of solutions whose objective vectors are not dominated among them. For two non-dominated solution sets $A$ and $B$ (in the objective space $\R^k$), we say that $A$ \emph{strictly dominates} $B$ (or that $A$ is \emph{better} than $B$) if, for every $\mathbf{z} \in B$, there is at least one $\mathbf{y} \in A$ such that $\mathbf{y}$ weakly dominates $\mathbf{z}$ (and this means that $A$ \emph{weakly dominates} $B$), but there exists at least one solution in $A$ that is not weakly dominated by any solution in $B$ (i.e. $B$ does not weakly dominate $A$).

\subsection{Notation}
CC can be computed as the sum of two components, the \emph{inherent component} and the \emph{nesting component}. The inherent component adds $+1$ for each occurrence of certain control flow structures and complex expressions, such as logical operators in conditionals and loops. The nesting component depends on the depth at which a specific control flow structure appears in the code related to the method declaration.

Let $s_i,s_j$ and $e_i,e_j$ be the start and end offsets of the $i$th and $j$th code extractions, respectively. We consider the $j$th extraction is nested in the $i$th extraction, denoted with $j \rightarrow i$, when $[s_j,e_j] \subset [s_i,e_i]$. In that case, we say that the $j$th extraction is the \emph{descendant} of the $i$th extraction, and that the $i$th extraction is the \emph{ancestor} of the $j$th extraction. Code that is nested inside another, which in turn is nested inside a third, is ultimately also nested within the third. Regarding the running example in Figure~\ref{fig:initial-running-example}, the first \texttt{for} loop in Lines 8--10 is nested in the second conditional in Lines 5--22. As we are working at method level, the code associated with the body of the original method can be defined as the $0$th extraction. Thus, $i \rightarrow 0$ can be written for all $i \geq 1$, which means that all extractions are nested in the body of the original method.

We consider that the $i$th extraction is in conflict with the $j$th extraction, denoted with $i \not\leftrightarrow j$ when $i$ and $j$ are not nested one in the other, and $[s_j,e_j] \cap [s_i,e_i] \neq \varnothing$. Two extractions cannot be extracted simultaneously if they are in conflict. In the running example, the extraction between Lines 16--24 conflicts with the extraction between Lines 17--27, and therefore, these two extractions cannot be extracted simultaneously.

Next, we define new variables to facilitate formal CC reduction. To better understand the use of each variable, Figure~\ref{fig:ccExampleInCode} shows them as comments in the running example presented in Figure~\ref{fig:initial-running-example}. We only consider the variables of those feasible extractions that contribute to the CC of the method. With this notation, the $0$th extraction is the complete method that we want to refactor. Those variables are defined as follows.

The nesting component of the $i$th feasible extraction is denoted by $\lambda_i$. Note that $\lambda_i = 0$ when no nesting exists in extraction $i$. $\lambda_0 = 0$ is the depth of the complete method. Referring to the running example, the conditional sentence in Line 5 of Figure~\ref{fig:ccExampleInCode} has $\lambda=0$ because its nesting level is zero, but the conditional sentence in Line 16 has $\lambda=2$ because it is nested in the \texttt{for} loop in Line 14, and that \texttt{for} loop is in turn nested in the conditional sentence in Line 10.

The accumulated inherent component of the $i$th extraction and the ones contained by it are denoted by $\iota_i$. Referring to the running example, the conditional sentence in Line 37 found in Figure~\ref{fig:accumulated_inherent_comp} has $\iota=1$ because the unique contribution to the inherent component inside that sentence is itself. But the conditional sentence in Line 21 has $\iota=5$ because inside that sentence there are the \texttt{else} sentence of Line 24 (with $\iota=4$), the \texttt{for} loop in Line 29 (with $\iota=2$), the conditional sentence in Line 31 (with $\iota=1$), and the conditional sentence in Line 37 (with $\iota=1$).

The accumulated nesting component of the extractions contained by the $i$th extraction (considering nesting $0$ for the $i$th extraction) is denoted as $\nu_i$. The total accumulated nesting component of the complete method, $\nu_0$, is obtained by adding every contribution $\nu_i$ to the zero extraction (that is, the complete method). Referring to the running example, the \texttt{for} loop in Line 29 found in Figure~\ref{fig:accumulated_nesting_comp} contributes to the accumulated nesting component of the method with $+2$ because its nesting level is $\lambda = 2$. On the other hand, the conditional sentence in Line 31 contributes to the accumulated nesting component of the method in the running example with $+3$ because its nesting level is $\lambda = 3$. Both sentences contribute to the accumulated nesting component of the method with $ \nu_i = +2+3 = +5$.

The number of extractions that contribute to the accumulated nesting component of the $i$th feasible extraction, including itself when its nesting component is not $0$, is denoted by $\mu_i$. Referring to the running example, the \texttt{for} loop in Line~29 found in Figure~\ref{fig:mu_contribution_to_nesting_comp} has $\mu=2$ because both the conditional sentence inside it (Line 31) and the \texttt{for} loop itself contribute to the nesting level. Hence, they are two sentences contributing, then $\mu=2$. In the case of the \texttt{else} sentence in Line 24 in Figure~\ref{fig:mu_contribution_to_nesting_comp}, $\mu=3$ because it has three sentences that contribute to the nesting level. They are the \texttt{for} loop in Line 29, the conditional sentence in Line~31, and the conditional sentence in Line 37. Note that the \texttt{else} sentence does not contribute to the nesting level~\cite{SSCC_sonar}, even though this sentence nesting level is different from $0$. That is why there is no contribution to $\mu$ in this \texttt{else} sentence, as it can be observed in Figure~\ref{fig:mu_contribution_to_nesting_comp}. In the case of the conditional sentence in Line~5, its $\mu$ value is $0$ because it does not contribute to the nesting level and it does not have any feasible extractions that contribute to the nesting level inside it.

Once every variable is obtained as explained above, they can be combined to define $CCR_{j \to i}$ and $NMCC_i$ as the cognitive complexity reduction of sequence $i$ when extracting sequence $j$, with $j \to i$,  and the cognitive complexity of the new method when extracting sequence $i$, respectively. The formal definition of these two last variables is, respectively, $CCR_{j \rightarrow i} = \iota_j + \nu_j + (\lambda_j - \lambda_i) \mu_j$ and $NMCC_i = \iota_i + \nu_i$.

Note that if one extraction is not nested, then $NMCC_j = CCR_{j \to i}$, since $\lambda_j$ would be $0$ in that case. That is why one would like to extract a sequence of lines of code that is deeply nested but also has a high $NMCC$, because in that case, both the original method and the extracted method would experience a large CC reduction.


\begin{figure}[!ht]
\begin{lstlisting}[columns=fullflexible, numbers=left,xleftmargin=0.5cm,xrightmargin=0cm,frame=single,framexleftmargin=0em,escapechar=ñ,stepnumber=1,
    showstringspaces=false,
    tabsize=1,
    breaklines=true,
    breakatwhitespace=false,
    basicstyle=\scriptsize,
    language=Java,
    mathescape=true]{}
// Cognitive complexity 20
public void routeAPacketTo(IPAddress ip, int ttl, List<Host> usedHosts) throws Exception {
    usedHosts.add(this);
    // [$\color{pgreen} \lambda$=0, $\color{pgreen} \iota$=1, $\color{pgreen} \nu$=0, $\color{pgreen} \mu$=0, CCR=1, NMCC=1]
    if (ttl == 0) {//Problem in routing
        throw new Exception("Routing problem, TTL is null : packet to " 
                            + ip + " deleted on host " + this.getName());
    }
    // [$\color{pgreen} \lambda$=0, $\color{pgreen} \iota$=8, $\color{pgreen} \nu$=11, $\color{pgreen} \mu$=6, CCR=19, NMCC=19]
    if (!hasIP(ip)) { //Packet not arrived
        Host nextHost = null;
        List<Host> directlyAccessibleHosts = getDirectlyAccessibleHosts();
        // [$\color{pgreen} \lambda$=1, $\color{pgreen} \iota$=2, $\color{pgreen} \nu$=1, $\color{pgreen} \mu$=2, CCR=5, NMCC=3]
        for (Host directlyAccessibleHost: directlyAccessibleHosts) {
            // [$\color{pgreen} \lambda$=2, $\color{pgreen} \iota$=1, $\color{pgreen} \nu$=0, $\color{pgreen} \mu$=1, CCR=3, NMCC=1]
            if (directlyAccessibleHost.hasIP(ip)) // If the packet is for a neighbour,
                                                  // We send it to him
                nextHost = directlyAccessibleHost;
        }
        // [$\color{pgreen} \lambda$=1, $\color{pgreen} \iota$=5, $\color{pgreen} \nu$=4, $\color{pgreen} \mu$=4, CCR=13, NMCC=9]
        if (nextHost != null) {
            nextHost.routeAPacketTo(ip, ttl - 1, usedHosts);
        // [$\color{pgreen} \lambda$=1, $\color{pgreen} \iota$=4, $\color{pgreen} \nu$=4, $\color{pgreen} \mu$=4, CCR=13, NMCC=9]
        } else {//We have to look in the routing table
            List<Host> directlyAccessible = getDirectlyAccessibleHosts();
            IPAddress nextIP = this.getRoutingTable().getNextHop(ip);
            boolean nextHostFound = false;
            // [$\color{pgreen} \lambda$=2, $\color{pgreen} \iota$=2, $\color{pgreen} \nu$=1, $\color{pgreen} \mu$=2, CCR=7, NMCC=3]
            for (Host aDirectlyAccessible: directlyAccessible) {
                // [$\color{pgreen} \lambda$=3, $\color{pgreen} \iota$=1, $\color{pgreen} \nu$=0, $\color{pgreen} \mu$=1, CCR=4, NMCC=1]
                if (aDirectlyAccessible.hasIP(nextIP)) { //Search the nextHop host object
                    aDirectlyAccessible.routeAPacketTo(ip, ttl - 1, usedHosts);
                    nextHostFound = true;
                }
            }
            // [$\color{pgreen} \lambda$=2, $\color{pgreen} \iota$=1, $\color{pgreen} \nu$=0, $\color{pgreen} \mu$=1, CCR=3, NMCC=1]
            if (!nextHostFound) { //Routing problem
                throw new Exception("Routing problem, there is no route corresponding 
                                     to the packet or the destination host is on the internet");
            }
        }
    }
}
\end{lstlisting}
\caption[Method \texttt{routeAPacketTo} in the \texttt{Host} class of the project \texttt{cybercaptor-server}.]{Method \texttt{routeAPacketTo} in the \texttt{Host} class of the open-source project \texttt{cybercaptor-server}, used as running example in the project. Comments show CC metrics of those statements contributing to the cognitive complexity of the method.}
\label{fig:ccExampleInCode}
\end{figure}

\section{Multi-objective ILP model and resolution algorithms}
\label{sec:modifiedILPproblem}
The formulation of the multi-objective ILP model begins with the model developed by Saborido et al.~\cite{ASCCRthroughtILP}. They use a zero-one program, that is, an integer program in which the variables can only take values 0 or 1. This initial ILP problem is enhanced by adding new objectives and constraints to improve its operation on a code. The reduction of CC to a given threshold is then formulated as a multi-objective optimization problem. The original model minimizes the number of extractions from a method while ensuring that all extractions in the code, as well as the original method, have CC lower than or equal to a given threshold $\tau$ after the application of the approach. SonarQube suggests that $\tau = 15$. New objectives added to the initial problem allow the developer to refactor the code by reducing the CC while balancing the LOC and the CC of the extracted methods. It is usually impossible to find a single solution that optimizes all objectives simultaneously, and much attention is given to the so-called Pareto optimal solutions, for which none of the objectives can be improved without worsening at least one of the others.

The vector of decision variables in the multi-objective ILP problem is the vector formed by every feasible extraction, and it is described as follows:

\begin{align}
& \label{eqn:extractions_objective}\min \sum_{i=1}^n x_i \\
\label{eqn:cc_objective} & \min c_M - c_m \\
\label{eqn:loc_objective} & \min t_M - t_m \\
\intertext{subject to:}
\label{eqn:conflict} & x_i + x_j \leq 1 \;\; \forall i \nleftrightarrow j  \\
\label{eqn:limit} &  \underbrace{NMCC_i \cdot x_i}_{\text{Term A}} - \underbrace{\sum_{j, j \rightarrow i} CCR_{j \to i} \cdot z_{ji}}_{\text{Term B}}  \leq \tau \;\; \forall i=0,1, \ldots, n \\
\label{eqn:zdef} & z_{ji} + \abs{\left\{l\middle|j \rightarrow l \rightarrow i \right\}} (z_{ji}-1) \leq x_j - \sum_{l, j \rightarrow l \rightarrow i} x_l   \;\; \forall j \rightarrow i  \\
\label{eqn:cMaxdef} & c_M \geq NMCC_i \cdot x_i - \sum_{j, j \rightarrow i} CCR_{j \to i} \cdot z_{ji} \;\; \forall i=0,1, \ldots, n \\
\label{eqn:cmindef} & c_m \leq \tau (1 - x_i) + NMCC_i \cdot x_i - \sum_{j, j \rightarrow i} CCR_{j \to i} \cdot z_{ji} \;\; \forall i=0,1, \ldots, n \\
\label{eqn:tMaxdef} & t_M \geq LOC_i \cdot x_i - \sum_{j, j \rightarrow i} LOC_j \cdot z_{ji} \;\; \forall i=0,1, \ldots, n \\
\label{eqn:tmindef} & t_m \leq LOC_{0} (1 - x_i) + LOC_i \cdot x_i - \sum_{j, j \rightarrow i} LOC_j \cdot z_{ji} \;\; \forall i=0,1, \ldots, n \\
\label{eqn:x0} & x_0 = 1  \\
\label{eqn:domain-x} & x_i \in \{0,1\} \;\; \forall i=0, 1, 2, \ldots, n  \\
\label{eqn:domain-z} & z_{ji} \in \{0,1\} \;\; \forall j \rightarrow i  \\
\label{eqn:domain-tc}& t_M, t_m, c_M, c_m \in \R^{+}
\end{align}
where the goal is to minimize the objective functions defined by Eq.~\eqref{eqn:extractions_objective}, that is, the number of extractions, Eq.~\eqref{eqn:cc_objective}, that describes the differences between the highest and the lowest CC of all the extracted methods, and Eq.~\eqref{eqn:loc_objective}, that describes the differences between the highest and the lowest LOC of all the extracted methods. For Eqs.~\eqref{eqn:cc_objective} and~\eqref{eqn:loc_objective}, $c_M$, $t_M$ (respectively $c_m$, $t_m$), are the maximum (respectively minimum) CC and LOC reached by all the extracted methods, respectively. These two new objectives ensure that both the amount of CC and LOC are balanced after the extract method refactoring is performed.

From the base model~\cite{ASCCRthroughtILP}, we know that $n$ is the number of feasible extractions and $x_i$ is a binary variable that takes the value $1$ if the $i$-th extraction is extracted, and $0$ otherwise. Eq.~\eqref{eqn:conflict} ensures that at most one of two feasible extractions in conflict will be selected in the final solution. It avoids choosing two feasible extractions that are in conflict simultaneously. Eq.~\eqref{eqn:limit} limits the CC of all the extractions in the code after applying the refactoring operations to be lower or equal to $\tau$. On the left-hand side, we find the CC of the $i$th extraction when extracted as a new function/method (Term A) minus the CC removed from extraction $i$ due to code extractions $j$ contained in $i$ (Term B). If there were two feasible extractions $k$ and $j$ with $k \to j \to i$ that are extracted, one should only subtract the $j$th extraction CC contribution, and not $k$. The reason is that the contribution of extraction $k$ is already considered in the contribution of $j$. For this reason, $x$ variables are not used in the expression of the CC reduction (Term~B). Instead, new variables $z_{ji}$ are defined for $j \to i$ that are one if and only if the $j$th feasible extraction is selected and no other extraction strictly between $j$ and $i$ is selected. If $z_{ji} = 1$, then one should subtract from the CC of the $i$th extraction the CC of the $j$th extraction. This is what Eq.~\eqref{eqn:limit} does. Variables $z$ are completely determined by the values of the $x$ variables. Eq.~\eqref{eqn:zdef} expresses the relationship between the $z$ variables and the $x$ variables. If $z_{ji} = 1$, then $x_j = 1$ and $x_l = 0$ for all $l$ with $j \to l \to i$. If $z_{ji} = 0$, then there is no constraint for the $x$ variables. Eq.~\eqref{eqn:x0} forces the CC reduction of the original function/method (represented with the $0$th extraction). Finally, Eqs.~\eqref{eqn:domain-x} and~\eqref{eqn:domain-z} define the binary domain of the variables.

Eqs.~\eqref{eqn:cc_objective}, \eqref{eqn:loc_objective}, \eqref{eqn:cMaxdef}, \eqref{eqn:cmindef}, \eqref{eqn:tMaxdef}, \eqref{eqn:tmindef}, and~\eqref{eqn:domain-tc} were added to the single-objective model to add the new objectives. Eq.~\eqref{eqn:cc_objective} is the objective of balancing the CC throughout all the extractions after the extract method refactoring is performed, and Eq.~\eqref{eqn:loc_objective} does the same for balancing the LOC. Constraints~\eqref{eqn:cMaxdef} and~\eqref{eqn:cmindef} make $c_M$ the upper bound of the CC of the extracted methods and $c_m$ the lower bound. Similarly, Constraints~\eqref{eqn:tMaxdef} and~\eqref{eqn:tmindef} make $t_M$ and $t_m$ the upper and lower bounds of the LOC of the extracted methods, respectively. Minimizing the difference between both variables, both for CC and LOC, means that in an optimal solution, these variables become maximums and minimums, respectively.

This model allows a family of seven optimization problems to be solved for each non-empty combination of objectives from Eqs.~\eqref{eqn:extractions_objective}~-~\eqref{eqn:cc_objective}. If Objective~\eqref{eqn:cc_objective} is not used in the model, then Constraints~\eqref{eqn:cMaxdef} and~\eqref{eqn:cmindef} might be excluded from it. Similarly, if Objective~\eqref{eqn:loc_objective} is excluded, then Constraints~\eqref{eqn:tMaxdef} and~\eqref{eqn:tmindef} would not be added to this model.

\subsection{Multi-objective problems resolution algorithms}
\label{subsec:mo-algorithms}
The algorithms implemented to solve the ILP problem are presented below. First, a weighted sum algorithm has been developed to find the supported solution of each PF. Second, an augmented $\varepsilon$-constraint and a hybrid method have been designed to obtain all efficient solutions of the multi-objective ILP problem.

The weighted sum algorithm solves a single-objective problem where the objective to optimize is a weighted sum of the original objectives in the multi-objective problem. Unsupported solutions, i.e., solutions that don't belong to the convex envelope of the PF, are not reached with this algorithm. Furthermore, with the weighted sum algorithm, many runs may be redundant, as numerous combinations of weights can yield the same efficient extreme solution~\cite{MAVROTAS2009455}. Algorithm~\ref{alg:weighted-sum-algorithm} shows the implementation of the weighted sum algorithm for $p$ objectives.

An efficient algorithm to solve the multi-objective ILP problem is the one proposed by Mavrotas~\cite{MAVROTAS2009455}. It is known as \emph{augmented $\varepsilon$-constraint} or AUGMECON, and it is shown in Algorithm~\ref{alg:augmecon}. This algorithm solves one sub-problem in each iteration, and each solution obtained is guaranteed to be efficient. 

Finally, an efficient algorithm to solve the multi-objective ILP problem proposed in Section~\ref{sec:modifiedILPproblem} is a simplification that we made from the MOCO problem-solving algorithm using full p-split~\cite{TPA-MiguelAngel}. This simplification is illustrated in Algorithm~\ref{alg:hybrid-method}, which utilizes a filter that identifies boxes contained within other boxes, as shown in Algorithm~\ref{alg:redundancy-elimination}. Full p-split is a method for decomposing the search space based on the objective space, utilizing an algorithm called the \emph{hybrid method}~\cite{guddat1985multiobjective}.
\section{ILP CC reducer approach}
We proposed an \texttt{ILP CC reducer tool} consisting of a solver that first takes a CSV file containing the \emph{refactoring cache of a method}, that is, a list of extract method refactoring opportunities associated with a method, to reduce its CC. This file contains relevant information about candidate code extractions over the original method. The refactoring cache of a method is generated from the tool developed by Saborido et al.~\cite{ASCCR}.

Second, the \texttt{refactoring cache processing module} automatically generates the following four CSV files, which are needed for the \texttt{ILP CC reducer module}, (i) \texttt{MethodName\footnote{\texttt{MethodName} is the name of the method under processing and not the literal ``MethodName''.}\_extractions.csv} stores the index of each feasible extraction ($i$), the $LOC_i$ of the extraction, and the $NMCC_i$ of each extraction. There is a fourth column that is currently unused, but could be helpful for future work. That column stores the number of parameters that each feasible extraction would need when called, (ii) \texttt{MethodName\footnotemark[\value{footnote}]\_nested.csv} stores the indices of the child extractions ($j$), the indices of their parents ($i$), and the $CCR_{j\rightarrow i}$ for each relationship, (iii) \texttt{MethodName\footnotemark[\value{footnote}]\_conflict.csv} stores the extractions in conflict. The conflict extractions are related in pairs in the two columns of the file, and (iv) \texttt{MethodName\footnotemark[\value{footnote}]\_feasible\_extractions\_offsets.csv} stores the index of each extraction ($i$), its start offset, and its end offset. These offsets are used to indicate the start and end positions of code extractions within the original file.

Thirdly, using the generated CSV files, the \texttt{ILP CC reducer tool} models the ILP optimization problem and solves it. This module takes as inputs from the user (i) the number of objectives, which sets the number of objectives to consider (from one to three), (ii) the algorithms selected for solving the ILP problem, that is \texttt{ObtainResultsAlgorithm} for the single-objective ILP problem, and \texttt{WeightedSumAlgorithm} (the user can specify the weights for each objective or the number of combinations of weights if desired), \texttt{EpsilonConstraintAlgorithm}, and \texttt{HybridMethodAlgorithm} for multi-objective optimization ILP problems, (iii) the threshold, i.e., the maximum CC that all extractions must have after the extraction has been performed, including the original method, and finally (iv) an objectives list, which is an optional input, and that is an ordered list of the objectives, between $EXTRACTIONS$, $CC$ and $LOC$ (the default order is \texttt{extractions}, \texttt{extractions} and \texttt{CC}, or \texttt{extractions}, \texttt{CC} and \texttt{LOC}, for one, two or three objectives, respectively).

\begin{figure}[!ht]
\centering
\includegraphics[width=\linewidth]{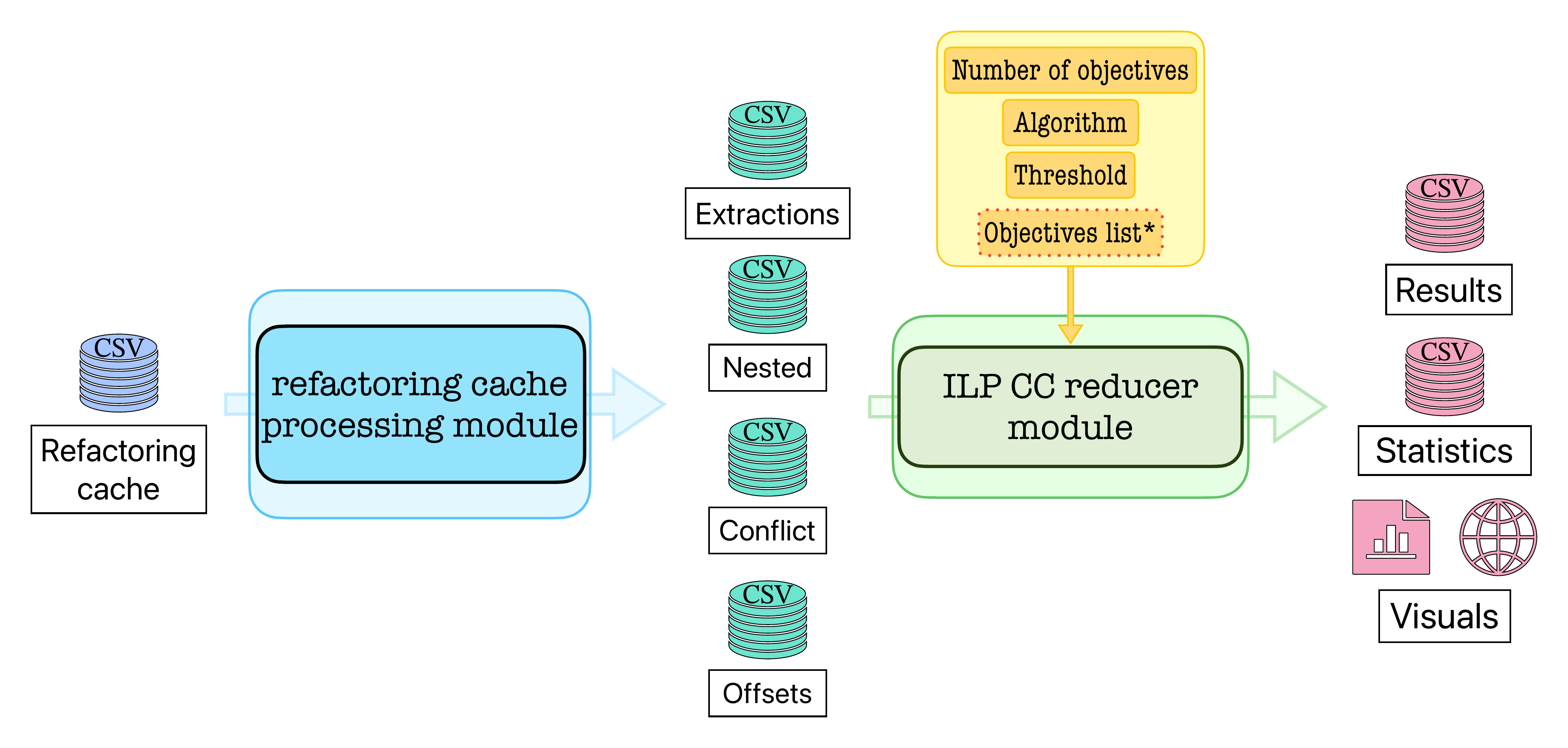}
\caption{Overview of the proposed ILP CC reducer tool. *Objectives list is optional.}
\label{fig:overview}
\end{figure}

The tool generates the corresponding model based on the parameters specified in the input. As shown in Figure~\ref{fig:overview}, the output is a CSV file containing the set of optimal solutions, statistics for each solution, and various visual content depending on the selected algorithm. One optimal solution to the ILP problem is a sequence of extractions that has a given impact on the user-specified objectives and that reduces the CC of the initial method below the given threshold.

The ILP CC reducer tool is developed in Python, using Pyomo to implement the ILP problem modeling. The solver used for resolving the built ILP model is CPLEX. Pyomo\footnote{\url{https://pyomo.readthedocs.io/en/stable/} (last accessed November 2025)} is an open-source Algebraic Modeling Language (AML) used for mathematical optimization. An AML is a language for expressing optimization models concisely and naturally, avoiding implementation details. It aims to standardize the way optimization problems are modeled, making it easier to exchange models between different users and communities. CPLEX\footnote{\url{https://www.ibm.com/es-es/products/ilog-cplex-optimization-studio} (last accessed November 2025)} is a prescriptive analytics solution that accelerates the development and implementation of decision optimization models using mathematical and constraint programming. It has excellent compatibility with Python. Moreover, it is a software with a long history of development, which has enabled it to achieve a high degree of maturity, robustness, and efficiency in resolving optimization problems.

\begin{figure}[!ht]
\centering
\includegraphics[width=\linewidth]{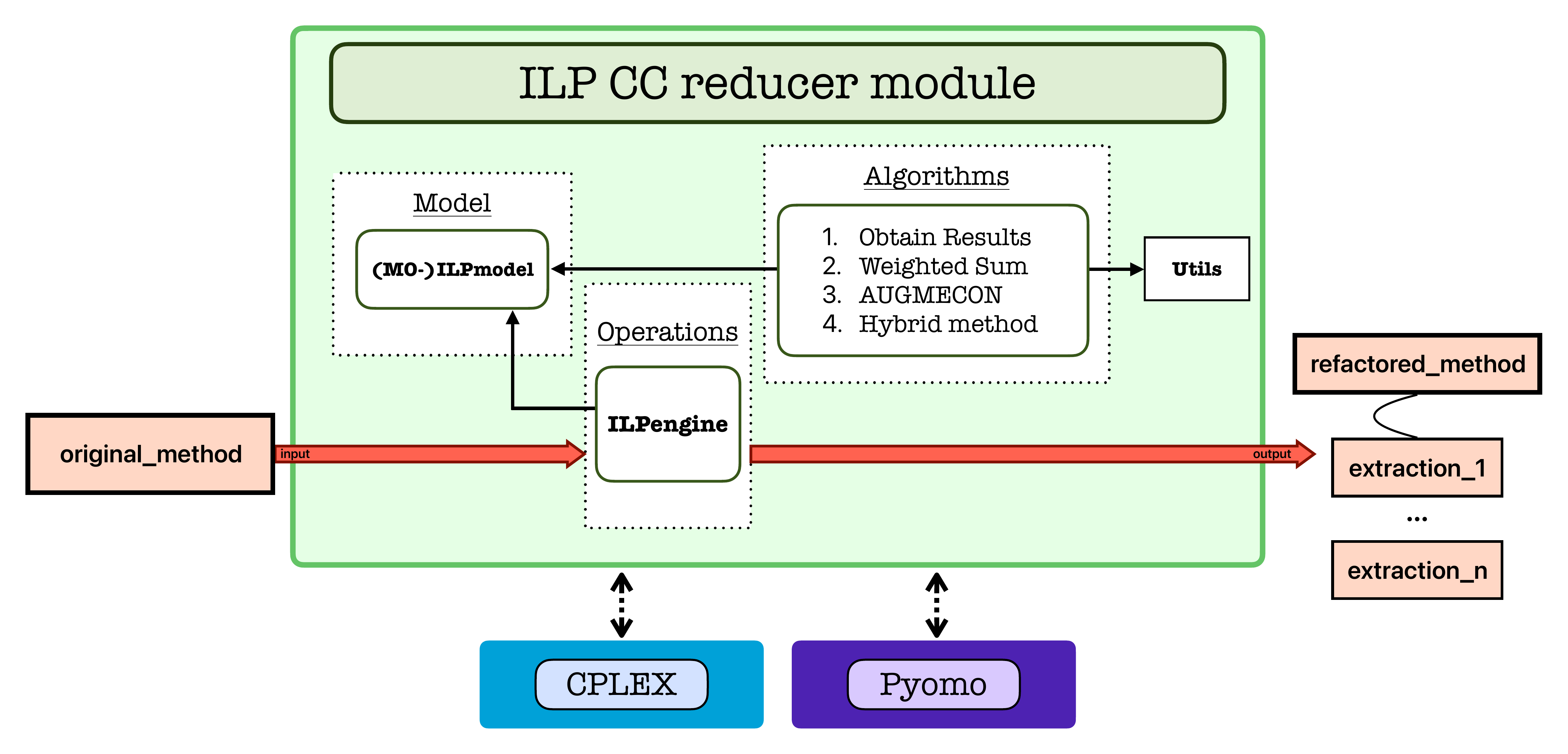}
\caption{Architecture of the ILP CC reducer module.}
\label{fig:architecture-diagram}
\end{figure}

The core of the ILP CC reducer tool is the \texttt{ILP CC reducer module}. Its architecture is shown in Figure~\ref{fig:architecture-diagram}, which is organized into three main modules that are (i) model, defining the corresponding combination of objective or objectives and constraints, (ii) ILPengine, that is, the main component that connects models and algorithms and manages the execution flow, and (iii) algorithms, implementing the optimization strategies explained in Subsection~\ref{subsec:mo-algorithms}. The rest of the modules presented in Figure~\ref{fig:architecture-diagram} are (i) utils, i.e., auxiliary tools for managing input and output data, (ii) \texttt{original\_method}/\texttt{refactored\_method} contain the original and refactored methods from a functional perspective, and that are the input and output of the module, respectively, (iii) Pyomo, which is the modeling framework used to build the mathematical models, and (iv) CPLEX, used as a solver for ILP problems.
\section{Empirical validation}
Since the ILP problem is parametrizable, different solutions can be obtained depending on the combination of objectives and the number of objectives selected. In this section, only the combination of \texttt{extractions} and \texttt{CC} objectives is solved for two objectives, due to space constraints. The remaining combinations of objectives can be found in our replication package. For three objectives, the problem is solved in the order \texttt{extractions}, \texttt{CC}, and \texttt{LOC}, but the order of objectives does not affect the result.

This section presents the results obtained for the hybrid method algorithm with \emph{full p-split}, which is chosen for its accuracy and versatility. It was used to solve the multi-objective ILP problem for 10 methods from an industry project at Ayesa Advanced Digital Services Limited (ASDA) and 1050 methods from 10 open-source projects. All result plots and three-dimensional Pareto fronts (PF) can also be viewed interactively in the replication package.

\subsection{Replication package}
\label{subsec:replication-package}
The complete tool resulting from this paper can be found on GitHub\footnote{\url{https://github.com/anhe2p9/M2I-TFM-Adriana.git}}. Similarly, the full replication package can be found on Zenodo\footnote{\url{https://doi.org/10.5281/zenodo.15682834}\label{zenodo-footnote}}. This full replication package includes (i) all the source code developed, (ii) an interactive \emph{full p-split} example's graph, (iii) an illustrating example's conflict graph, (iv) open source projects' results, and (v) the results' statistics. All the studies presented in this paper can be replicated using the materials provided in this replication package.

\subsection{Experimental setup}
\label{subsec:experimental_setup}
We conducted the experiments on a Doom Intel(R) Xeon(R) Gold 6240R CPU machine with 220~GB of RAM and 96 cores, running Ubuntu. We used PyCharm Community Edition 2025.1. We set the cognitive complexity threshold to the default value proposed by SonarQube ($\tau = 15$). AST processing and Extract Method refactorings were performed through Eclipse JDT version 3.31.0 and Eclipse LTK version 3.10.0. To solve ILP optimization problems, we used the Pyomo library for modeling and IBM ILOG CPLEX Optimization Studio V22.1.1, accessed through the IBM Academic Initiative, to solve each ILP model.

\subsection{Objects of study}
\label{subsec:object-of-study}
To evaluate the performance of the proposed approach, several methods, both from open-source and industry projects at Ayesa Advanced Digital Services Limited (ASDA), have been selected.

\subsubsection{Ayesa projects}
Ayesa Advanced Digital Services Limited (hereafter referred to as Ayesa) is a global provider of technological and engineering services, with 13,000 professionals and a direct presence in 24 countries across Europe, the Americas, Africa, Asia, and Oceania. They develop and implement digital solutions for companies and public administrations, applying the latest technologies to design and supervise infrastructures. They have teams specialized in more than 70 disciplines and certified in market-leading technologies, working across digital administration, health, industry, consumer, banking, insurance, telecommunications, media, energy and utilities, transportation, building and urban planning, resources, and the environment. In its vocation to be a global, creative, technology-driven, and human-centric company, Ayesa is committed to talent, through diversity and inclusion, as well as sustainability, as hallmarks and levers for innovation. Ten methods from an industry project from Ayesa are studied. As Ayesa's code is confidential, the PF's graphs are shown, but no more information about the code is provided.

\subsubsection{Open-source projects}
We utilize a diverse set of 10 open-source projects, selected from the benchmark by Saborido et al.~\cite{ASCCR}. These projects comprise two widely used frameworks for multi-objective optimization, five platform components designed to accelerate smart-solution development, and three popular open-source projects, each with over 10,000 stars and more than 900 forks. These open-source projects are shown in Table~\ref{tab:projects_and_commits}. It is also specified the commit hash for each open-source project.

\begin{table}[ht]
    \centering
    \caption{Open-source projects and their respective commits used for experimentation.}
    \label{tab:projects_and_commits}
    \begin{tabular}{ccc}
        \hline
        Project & Commit & Number of issues considered \\
        \hline
        ByteCode & \href{https://github.com/Konloch/bytecode-viewer/tree/55bfc32185991a76fac1c475ab6475f47c8f8eea}{55bfc32} & 20 \\
        Cybercaptor & \href{https://github.com/fiware-cybercaptor/cybercaptor-server/tree/b6b1f1046d6ba6747d541e256a7d3879cb69849a}{b6b1f10} & 19 \\
        FastJson & \href{https://github.com/alibaba/fastjson/tree/93d8c01e907fe35a8ff0eb5fe1c3b279d2f30282}{93d8c01e9} & 21 \\
        Fiware-Commons & \href{https://github.com/telefonicaid/fiware-commons/tree/f83b3425a3910c61ab20eae1af8d6f5c4d1e989c}{f83b342} & 3 \\
        IoTBroker & \href{https://github.com/mobius-software-ltd/iotbroker.cloud-java-client/tree/98eecebc5d18bb6b0ed198b3c9e9c95fa3a47987}{98eeceb} & 10 \\
        Jedis & \href{https://github.com/redis/jedis/tree/cfc227f724a4ea7b87a19da162ad726bba31751b}{cfc227f7} & 3 \\
        jMetal & \href{https://github.com/jMetal/jMetal/tree/e6baf75aa9bf475359ca249096d0231eb61bd0c2}{e6baf75aa} & 14 \\
        Knowage-core & \href{https://github.com/KnowageLabs/Knowage-Server/tree/dfed28a869125c51e51c66e433acd6e12b359828}{dfed28a869} & 15 \\
        MOEA-Framework & \href{https://github.com/MOEAFramework/MOEAFramework/tree/223393fdbdcc003efd17e97f1fd18a5318077bde}{223393fd} & 16 \\
        \hline
    \end{tabular}
\end{table}

\section{Results}
\label{sec:results}
This section presents the results obtained for two objectives (\texttt{extractions} and \texttt{CC}) and for three objectives (\texttt{extractions}, \texttt{CC}, and \texttt{LOC}) in both Ayesa projects and the ten open-source projects under study. For two-objective problems, the complete PF in two dimensions is obtained and presented. For three-objective problems, the results are presented through parallel coordinate plots, tables showing numerical results, and three-dimensional PF plots. A parallel coordinates plot is an intuitive way to represent results for multivariate data, i.e., data with multiple variables or dimensions. It is a practical way of representing the solutions of the PF, as it displays the results for each objective function and links them with straight lines, allowing all solutions to be observed simultaneously. We used 3600 seconds as the stopping criterion. If the algorithm used ends, then the whole PF is obtained.

We compute the hypervolume (HV) as a reference for future applications that attempt to solve the same problem using heuristic algorithms, allowing for comparison between the two. HV is the $n$-dimensional space that is ``contained'' by a set of points~\cite{hypervolume-bradstreet}. It contains numerous mathematical properties and provides a single scalar value that simultaneously reflects both the diversity of solutions along the PF and their proximity (convergence) to the actual Pareto-optimal front. The HV has been normalized to ensure comparability across different instances and objective scales. As a result, reported values are bounded between zero and one. The normalization process used divided the absolute HV obtained with the \texttt{pymoo} library by the total volume obtained from the cube defined between the ideal and the reference points.

\subsection{Two objectives' results}
\label{subsec:two-objs-results}
Every solution in the PF obtained using the hybrid method algorithm for two objectives (specifically, \texttt{extractions} and \texttt{CC}) can be represented with a 2D representation. The order of the objectives does not change the solution.

The results for the remaining objective combinations, including the number of extractions and LOC difference, as well as the combination of CC difference and LOC difference objectives, are not presented in this document to reduce its size. Still, they can be found in Zenodo\footnote{\url{https://doi.org/10.5281/zenodo.15682834}}.

\subsubsection{Ayesa's results for two objectives}
\label{subsubsec:ayesa_results_two_objs}
Table~\ref{tab:resultados-hipervolumen-Ayesa-2objs} presents a summary of the performance of the hybrid method algorithm for each Ayesa method according to the HV metric. These include the identifier assigned to each method, the total number of solutions obtained, and the reference point employed during evaluation. To provide a reference of the solution quality for future results, we also report the normalized HV covered by the solutions obtained for each method under study, along with the median and inter-quartile range (\emph{IQR}) obtained for each objective under study. The specified reference point in Table~\ref{tab:resultados-hipervolumen-Ayesa-2objs} is the nadir point of the obtained PF for each method after adding one unit to the objective function values, $Reference = (na_1 + 1, ..., na_n + 1)$. The HV obtained for the method that has only one solution is 1.

\begin{table}[!ht]
    \centering
    \caption[Results obtained for two objectives for each Ayesa's method]{Results for two objectives obtained for methods under study in industry projects from Ayesa. Each column represents, respectively, the order of each method, the number of solutions obtained for the method, and the reference point used for the method (\emph{Extractions~(E), CC~(C)}) to compute the HV, normalized HV (N-HV), and the median and inter-quartile range (\emph{IQR})  for the two objective functions.}
    \label{tab:resultados-hipervolumen-Ayesa-2objs}
    \begin{tabular}{rcccccccc}
        \footnotesize
        \\
        \hline
        \multirow{2}{2cm}{\centering\textbf{Ayesa's method}} & \multirow{2}{1.5cm}{\textbf{No. sol}} & \textbf{Reference} & \multirow{2}{2cm}{\centering\textbf{N-HV}} & \multicolumn{2}{c}{\textbf{Extractions}} & \multicolumn{2}{c}{\textbf{CC}} \\
        & & \emph{(E,C)} & & \emph{Median} & \emph{IQR} & \emph{Median} & \emph{IQR} \\ 
        \hline
         1st & 1 & (3,14) & 1.00 & 2.0 & 0.0 & 13.0 & 0.0 \\
         2nd & 2 &  (4,9) & 0.57 & 2.5 & 0.5 &  5.0 & 3.0 \\
         3rd & 2 &  (4,2) & 0.75 & 2.5 & 0.5 &  0.5 & 0.5 \\
         4th & 2 &  (5,6) & 0.67 & 3.0 & 1.0 &  4.5 & 0.5 \\
         5th & 3 &  (7,3) & 0.53 & 5.0 & 2.0 &  1.0 & 1.0 \\
         6th & 3 &  (8,4) & 0.67 & 3.0 & 2.5 &  2.0 & 1.0 \\
         7th & 3 &  (7,6) & 0.72 & 3.0 & 2.0 &  2.0 & 2.0 \\
         8th & 2 &  (6,4) & 0.50 & 3.5 & 1.5 &  2.0 & 1.0 \\
         9th & 3 &  (8,6) & 0.60 & 5.0 & 2.0 &  3.0 & 1.5 \\
        10th & 2 &  (8,3) & 0.58 & 4.5 & 2.5 &  1.5 & 0.5 \\
        \hline
    \end{tabular}
\end{table}

\begin{figure}[!ht]
\centering
\includegraphics[width=0.8\linewidth]{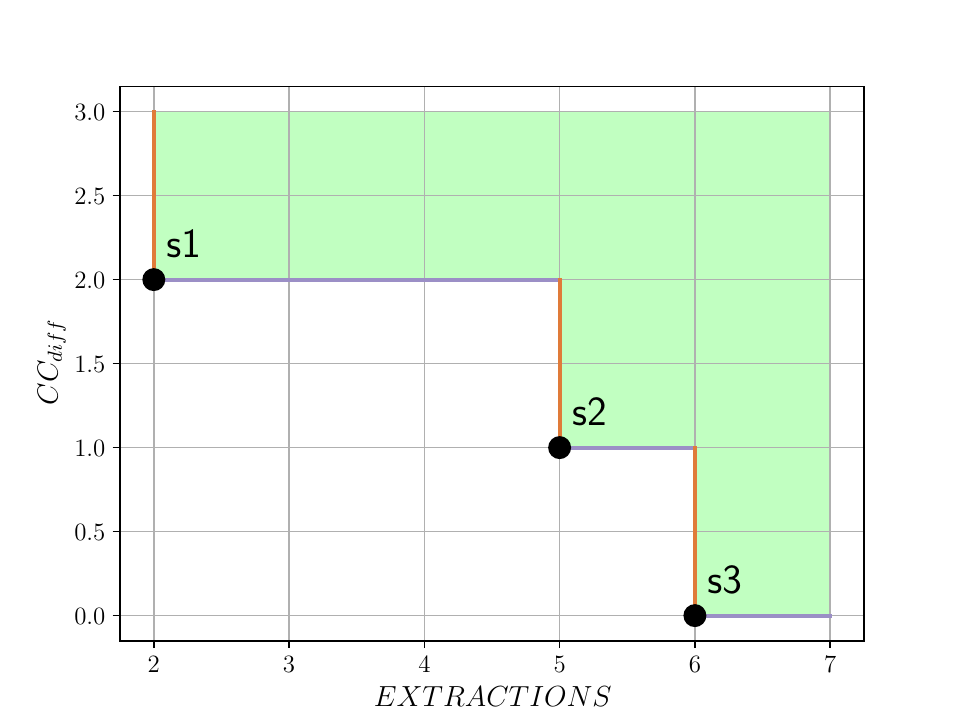}
\caption{2D PF obtained by the hybrid method algorithm for two objectives for the fifth Ayesa's method.}
\label{fig:ayesa_2dPF}
\end{figure}

The fifth Ayesa's method studied for two objectives is a non-convex PF. It is depicted in Figure~\ref{fig:ayesa_2dPF}, and as can be observed, the second solution, \textbf{s2}, could not be reached with linear combinations of the two objectives studied. The number of extractions varies from 2 to 6 across the different solutions obtained, whereas for the CC difference, the variation ranges from 0 to 2. This means that, for the worst-case scenario in the CC difference objective, the difference in CC among the code extractions varies at most $2$, and the CC remains constant for the best-case scenario.

\subsubsection{Open-source methods' results for two objectives}
\label{subsubsec:results_open_source_two_objs}
Table~\ref{tab:hipervolumen-open-source-tres-objs} in Appendix~\ref{ch:ApendiceC_tablas_open-source} presents the same summary as the one for Ayesa's project, but in this case, the results are the ones obtained for the nine open-source projects of Table~\ref{tab:projects_and_commits}.

\begin{figure}[!ht]
\centering
\includegraphics[width=0.8\linewidth]{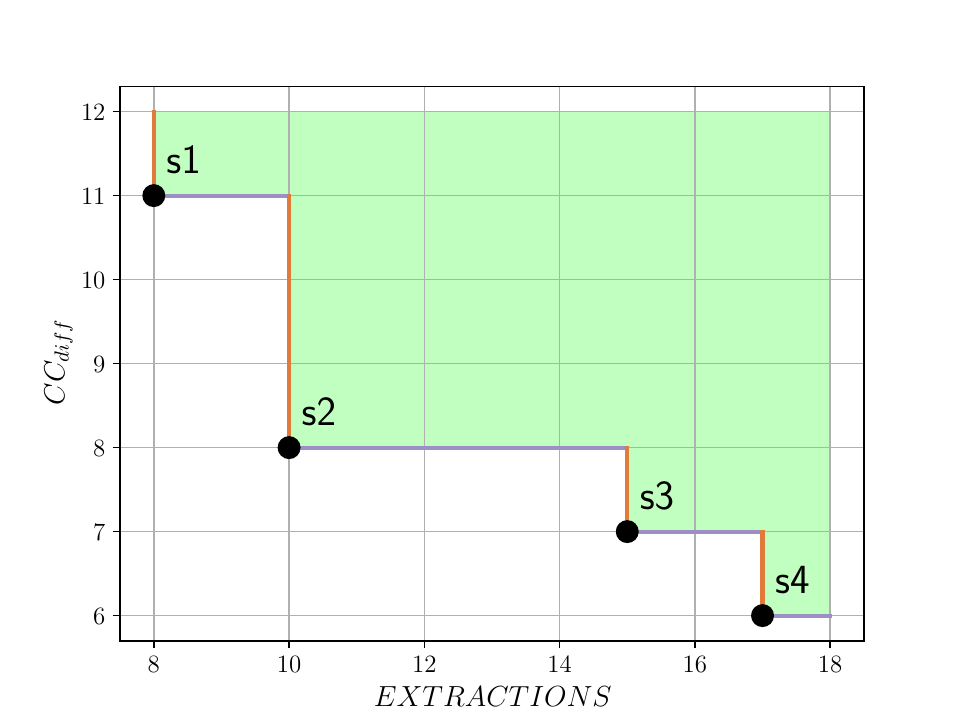}
\caption[\texttt{seekObjectToField(long[])} method result from \texttt{JSONScanner} class in \texttt{fastjson} project]{PF obtained by the epsilon constraint algorithm for the \texttt{seekObjectToField(long[])} method from the \texttt{JSONScanner} class in the \texttt{fastjson} project.}
\label{fig:seekObjectToField_2dPF}
\end{figure}

The two-dimensional PF selected to be shown in this case (out of the $121$ methods under study from open-source projects) is depicted in Figure~\ref{fig:seekObjectToField_2dPF}. Again, this PF has an unsupported solution, that is the solution \textbf{s3}. The number of extractions for each solution is $8$, $10$, $15$, and $17$ extractions, with a difference od CC of $11$, $8$, $7$ and $6$, respectively.

\subsection{Three objectives' results}
\label{sec:three-results}
Every solution from the PF obtained from the hybrid method algorithm for three objectives (\texttt{extractions}, \texttt{CC}, and \texttt{LOC}) can be represented with parallel coordinates, as well as directly in a 3D representation. In any case, the order of the objectives does not change the solution.

\subsubsection{Ayesa's results for three objectives}
\label{subsubsec:ayesa_results_three_objs}
Table~\ref{tab:resultados-hipervolumen-Ayesa-3objs} presents a summary of the performance of the hybrid method algorithm for each Ayesa method according to the HV metric. In the same way that the results were presented for two objectives, they are now shown for three objectives, also including the results for LOC balancing.

\begin{table}[!ht]
    \centering
    \caption[Results obtained for three objectives for each Ayesa's method]{Results obtained for three objectives for methods under study in industry projects from Ayesa. Each column represents, respectively, the order of each method, the number of solutions obtained for the method, and the reference point used for the method (\emph{Extractions~(E), CC~(C), LOC~(L)}) to compute the HV, normalized HV (N-HV), and the median and inter-quartile range (\emph{IQR})  for the three objective functions.}
    \label{tab:resultados-hipervolumen-Ayesa-3objs}
    \begin{tabular}{rccccccccc}
        \footnotesize
        \\
        \hline
        \multirow{2}{1.4cm}{\centering\textbf{Ayesa's method}} & \multirow{2}{1.1cm}{\textbf{No. sol}} & \textbf{Reference} & \multirow{2}{1.3cm}{\centering\textbf{N-HV}} & \multicolumn{2}{c}{\textbf{Extractions}} & \multicolumn{2}{c}{\textbf{CC}} & \multicolumn{2}{c}{\textbf{LOC}} \\
        & & \emph{(E,C,L)} & & \emph{Median} & \emph{IQR} & \emph{Median} & \emph{IQR} & \emph{Median} & \emph{IQR} \\ 
        \hline
         1st &  1 &  (3,14,19) & 1.00 & 2.0 & 0.00 & 13.0 & 0.00 & 18.0 &  0.00 \\
         2nd &  2 &   (4,9,26) & 0.51 & 2.5 & 0.50 &  5.0 & 3.00 & 18.5 &  6.50 \\
         3rd &  2 &   (4,2,11) & 0.52 & 2.5 & 0.50 &  0.5 & 0.50 &  5.5 &  4.50 \\
         4th &  4 &   (8,6,21) & 0.57 & 5.0 & 2.75 &  4.0 & 0.25 & 16.5 &  5.25 \\
         5th &  4 &   (9,3,13) & 0.41 & 5.5 & 2.25 &  1.0 & 0.50 &  8.0 &  5.75 \\
         6th &  5 &    (9,4,6) & 0.49 & 4.0 & 4.00 &  2.0 & 0.00 &  4.0 &  1.00 \\
         7th &  6 &   (8,6,25) & 0.33 & 3.5 & 2.50 &  2.5 & 1.75 & 18.5 &  4.75 \\
         8th &  7 &   (9,6,12) & 0.66 & 5.0 & 1.50 &  2.0 & 2.00 &  5.0 &  3.00 \\
         9th &  8 &  (11,6,13) & 0.49 & 7.0 & 2.50 &  3.0 & 0.25 &  5.5 &  7.00 \\
        10th & 20 & (12,11,45) & 0.79 & 5.5 & 3.25 &  3.5 & 3.25 &  9.5 & 10.00 \\
        \hline
    \end{tabular}
\end{table}

\begin{figure}[!ht]
    \centering
    \begin{subfigure}[t]{0.49\textwidth}
        \centering
        \includegraphics[width=\linewidth]{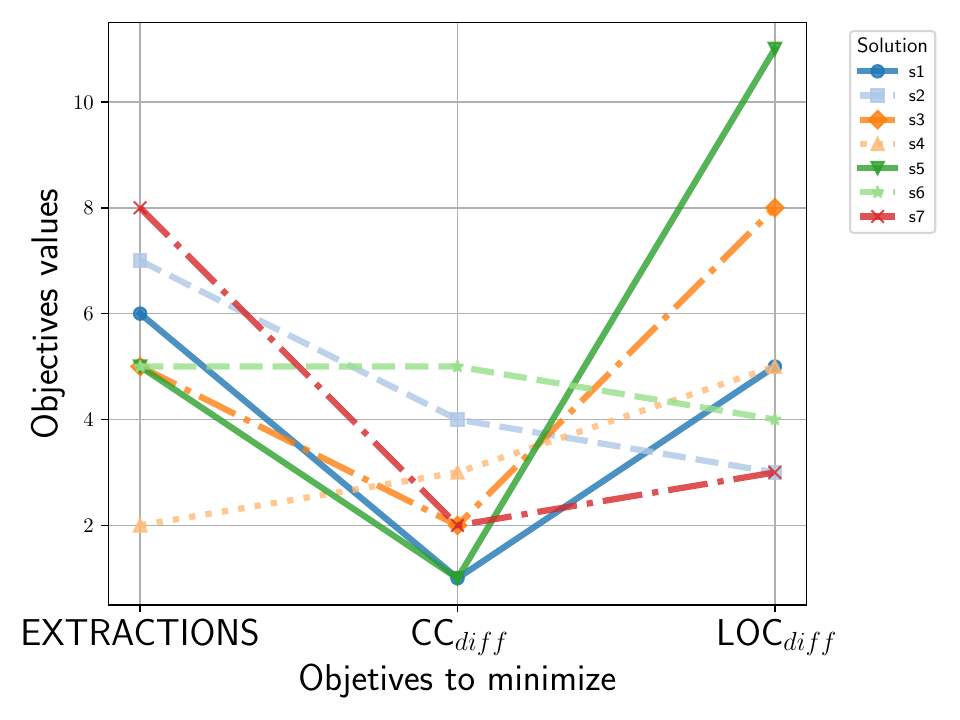}
        \caption{Parallel coordinates.}
        \label{fig:ayesa8_parallel_coordinates}
    \end{subfigure}
    \hfill
    \begin{subfigure}[t]{0.49\textwidth}
        \vspace{-3.8cm}
        \centering
        \resizebox{\textwidth}{!}{
            \begin{tabular}{|c|c|c|c|}
                \hline
                \diagbox[width=3cm]{\textbf{Solution}}{\textbf{Objective}} & $Extractions$ & $CC_{diff}$ & $LOC_{diff}$ \\
                \hline
                s1 & 6 &  1 &   5 \\
                \hline
                s2 & 7 &  4 &   3 \\
                \hline
                s3 & 5 &  2 &   8 \\
                \hline
                s4 & 2 &  3 &   5 \\
                \hline
                s5 & 5 &  1 &  11 \\
                \hline
                s6 & 5 &  5 &   4 \\
                \hline
                s7 & 8 &  2 &   3 \\
                \hline
            \end{tabular}
        }
        \caption{Table.}
        \label{tab:ayesa_8_tabla}
    \end{subfigure}
    \hfill
    \begin{subfigure}[t]{0.7\textwidth}
        \centering
        \includegraphics[width=0.8\linewidth]{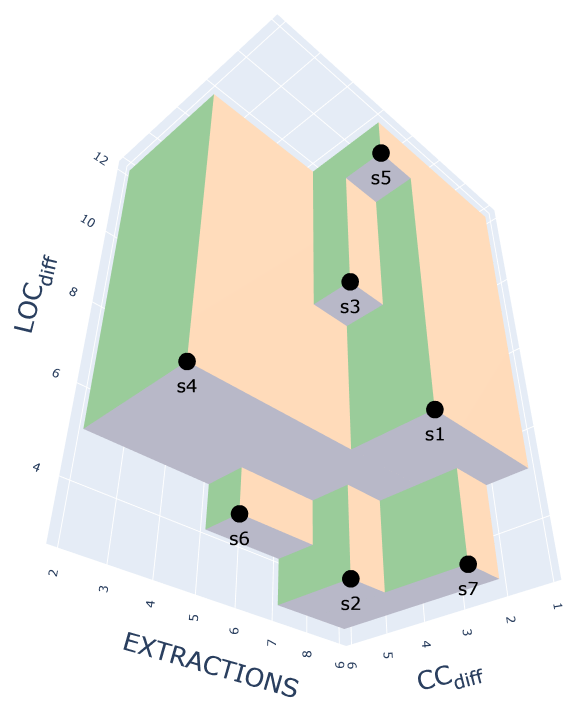}
        \caption{3D PF visualization.}
        \label{fig:ayesa_8_3dPF}
    \end{subfigure}
\caption{3D PF obtained by the hybrid method algorithm for three objectives for the eighth Ayesa's method.}
\label{fig:ayesa_8_general}
\end{figure}

\begin{figure}[!ht]
    \centering
    \vspace{3.2cm}
    \raisebox{-3cm}{
    \begin{subfigure}[t]{0.485\textwidth}
        \centering
        \includegraphics[width=\linewidth]{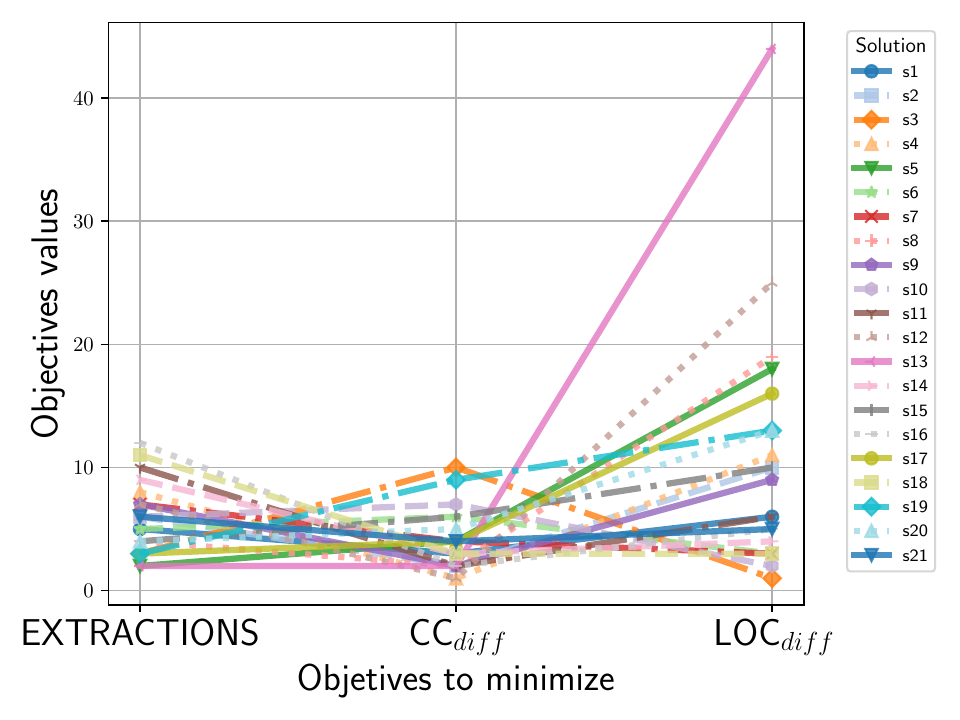}
        \caption{Parallel coordinates.}
        \label{fig:ayesa10_parallel_coordinates}
    \end{subfigure}
    }
    \hfill
    \begin{subfigure}[t]{0.485\textwidth}
        \vspace{-2.8cm}
        \centering
        \resizebox{\textwidth}{!}{
            \begin{tabular}{|c|c|c|c|}
                \hline
                \diagbox[width=3cm]{\textbf{Solution}}{\textbf{Objective}} & $Extractions$ & $CC_{diff}$ & $LOC_{diff}$ \\
                \hline
                 s1 &  9 &  3 &   4 \\
                \hline
                 s2 &  2 &  4 &  18 \\
                \hline
                 s3 &  5 &  3 &   6 \\
                \hline
                 s4 & 10 &  2 &   6 \\
                \hline
                 s5 &  4 &  2 &  19 \\
                \hline
                 s6 &  4 &  6 &  10 \\
                \hline
                 s7 &  3 &  4 &  16 \\
                \hline
                 s8 &  6 &  7 &   2 \\
                \hline
                 s9 &  2 &  2 &  44 \\
                \hline
                s10 &  7 &  4 &   3 \\
                \hline
                s11 &  7 &  2 &   9 \\
                \hline
                s12 &  3 & 10 &   1 \\
                \hline
                s13 &  7 &  1 &  25 \\
                \hline
                s14 & 11 &  3 &   3 \\
                \hline
                s15 &  6 &  4 &   5 \\
                \hline
                s16 &  8 &  1 &  11 \\
                \hline
                s17 &  5 &  6 &   3 \\
                \hline
                s18 &  4 &  5 &  13 \\
                \hline
                s19 &  6 &  2 &  10 \\
                \hline
                s20 &  3 &  9 &  13 \\
                \hline
            \end{tabular}
        }
        \caption{Table.}
        \label{tab:ayesa_10_tabla}
    \end{subfigure}
    \hfill
    \begin{subfigure}[t]{0.9\textwidth}
        \includegraphics[width=0.9\linewidth]{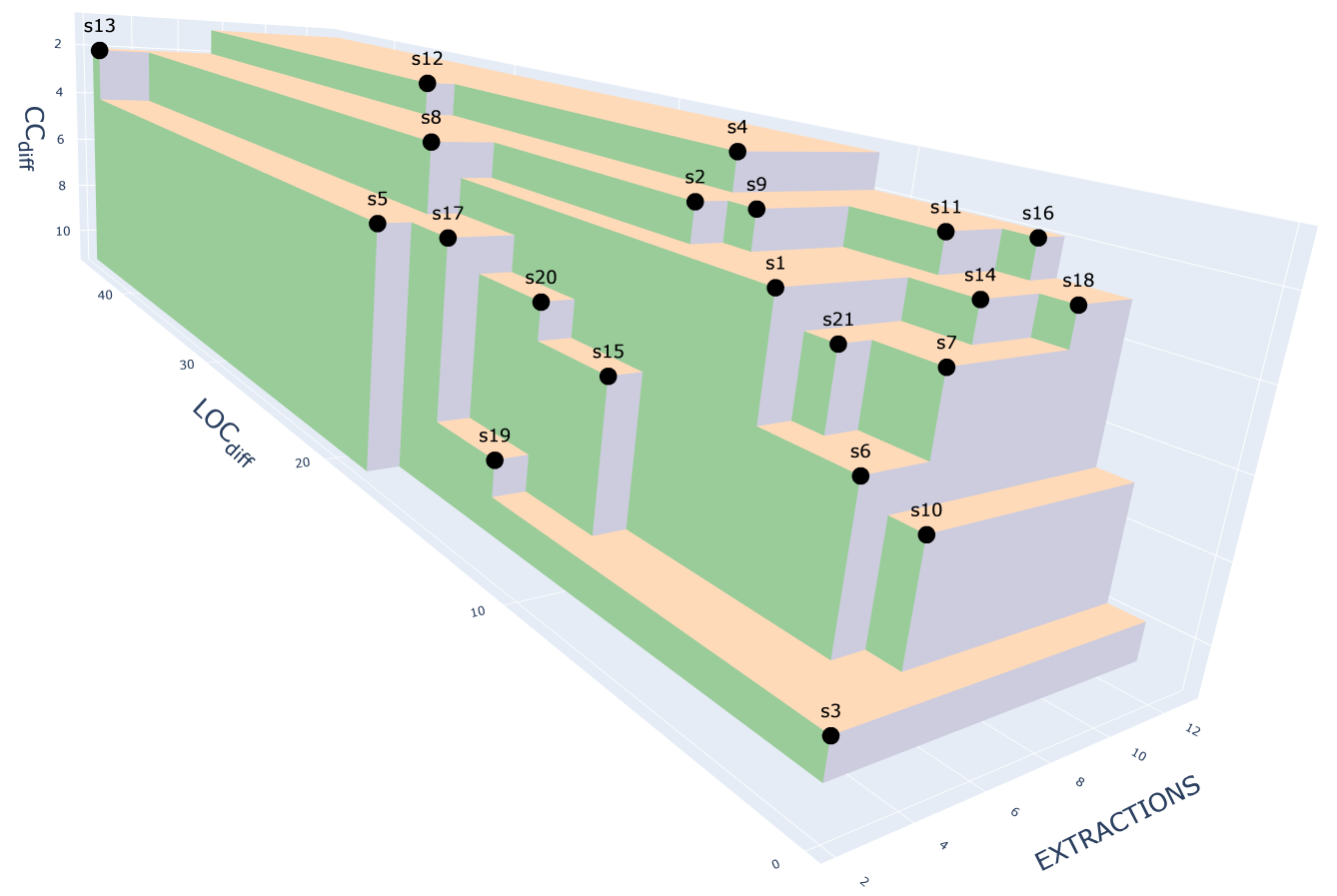}
        \caption{3D PF visualization.}
        \label{fig:ayesa_10_3dPF}
    \end{subfigure}
\caption{3D PF obtained by the hybrid method algorithm for three objectives for the tenth Ayesa's method.}
\label{fig:ayesa_10_general}
\end{figure}

Two instances out of the ten studied methods from Ayesa's project are shown in Figures~\ref{fig:ayesa_8_general} and~\ref{fig:ayesa_10_general}. For each figure, the first subfigure corresponds to the parallel coordinates plot of each PF, the second one is the list of Pareto optimal solutions obtained, and the third is a visualization of the obtained PF in three dimensions. All plots can be found in Zenodo\footnote{\url{https://doi.org/10.5281/zenodo.15682834}}.

As can be observed in Figure~\ref{fig:ayesa_8_general}, the number of extractions varies from $2$ to $8$ across the different solutions obtained. For CC difference, the variation goes from $1$ to $5$. Similarly, the LOC difference between the methods obtained after extraction varies by at most $3$ in the best case and $11$ in the worst case.

As shown in Figure~\ref{fig:ayesa_10_general}, this method is the one with the maximum number of solutions obtained of all those that have been depicted from Ayesa. Its normalized HV is also the highest in Table~\ref{tab:resultados-hipervolumen-Ayesa-3objs}.

\subsubsection{Open-source methods' results for three objectives}
\label{subsubsec:results_open_source_three_objs}
Finally, the results obtained for the three objectives using the selected methods from the nine open-source projects are shown in Table~\ref{tab:hipervolumen-open-source-tres-objs}. The structure is the same as the one in Table~\ref{tab:resultados-hipervolumen-Ayesa-3objs}. For these methods, the number of solutions shown is significant because $12\%$ of the studied methods have only one solution. Therefore, in these cases, there is no conflict between the three objectives under study. For these cases (with only one solution), the HV obtained is one.

\begin{figure}[!ht]
    \centering
    \begin{subfigure}[t]{0.49\textwidth}
        \centering
        \includegraphics[width=\linewidth]{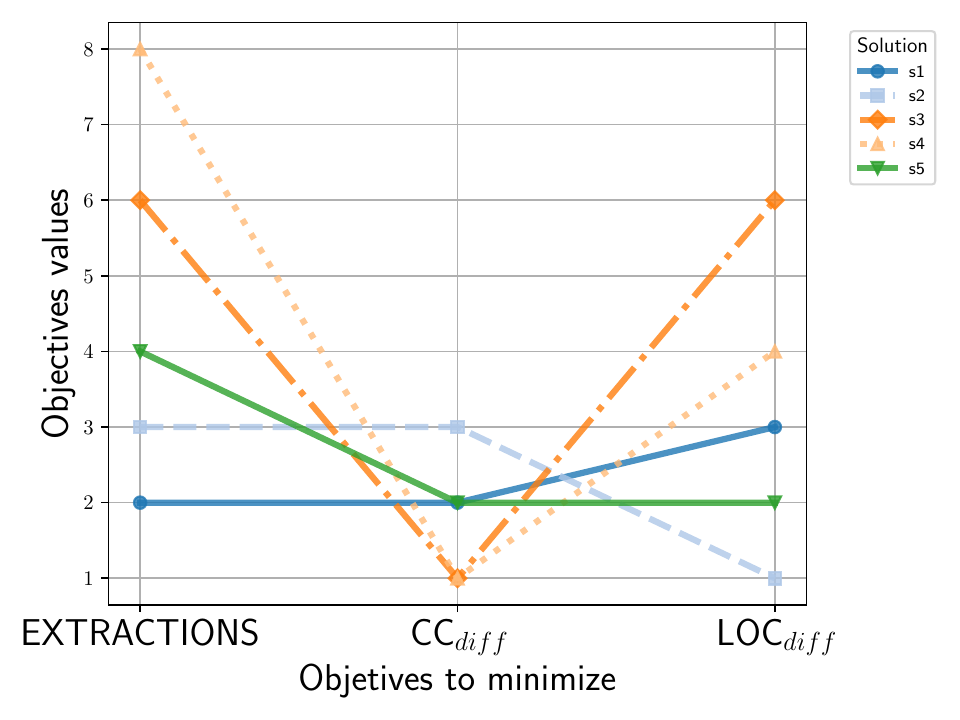}
        \caption{Parallel coordinates.}
        \label{fig:routeAPacketTo_result}
    \end{subfigure}
    \hfill
    \begin{subfigure}[t]{0.49\textwidth}
        \vspace{-3.5cm}
        \centering
        \resizebox{\textwidth}{!}{
            \begin{tabular}{|c|c|c|c|}
                \hline
                \diagbox[width=3cm]{\textbf{Solution}}{\textbf{Objective}} & $Extractions$ & $CC_{diff}$ & $LOC_{diff}$ \\
                \hline
                s1 & 2 & 2 & 3 \\
                \hline
                s2 & 3 & 3 & 1 \\
                \hline
                s3 & 6 & 1 & 6 \\
                \hline
                s4 & 8 & 1 & 4 \\
                \hline
                s5 & 4 & 2 & 2 \\
                \hline
            \end{tabular}
        }
        \caption{Table.}
        \label{tab:routeAPacketTo_tabla}
    \end{subfigure}
    \hfill
    \begin{subfigure}[t]{0.6\textwidth}
        \centering
        \includegraphics[width=0.9\linewidth]{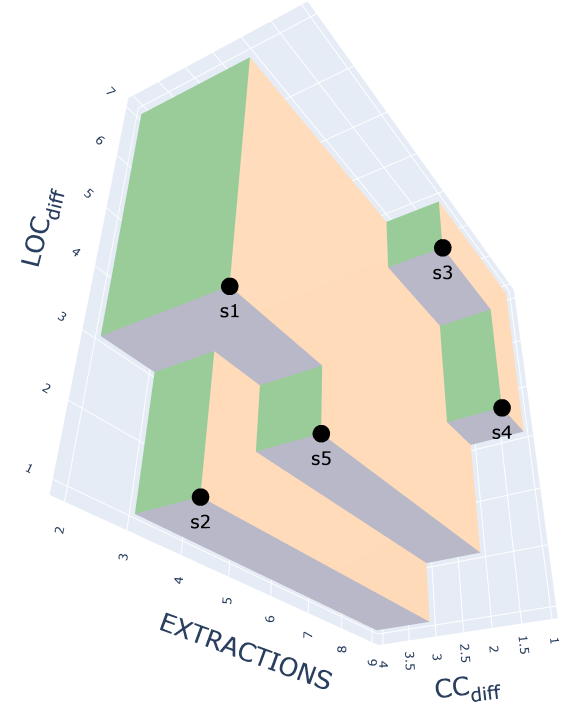}
        \caption{3D PF visualization.}
        \label{fig:routeAPacketTo_3dPF}
    \end{subfigure}
\caption[\texttt{routeAPacketTo} method result from \texttt{Host} class in \texttt{cybercaptor-server} project]{\texttt{routeAPacketTo} method result from \texttt{Host} class in \texttt{cybercaptor-server} project. This is the solution for the method used for the running example in Figure~\ref{fig:ccExampleInCode}.}
\label{fig:routeAPacketTo_general}
\end{figure}

The solutions for the three objectives obtained for the running example in Figure~\ref{fig:initial-running-example} are presented in Figure~\ref{fig:routeAPacketTo_general}. 

If the solution that best balances LOC is chosen, i.e., the second solution (\textbf{s2} in Figure~\ref{fig:routeAPacketTo_general}), the corresponding extractions are the ones in Figure~\ref{fig:refactoring_solution_multi-objective}. As shown in Figure~\ref{fig:refactoring_solution_multi-objective}, there are three extractions that take into account the original method, and they are nested within one another. The CC of each method, respectively, is $2$, $5$, and $4$, so the difference between the maximum and minimum CC of all methods after the extractions have been performed is $5 - 2 = 3$. Similarly, the LOC of each resulting method is $7$, $8$, and $8$, so the difference between the maximum and minimum LOC of all methods after the refactoring is $8 - 7 = 1$.

\section{Related work}
\label{sec:SOTA}
A significant portion of the time spent on software project development is dedicated to writing code that performs specific tasks and fulfills initial requirements. Then, some time is needed to modify that code and make it more efficient, less complex, more legible, and more maintainable. This modification is known as refactoring, and it might require a significant amount of time. Existing literature suggests the existence of poor reviewing practices, i.e., \emph{anti-patterns}, that may contribute to a slowdown in integration and affect the overall sustainability of the project~\cite{chouchenRefactorings}, which should be avoided. If design patterns are solutions to common problems encountered in software development, anti-patterns are solutions that apparently solve the problem, but the resulting code is hardly maintainable and reusable. To avoid the use of anti-patterns, developers apply changes to code that enhance the design without altering its functionality, with the so-called refactorings. When one refactors code, the developer has the opportunity to improve legibility without altering the code semantics. By definition, refactoring operations preserve the code semantics. The refactoring research effort is fragmented over several research communities, various domains, and objectives~\cite{30YofRefactoringResearch}. Opdyke and Johnson first used the term \emph{Refactoring} in September 1990~\cite{Opdyke1990}. Two years later, they defined a set of refactorings, which they described as ``program restructuring operations'', that they claimed supported the design, evolution, and reuse of object-oriented application frameworks~\cite{Opdyke1992}. The first book about refactoring was published by Fowler et al.~\cite{fowler1999improving}, and the use of refactoring gained popularity, defining the refactoring process as ``small changes and testing after each change''. Later, Stroggylos and Spinellis analyzed different source codes to evaluate software enhancement due to refactoring, because they noticed that measurements on real life did not indicate an increase in quality achieved via code refactoring~\cite{Strogglyos2007}. Kaur, Amandeep, and Kaur, Manpreet also carried out an analysis about code refactoring impact on Software Quality, where they define it as ``a way to clean up code that minimizes the chances of introducing bugs''~\cite{Kaur2021AnalysisOC}. They concluded that refactoring reduces software complexity, making it easily understandable. Developers' productivity is improving due to the long-term effect of refactoring, specifically increasing two crucial factors: understandability and maintainability of the code. Extract method refactoring can help avoid cloned code, a code smell that can lead to bugs. If there are code clones, a solution to this code smell is to put that piece of repeated code inside a new method, and to replace each appearance of that code clone with a call to that new method. Several tools can automatically repair this code smell. Extract method refactoring can also be used to reduce the CC of a code, as it ``postpones'' the effort of understanding a particular piece of code. Thus, the developer can focus on the code's functionality and check the new method if more details are needed.

The problem of reducing CC has been approached in recent years in various ways. Saborido et al. analyzed the difficulties faced by developers when deciding on the best method to extract in order to reduce the CC of Java code and proposed enumerative approaches for CC reduction~\cite{ASCCR}. Recently, Saborido et al. modeled the software CC reduction task as an ILP optimization problem~\cite{ASCCRthroughtILP}. The purpose was to minimize the number of sentences extracted from a method, so that there is not a high quantity of extractions obtained. For each method with CC greater than a given threshold, their approach searches for sequences of applicable extract method refactoring operations. Finally, it outputs the changes to perform on each method. They were able to reduce the CC to or below the default threshold in the majority of the methods under study, thereby enhancing the performance of the search compared to the results found in their previous work, where they used heuristic approaches.

We model the source code CC reduction as a multi-objective ILP optimization problem. Multi-objective optimization problems aim to simultaneously optimize multiple conflicting objectives, which commonly arise in real-world scenarios. In recent years, numerous metaheuristic algorithms have been developed to tackle these problems. These approaches seek to approximate the PF, which consists of all Pareto optimal solutions, by generating a set of non-dominated solutions. Despite the previous, there exist exact algorithms to find the PF. Chicano et al. proposed two strategies based on ILP that can find the complete PF~\cite{augmecon2016chicano}, and the solutions obtained are well-distributed at any time of execution, i.e., \emph{anytime} algorithms. They developed an augmented $\varepsilon$-constraint algorithm \emph{(AUGMECON)}, that solves one sub-problem on each iteration, and guarantees that the non-dominated point obtained is efficient. They also improved existing anytime algorithms. As a result of his PhD, Miguel Ángel Domínguez Ríos proposed novel algorithms that compute the complete PF for combinatorial problems or problems with integer variables, dividing the objective space into several subspaces via specific constraints~\cite{TPA-MiguelAngel}. Ángel Domínguez Ríos et al. designed an exact algorithm that solves any multi-objective combinatorial optimization (MOCO) problem, and whose solutions obtained are well distributed over the objective space anytime~\cite{anytimeAlgorithmMiguelAngelYChicano}. They also designed an anytime hybrid algorithm, called \emph{MOFeLS}, that uses an ILP solver and local search to solve any multi-objective problem with binary variables and linear objective functions and constraints~\cite{MOFeLS}. Moreover, they implemented an anytime approximated algorithm, called \emph{HBOXES}, that can become an exact algorithm depending on the input parameters, and that solves any MOCO problem~\cite{TPA-MiguelAngel}.

The unique techniques that can automatically reduce methods' CC to a given threshold were proposed by Saborido et al.~\cite{ASCCR}\cite{ASCCRthroughtILP}. They reduce the CC of methods by minimizing the number of code extractions. However, they do not take into account the balance of CC among the resulting methods or the length (LOC) of code extractions. That is why the approach presented in this paper is novel, as it balances CC and LOC, thereby allowing developers more operational leeway to further edit the code after applying the tool presented in this paper. While in related work the focus was solely on reducing cognitive complexity by minimizing the number of extracted methods, our proposal introduces a multi-objective perspective that integrates the balance of CC and LOC among the resulting methods, i.e., the original method and all its extractions performed. This contribution is particularly relevant in collaborative and long-term projects, where readability and maintainability of code are highly relevant. If no single method remains with high CC and none of them is particularly longer than the others, then the refactored code becomes more uniform and manageable. This approach facilitates future modifications and extensions. Furthermore, this balance provides developers with the flexibility to continue editing the code without compromising quality or easily exceeding CC thresholds again.

To the best of our knowledge, no existing works or tools in the industry explicitly address this balance using a multi-objective ILP formulation, which positions this contribution as a significant advancement in the field of automated refactoring.
\section{Conclusion and future work}
\label{sec:Conclusion}
We proposed a new multi-objective ILP formulation for reducing software CC has been developed, allowing for the minimization of the number of code extractions while balancing the LOC and CC of extracted methods.

A novel software tool has also been designed and developed to ease the CC reduction of software. This tool integrates several multi-objective ILP problem-solving algorithms, including an $\varepsilon$-constraint algorithm and a hybrid method algorithm, which traverse the entire objective space and can approximate the complete Pareto front. The proposed multi-objective ILP model and the software tool have been validated using a benchmark of 121 methods from nine open-source projects and 10 from an industrial software project in Java. A set of efficient solutions has been obtained for each of them. Regarding the solver used, CPLEX boasts high computational efficiency and employs advanced algorithms that enable the solution of large and complex problems quickly and accurately. This advantage enables the developed tool to be integrated into continuous software integration processes. Thus, the objectives set for this work have been met.

In contrast, one limitation of the work is that CPLEX is not free, and its commercial license is expensive. 

In some cases, such as for method number 97 from open-source projects, which aim to solve two objectives, no solution has been obtained. For now, no exhaustive record of the methods that cannot be solved has been made.

As future work, three main lines of development are proposed to extend and enrich this work. First, as assigning a representative name to the extractions is crucial, we can integrate the use of LLMs for method naming. The more extractions are performed, the more challenging the method name assignment is. The key is how to prompt the model to indicate all the requirements needed to infer method names adequately. The application of changes to the source code has not yet been automated. This task will be included as future work and completed in the ILP CC reducer tool. Theoretically, we will have a Java program that reads the output CSV to apply the extractions. Second, it is planned to design and conduct a series of surveys targeting expert or semi-expert users to assess the perceived quality and usefulness of the results generated by the proposed multi-objective ILP approach. The responses will be systematically compared with those from previous evaluations based on the single-objective ILP approach developed by Saborido et al.~\cite{ASCCRthroughtILP}. The purpose of this comparative study is not only to validate the improvements introduced by the multi-objective ILP formulation but also to gather qualitative insights into user preferences and expectations, to guide future iterations of the system. Finally, the study aims to investigate the scalability of the multi-objective ILP model, as adding numerous constraints can lead to an increase in execution time that makes the model unrunnable. In this case, the possibility of solving the model presented in this paper using alternative optimization techniques will be explored.

\appendix
\newpage
\clearpage

\pagenumbering{roman}
\section{Figures used for CC reduction problem definitions}

\begin{figure}[!ht]
\centering

\begin{tikzpicture}
\node[inner sep=0pt] (code) {
\begin{minipage}{\textwidth}
\begin{lstlisting}[columns=fullflexible, numbers=left,xleftmargin=0.5cm,xrightmargin=0cm,frame=single,framexleftmargin=0em,escapechar=ñ,stepnumber=1,
    showstringspaces=false,
    tabsize=1,
    breaklines=true,
    breakatwhitespace=false,
    basicstyle=\scriptsize,
    language=Java,
    mathescape=true]{}
public void routeAPacketTo(IPAddress ip, int ttl, List<Host> usedHosts) throws Exception {
    usedHosts.add(this);
    if (ttl == 0) {//Problem in routing
        throw new Exception("Routing problem...");
    }
    if (!hasIP(ip)) { //Packet not arrived
        Host nextHost = null;
        List<Host> directlyAccessibleHosts = getDirectlyAccessibleHosts();
        for (Host directlyAccessibleHost: directlyAccessibleHosts) {
            if (...) // If the packet is for a neighbour, we send it to him
                nextHost = directlyAccessibleHost;
        }
        if (nextHost != null) {
            nextHost.routeAPacketTo(ip, ttl - 1, usedHosts);
        } else {//We have to look in the routing table
            List<Host> directlyAccessible = getDirectlyAccessibleHosts();
            IPAddress nextIP = this.getRoutingTable().getNextHop(ip);
            boolean nextHostFound = false;
            for (Host aDirectlyAccessible: directlyAccessible) {
                if (...) { //Search the nextHop host object
                    aDirectlyAccessible.routeAPacketTo(ip, ttl - 1, usedHosts);
                    nextHostFound = true;
                }
            }
            if (!nextHostFound) { //Routing problem
                throw new Exception("Routing problem...");
            }
        }
    }
}
\end{lstlisting}
\end{minipage}
};

\begin{scope}[shift={(code.south west)}]
  \draw[red, thick, fill=red, fill opacity=0.05] (0.8,7.95) rectangle (1.2,8.25);
\end{scope}

\begin{scope}[shift={(code.south west)}]
  \draw[red, thick, fill=red, fill opacity=0.05] (0.8,7.1) rectangle (01.2,7.4);
\end{scope}

\begin{scope}[shift={(code.south west)}]
  \draw[red, thick, fill=red, fill opacity=0.05] (1.2,6.27) rectangle (1.71,6.55);
\end{scope}

\begin{scope}[shift={(code.south west)}]
  \draw[red, thick, fill=red, fill opacity=0.05] (1.5,6) rectangle (1.9,6.28);
\end{scope}

\begin{scope}[shift={(code.south west)}]
  \draw[red, thick, fill=red, fill opacity=0.05] (1.2,5.1) rectangle (1.56,5.45);
\end{scope}

\begin{scope}[shift={(code.south west)}]
  \draw[red, thick, fill=red, fill opacity=0.05] (1.4,4.55) rectangle (2.05,4.9);
\end{scope}

\begin{scope}[shift={(code.south west)}]
  \draw[red, thick, fill=red, fill opacity=0.05] (1.6,3.45) rectangle (2.09,3.75);
\end{scope}

\begin{scope}[shift={(code.south west)}]
  \draw[red, thick, fill=red, fill opacity=0.05] (1.9,3.15) rectangle (2.3,3.45);
\end{scope}

\begin{scope}[shift={(code.south west)}]
  \draw[red, thick, fill=red, fill opacity=0.05] (1.6,1.75) rectangle (1.92,2.06);
\end{scope}

\begin{scope}[shift={(code.south west)}]
  \node[draw=red, fill=white, text=red, thick, rectangle, rounded corners, minimum size=0.01cm, font=\tiny] at (0.57,8.08) {1};
  \node[draw=red, fill=white, text=red, thick, rectangle, rounded corners, minimum size=0.01cm, font=\tiny] at (0.57,7.25) {1};
  \node[draw=red, fill=white, text=red, thick, rectangle, rounded corners, minimum size=0.01cm, font=\tiny] at (0.77,6.45) {1+1};
  \node[draw=red, fill=white, text=red, thick, rectangle, rounded corners, minimum size=0.01cm, font=\tiny] at (1.11,6.07) {1+2};
  \node[draw=red, fill=white, text=red, thick, rectangle, rounded corners, minimum size=0.01cm, font=\tiny] at (0.83,5.28) {1+1};
  \node[draw=red, fill=white, text=red, thick, rectangle, rounded corners, minimum size=0.01cm, font=\tiny] at (1.1,4.71) {1};
  \node[draw=red, fill=white, text=red, thick, rectangle, rounded corners, minimum size=0.01cm, font=\tiny] at (1.24,3.64) {1+2};
  \node[draw=red, fill=white, text=red, thick, rectangle, rounded corners, minimum size=0.01cm, font=\tiny] at (1.53,3.25) {1+3};
  \node[draw=red, fill=white, text=red, thick, rectangle, rounded corners, minimum size=0.01cm, font=\tiny] at (1.22,1.91) {1+2};
\end{scope}

\end{tikzpicture}

\caption[Count of Cognitive Complexity]{Example of the count of Cognitive Complexity (CC), from the example in Figure~\ref{fig:ccExampleInCode}. It is demonstrated that the inherent and nesting components are the sum of two components when a nesting component is present.}
\label{fig:count_of_CC}
\end{figure}

\begin{figure}[!ht]
\centering

\begin{tikzpicture}
\node[inner sep=0pt] (code) {
\begin{minipage}{\textwidth}
\begin{lstlisting}[columns=fullflexible, numbers=left,xleftmargin=0.5cm,xrightmargin=0cm,frame=single,framexleftmargin=0em,escapechar=ñ,stepnumber=1,
    showstringspaces=false,
    tabsize=1,
    breaklines=true,
    breakatwhitespace=false,
    basicstyle=\scriptsize,
    language=Java,
    mathescape=true]{}
// Cognitive complexity 20
public void routeAPacketTo(IPAddress ip, int ttl, List<Host> usedHosts) throws Exception {
    usedHosts.add(this);
    // [$\color{pgreen} \lambda$=0, $\color{pgreen} \iota$=1, $\color{pgreen} \nu$=0, $\color{pgreen} \mu$=0, CCR=1, NMCC=1]
    if (ttl == 0) {//Problem in routing
        throw new Exception("Routing problem, TTL is null : packet to " 
                            + ip + " deleted on host " + this.getName());
    }
    // [$\color{pgreen} \lambda$=0, $\color{pgreen} \iota$=8, $\color{pgreen} \nu$=11, $\color{pgreen} \mu$=6, CCR=19, NMCC=19]
    if (!hasIP(ip)) { //Packet not arrived
        Host nextHost = null;
        List<Host> directlyAccessibleHosts = getDirectlyAccessibleHosts();
        // [$\color{pgreen} \lambda$=1, $\color{pgreen} \iota$=2, $\color{pgreen} \nu$=1, $\color{pgreen} \mu$=2, CCR=5, NMCC=3]
        for (Host directlyAccessibleHost: directlyAccessibleHosts) {
            // [$\color{pgreen} \lambda$=2, $\color{pgreen} \iota$=1, $\color{pgreen} \nu$=0, $\color{pgreen} \mu$=1, CCR=3, NMCC=1]
            if (directlyAccessibleHost.hasIP(ip)) // If the packet is for a neighbour,
                                                  // We send it to him
                nextHost = directlyAccessibleHost;
        }
        // [$\color{pgreen} \lambda$=1, $\color{pgreen} \iota$=5, $\color{pgreen} \nu$=4, $\color{pgreen} \mu$=4, CCR=13, NMCC=9]
        if (nextHost != null) {
            nextHost.routeAPacketTo(ip, ttl - 1, usedHosts);
        // [$\color{pgreen} \lambda$=1, $\color{pgreen} \iota$=4, $\color{pgreen} \nu$=4, $\color{pgreen} \mu$=3, CCR=12, NMCC=9]
        } else {//We have to look in the routing table
            List<Host> directlyAccessible = getDirectlyAccessibleHosts();
            IPAddress nextIP = this.getRoutingTable().getNextHop(ip);
            boolean nextHostFound = false;
            // [$\color{pgreen} \lambda$=2, $\color{pgreen} \iota$=2, $\color{pgreen} \nu$=1, $\color{pgreen} \mu$=2, CCR=7, NMCC=3]
            for (Host aDirectlyAccessible: directlyAccessible) {
                // [$\color{pgreen} \lambda$=3, $\color{pgreen} \iota$=1, $\color{pgreen} \nu$=0, $\color{pgreen} \mu$=1, CCR=4, NMCC=1]
                if (aDirectlyAccessible.hasIP(nextIP)) { //Search the nextHop host object
                    aDirectlyAccessible.routeAPacketTo(ip, ttl - 1, usedHosts);
                    nextHostFound = true;
                }
            }
            // [$\color{pgreen} \lambda$=2, $\color{pgreen} \iota$=1, $\color{pgreen} \nu$=0, $\color{pgreen} \mu$=1, CCR=3, NMCC=1]
            if (!nextHostFound) { //Routing problem
                throw new Exception("Routing problem, there is no route corresponding 
                                     to the packet or the destination host is on the internet");
            }
        }
    }
}
\end{lstlisting}
\end{minipage}
};

\begin{scope}[shift={(code.south west)}]
  \draw[awesome, thick, fill=awesome, fill opacity=0.05] (1.97,11.3) rectangle (2.5,11.65);
\end{scope}

\begin{scope}[shift={(code.south west)}]
  \draw[awesome, thick, fill=awesome, fill opacity=0.05] (1.97,9.9) rectangle (2.5,10.2);
\end{scope}

\begin{scope}[shift={(code.south west)}]
  \draw[awesome, thick, fill=awesome, fill opacity=0.05] (2.37,8.8) rectangle (2.9,9.1);
\end{scope}

\begin{scope}[shift={(code.south west)}]
  \draw[awesome, thick, fill=awesome, fill opacity=0.05] (2.75,8.2) rectangle (3.27,8.5);
\end{scope}

\begin{scope}[shift={(code.south west)}]
  \draw[awesome, thick, fill=awesome, fill opacity=0.05] (2.37,6.8) rectangle (2.9,7.15);
\end{scope}

\begin{scope}[shift={(code.south west)}]
  \draw[awesome, thick, fill=awesome, fill opacity=0.05] (2.37,5.95) rectangle (2.9,6.25);
\end{scope}

\begin{scope}[shift={(code.south west)}]
  \draw[awesome, thick, fill=awesome, fill opacity=0.05] (2.75,4.55) rectangle (3.26,4.9);
\end{scope}

\begin{scope}[shift={(code.south west)}]
  \draw[awesome, thick, fill=awesome, fill opacity=0.05] (3.1,4) rectangle (3.67,4.3);
\end{scope}

\begin{scope}[shift={(code.south west)}]
  \draw[awesome, thick, fill=awesome, fill opacity=0.05] (2.75,2.32) rectangle (3.26,2.65);
\end{scope}

\begin{scope}[shift={(code.south west)}]
\draw[-, thick, awesome] (0.9,10.45) -- (0.9,11.08);
\end{scope}

\begin{scope}[shift={(code.south west)}]
\draw[-, thick, awesome] (0.9,0.95) -- (0.9,9.65);
\end{scope}

\begin{scope}[shift={(code.south west)}]
\draw[-, thick, awesome] (1.3,7.4) -- (1.3,8.5);
\end{scope}

\begin{scope}[shift={(code.south west)}]
\draw[-, thick, awesome] (1.7,7.4) -- (1.7,7.95);
\end{scope}

\begin{scope}[shift={(code.south west)}]
\draw[-, thick, awesome] (1.3,6.2) -- (1.3,6.55);
\end{scope}

\begin{scope}[shift={(code.south west)}]
\draw[-, thick, awesome] (1.3,1.2) -- (1.3,5.65);
\end{scope}

\begin{scope}[shift={(code.south west)}]
\draw[-, thick, awesome] (1.7,2.9) -- (1.7,4.3);
\end{scope}

\begin{scope}[shift={(code.south west)}]
\draw[-, thick, awesome] (2.05,3.15) -- (2.05,3.75);
\end{scope}

\begin{scope}[shift={(code.south west)}]
\draw[-, thick, awesome] (1.7,1.4) -- (1.7,2.05);
\end{scope}

\node[draw=awesome, fill=white, text=awesome, thick, circle, minimum size=0.01cm, font=\tiny] at (-5.2,4.45) {1}; 
\node[draw=awesome, fill=white, text=awesome, thick, circle, minimum size=0.01cm, font=\tiny] at (-5.2,2.8) {8}; 
\node[draw=awesome, fill=white, text=awesome, thick, circle, minimum size=0.01cm, font=\tiny] at (-4.83,1.8) {2}; 
\node[draw=awesome, fill=white, text=awesome, thick, circle, minimum size=0.01cm, font=\tiny] at (-4.4,1.3) {1}; 
\node[draw=awesome, fill=white, text=awesome, thick, circle, minimum size=0.01cm, font=\tiny] at (-4.83,-0.05) {5}; 
\node[draw=awesome, fill=white, text=awesome, thick, circle, minimum size=0.01cm, font=\tiny] at (-4.83,-1.5) {4}; 
\node[draw=awesome, fill=white, text=awesome, thick, circle, minimum size=0.01cm, font=\tiny] at (-4.4,-2.47) {2}; 
\node[draw=awesome, fill=white, text=awesome, thick, circle, minimum size=0.01cm, font=\tiny] at (-4.4,-4.59) {1}; 
\node[draw=awesome, fill=white, text=awesome, thick, circle, minimum size=0.01cm, font=\tiny] at (-4.06,-2.86) {1}; 

\end{tikzpicture}

\caption[Count of accumulated inherent component]{Count of accumulated inherent component, $\iota$, from the example in Figure~\ref{fig:ccExampleInCode}.}
\label{fig:accumulated_inherent_comp}
\end{figure}

\begin{figure}[!ht]
\centering

\begin{tikzpicture}
\node[inner sep=0pt] (code) {
\begin{minipage}{\textwidth}
\begin{lstlisting}[columns=fullflexible, numbers=left,xleftmargin=0.5cm,xrightmargin=0cm,frame=single,framexleftmargin=0em,escapechar=ñ,stepnumber=1,
    showstringspaces=false,
    tabsize=1,
    breaklines=true,
    breakatwhitespace=false,
    basicstyle=\scriptsize,
    language=Java,
    mathescape=true]{}
// Cognitive complexity 20
public void routeAPacketTo(IPAddress ip, int ttl, List<Host> usedHosts) throws Exception {
    usedHosts.add(this);
    // [$\color{pgreen} \lambda$=0, $\color{pgreen} \iota$=1, $\color{pgreen} \nu$=0, $\color{pgreen} \mu$=0, CCR=1, NMCC=1]
    if (ttl == 0) {//Problem in routing
        throw new Exception("Routing problem, TTL is null : packet to " 
                            + ip + " deleted on host " + this.getName());
    }
    // [$\color{pgreen} \lambda$=0, $\color{pgreen} \iota$=8, $\color{pgreen} \nu$=11, $\color{pgreen} \mu$=6, CCR=19, NMCC=19]
    if (!hasIP(ip)) { //Packet not arrived
        Host nextHost = null;
        List<Host> directlyAccessibleHosts = getDirectlyAccessibleHosts();
        // [$\color{pgreen} \lambda$=1, $\color{pgreen} \iota$=2, $\color{pgreen} \nu$=1, $\color{pgreen} \mu$=2, CCR=5, NMCC=3]
        for (Host directlyAccessibleHost: directlyAccessibleHosts) {
            // [$\color{pgreen} \lambda$=2, $\color{pgreen} \iota$=1, $\color{pgreen} \nu$=0, $\color{pgreen} \mu$=1, CCR=3, NMCC=1]
            if (directlyAccessibleHost.hasIP(ip)) // If the packet is for a neighbour,
                                                  // We send it to him
                nextHost = directlyAccessibleHost;
        }
        // [$\color{pgreen} \lambda$=1, $\color{pgreen} \iota$=5, $\color{pgreen} \nu$=4, $\color{pgreen} \mu$=4, CCR=13, NMCC=9]
        if (nextHost != null) {
            nextHost.routeAPacketTo(ip, ttl - 1, usedHosts);
        // [$\color{pgreen} \lambda$=1, $\color{pgreen} \iota$=4, $\color{pgreen} \nu$=4, $\color{pgreen} \mu$=3, CCR=12, NMCC=9]
        } else {//We have to look in the routing table
            List<Host> directlyAccessible = getDirectlyAccessibleHosts();
            IPAddress nextIP = this.getRoutingTable().getNextHop(ip);
            boolean nextHostFound = false;
            // [$\color{pgreen} \lambda$=2, $\color{pgreen} \iota$=2, $\color{pgreen} \nu$=1, $\color{pgreen} \mu$=2, CCR=7, NMCC=3]
            for (Host aDirectlyAccessible  directlyAccessible) {
                // [$\color{pgreen} \lambda$=3, $\color{pgreen} \iota$=1, $\color{pgreen} \nu$=0, $\color{pgreen} \mu$=1, CCR=4, NMCC=1]
                if (aDirectlyAccessible.hasIP(nextIP)) { //Search the nextHop host object
                    aDirectlyAccessible.routeAPacketTo(ip, ttl - 1, usedHosts);
                    nextHostFound = true;
                }
            }
            // [$\color{pgreen} \lambda$=2, $\color{pgreen} \iota$=1, $\color{pgreen} \nu$=0, $\color{pgreen} \mu$=1, CCR=3, NMCC=1]
            if (!nextHostFound) { //Routing problem
                throw new Exception("Routing problem, there is no route corresponding 
                                     to the packet or the destination host is on the internet");
            }
        }
    }
}
\end{lstlisting}
\end{minipage}
};

\begin{scope}[shift={(code.south west)}]
  \draw[blue, thick, fill=blue, fill opacity=0.05] (2.63,11.3) rectangle (3.2,11.6);
\end{scope}

\begin{scope}[shift={(code.south west)}]
  \draw[blue, thick, fill=blue, fill opacity=0.05] (2.63,9.9) rectangle (3.35,10.22);
\end{scope}

\begin{scope}[shift={(code.south west)}]
  \draw[blue, thick, fill=blue, fill opacity=0.05] (3,8.75) rectangle (3.6,9.1);
\end{scope}

\begin{scope}[shift={(code.south west)}]
  \draw[blue, thick, fill=blue, fill opacity=0.05] (3.4,8.2) rectangle (4,8.55);
\end{scope}

\begin{scope}[shift={(code.south west)}]
  \draw[blue, thick, fill=blue, fill opacity=0.05] (3,6.8) rectangle (3.6,7.15);
\end{scope}

\begin{scope}[shift={(code.south west)}]
  \draw[blue, thick, fill=blue, fill opacity=0.05] (3,5.95) rectangle (3.6,6.3);
\end{scope}

\begin{scope}[shift={(code.south west)}]
  \draw[blue, thick, fill=blue, fill opacity=0.05] (3.4,4.55) rectangle (4,4.9);
\end{scope}

\begin{scope}[shift={(code.south west)}]
  \draw[blue, thick, fill=blue, fill opacity=0.05] (3.75,4) rectangle (4.37,4.3);
\end{scope}

\begin{scope}[shift={(code.south west)}]
  \draw[blue, thick, fill=blue, fill opacity=0.05] (3.4,2.3) rectangle (4,2.6);
\end{scope}

\begin{scope}[shift={(code.south west)}]
\draw[dotted, thick, blue] (0.9,10.5) -- (0.9,11.1);
\end{scope}

\begin{scope}[shift={(code.south west)}]
\draw[dotted, thick, blue] (0.9,0.9) -- (0.9,9.6);
\end{scope}

\begin{scope}[shift={(code.south west)}]
\draw[dotted, thick, blue] (1.28,7.4) -- (1.28,8.5);
\end{scope}

\begin{scope}[shift={(code.south west)}]
\draw[dotted, thick, blue] (1.65,7.47) -- (1.65,7.97);
\end{scope}

\begin{scope}[shift={(code.south west)}]
\draw[dotted, thick, blue] (1.28,6.2) -- (1.28,6.6);
\end{scope}

\begin{scope}[shift={(code.south west)}]
\draw[dotted, thick, blue] (1.28,1.3) -- (1.28,5.7);
\end{scope}

\begin{scope}[shift={(code.south west)}]
\draw[dotted, thick, blue] (1.65,2.9) -- (1.65,4.3);
\end{scope}

\begin{scope}[shift={(code.south west)}]
\draw[dotted, thick, blue] (2.05,3.2) -- (2.05,3.75);
\end{scope}

\begin{scope}[shift={(code.south west)}]
\draw[dotted, thick, blue] (1.65,1.5) -- (1.65,2.1);
\end{scope}

\begin{scope}[shift={(code.south west)}] 
\draw[<-, thick, blue] (0.9,8.65) -- (1.25,8.65);
\end{scope}

\begin{scope}[shift={(code.south west)}] 
\draw[<-, thick, blue] (0.9,8.1) -- (1.25,8.1);
\end{scope}
\begin{scope}[shift={(code.south west)}] 
\draw[<-, thick, blue] (1.25,8.1) -- (1.61,8.1);
\end{scope}

\begin{scope}[shift={(code.south west)}] 
\draw[<-, thick, blue] (0.9,6.7) -- (1.25,6.7);
\end{scope}

\begin{scope}[shift={(code.south west)}] 
\draw[<-, thick, blue] (0.9,5.8) -- (1.25,5.8);
\end{scope}

\begin{scope}[shift={(code.south west)}] 
\draw[<-, thick, blue] (0.9,4.45) -- (1.25,4.45);
\end{scope}
\begin{scope}[shift={(code.south west)}] 
\draw[<-, thick, blue] (1.25,4.45) -- (1.61,4.45);
\end{scope}

\begin{scope}[shift={(code.south west)}] 
\draw[<-, thick, blue] (0.9,3.87) -- (1.25,3.87);
\end{scope}
\begin{scope}[shift={(code.south west)}] 
\draw[<-, thick, blue] (1.25,3.87) -- (1.61,3.87);
\end{scope}
\begin{scope}[shift={(code.south west)}] 
\draw[<-, thick, blue] (1.61,3.87) -- (2,3.87);
\end{scope}

\begin{scope}[shift={(code.south west)}] 
\draw[<-, thick, blue] (0.9,2.2) -- (1.25,2.2);
\end{scope}
\begin{scope}[shift={(code.south west)}] 
\draw[<-, thick, blue] (1.25,2.2) -- (1.61,2.2);
\end{scope}

\node[blue] at (-5,2.6) {1}; 

\node[blue] at (-5,2) {1}; 
\node[blue] at (-4.65,2) {1}; 

\node[blue] at (-5,0.6) {1}; 

\node[blue] at (-5,-0.3) {0}; 

\node[blue] at (-5,-1.6) {1}; 
\node[blue] at (-4.65,-1.6) {1}; 

\node[blue] at (-5,-2.25) {1}; 
\node[blue] at (-4.65,-2.25) {1}; 
\node[blue] at (-4.25,-2.25) {1}; 

\node[blue] at (-5,-3.9) {1}; 
\node[blue] at (-4.65,-3.9) {1}; 

\end{tikzpicture}

\caption[Count of accumulated nested component]{Count of accumulated nested component, $\nu$, from the example in Figure~\ref{fig:ccExampleInCode}. Note that the \texttt{else} sentence does not increment the nested component, just the inherent one.}
\label{fig:accumulated_nesting_comp}
\end{figure}

\begin{figure}[!ht]
\centering

\begin{tikzpicture}
\node[inner sep=0pt] (code) {
\begin{minipage}{\textwidth}
\begin{lstlisting}[columns=fullflexible, numbers=left,xleftmargin=0.5cm,xrightmargin=0cm,frame=single,framexleftmargin=0em,escapechar=ñ,stepnumber=1,
    showstringspaces=false,
    tabsize=1,
    breaklines=true,
    breakatwhitespace=false,
    basicstyle=\scriptsize,
    language=Java,
    mathescape=true]{}
// Cognitive complexity 20
public void routeAPacketTo(IPAddress ip, int ttl, List<Host> usedHosts) throws Exception {
    usedHosts.add(this);
    // [$\color{pgreen} \lambda$=0, $\color{pgreen} \iota$=1, $\color{pgreen} \nu$=0, $\color{pgreen} \mu$=0, CCR=1, NMCC=1]
    if (ttl == 0) {//Problem in routing
        throw new Exception("Routing problem, TTL is null : packet to " 
                            + ip + " deleted on host " + this.getName());
    }
    // [$\color{pgreen} \lambda$=0, $\color{pgreen} \iota$=8, $\color{pgreen} \nu$=11, $\color{pgreen} \mu$=6, CCR=19, NMCC=19]
    if (!hasIP(ip)) { //Packet not arrived
        Host nextHost = null;
        List<Host> directlyAccessibleHosts = getDirectlyAccessibleHosts();
        // [$\color{pgreen} \lambda$=1, $\color{pgreen} \iota$=2, $\color{pgreen} \nu$=1, $\color{pgreen} \mu$=2, CCR=5, NMCC=3]
        for (Host directlyAccessibleHost: directlyAccessibleHosts) {
            // [$\color{pgreen} \lambda$=2, $\color{pgreen} \iota$=1, $\color{pgreen} \nu$=0, $\color{pgreen} \mu$=1, CCR=3, NMCC=1]
            if (directlyAccessibleHost.hasIP(ip)) // If the packet is for a neighbour,
                                                  // We send it to him
                nextHost = directlyAccessibleHost;
        }
        // [$\color{pgreen} \lambda$=1, $\color{pgreen} \iota$=5, $\color{pgreen} \nu$=4, $\color{pgreen} \mu$=4, CCR=13, NMCC=9]
        if (nextHost != null) {
            nextHost.routeAPacketTo(ip, ttl - 1, usedHosts);
        // [$\color{pgreen} \lambda$=1, $\color{pgreen} \iota$=4, $\color{pgreen} \nu$=4, $\color{pgreen} \mu$=3, CCR=12, NMCC=9]
        } else {//We have to look in the routing table
            List<Host> directlyAccessible = getDirectlyAccessibleHosts();
            IPAddress nextIP = this.getRoutingTable().getNextHop(ip);
            boolean nextHostFound = false;
            // [$\color{pgreen} \lambda$=2, $\color{pgreen} \iota$=2, $\color{pgreen} \nu$=1, $\color{pgreen} \mu$=2, CCR=7, NMCC=3]
            for (Host aDirectlyAccessible: directlyAccessible) {
                // [$\color{pgreen} \lambda$=3, $\color{pgreen} \iota$=1, $\color{pgreen} \nu$=0, $\color{pgreen} \mu$=1, CCR=4, NMCC=1]
                if (aDirectlyAccessible.hasIP(nextIP)) { //Search the nextHop host object
                    aDirectlyAccessible.routeAPacketTo(ip, ttl - 1, usedHosts);
                    nextHostFound = true;
                }
            }
            // [$\color{pgreen} \lambda$=2, $\color{pgreen} \iota$=1, $\color{pgreen} \nu$=0, $\color{pgreen} \mu$=1, CCR=3, NMCC=1]
            if (!nextHostFound) { //Routing problem
                throw new Exception("Routing problem, there is no route corresponding 
                                     to the packet or the destination host is on the internet");
            }
        }
    }
}
\end{lstlisting}
\end{minipage}
};

\begin{scope}[shift={(code.south west)}]
  \draw[green(munsell), thick, fill=green(munsell), fill opacity=0.05] (3.3,11.3) rectangle (3.95,11.6);
\end{scope}

\begin{scope}[shift={(code.south west)}]
  \draw[green(munsell), thick, fill=green(munsell), fill opacity=0.05] (3.45,9.9) rectangle (4.1,10.22);
\end{scope}

\begin{scope}[shift={(code.south west)}]
  \draw[green(munsell), thick, fill=green(munsell), fill opacity=0.05] (3.68,8.75) rectangle (4.3,9.1);
\end{scope}

\begin{scope}[shift={(code.south west)}]
  \draw[green(munsell), thick, fill=green(munsell), fill opacity=0.05] (4.05,8.2) rectangle (4.67,8.55);
\end{scope}

\begin{scope}[shift={(code.south west)}]
  \draw[green(munsell), thick, fill=green(munsell), fill opacity=0.05] (3.7,6.8) rectangle (4.3,7.15);
\end{scope}

\begin{scope}[shift={(code.south west)}]
  \draw[green(munsell), thick, fill=green(munsell), fill opacity=0.05] (3.7,5.95) rectangle (4.3,6.3);
\end{scope}

\begin{scope}[shift={(code.south west)}]
  \draw[green(munsell), thick, fill=green(munsell), fill opacity=0.05] (4.05,4.55) rectangle (4.7,4.9);
\end{scope}

\begin{scope}[shift={(code.south west)}]
  \draw[green(munsell), thick, fill=green(munsell), fill opacity=0.05] (4.45,4) rectangle (5.05,4.3);
\end{scope}

\begin{scope}[shift={(code.south west)}]
  \draw[green(munsell), thick, fill=green(munsell), fill opacity=0.05] (4.05,2.3) rectangle (4.66,2.6);
\end{scope}

\begin{scope}[shift={(code.south west)}]
\draw[-, thick, green(munsell)] (0.9,10.45) -- (0.9,11.08);
\end{scope}

\begin{scope}[shift={(code.south west)}]
\draw[-, thick, green(munsell)] (0.9,0.95) -- (0.9,9.65);
\end{scope}

\begin{scope}[shift={(code.south west)}]
\draw[-, thick, green(munsell)] (1.3,7.4) -- (1.3,8.5);
\end{scope}

\begin{scope}[shift={(code.south west)}]
\draw[-, thick, green(munsell)] (1.7,7.4) -- (1.7,7.95);
\end{scope}

\begin{scope}[shift={(code.south west)}]
\draw[-, thick, green(munsell)] (1.3,6.2) -- (1.3,6.55);
\end{scope}

\begin{scope}[shift={(code.south west)}]
\draw[-, thick, green(munsell)] (1.3,1.2) -- (1.3,5.65);
\end{scope}

\begin{scope}[shift={(code.south west)}]
\draw[-, thick, green(munsell)] (1.7,2.9) -- (1.7,4.3);
\end{scope}

\begin{scope}[shift={(code.south west)}]
\draw[-, thick, green(munsell)] (2.05,3.15) -- (2.05,3.75);
\end{scope}

\begin{scope}[shift={(code.south west)}]
\draw[-, thick, green(munsell)] (1.7,1.4) -- (1.7,2.05);
\end{scope}

\node[draw=green(munsell), fill=white, text=green(munsell), thick, circle, minimum size=0.01cm, font=\tiny] at (-4.83,1.8) {1}; 
\node[draw=green(munsell), fill=white, text=green(munsell), thick, circle, minimum size=0.01cm, font=\tiny] at (-4.4,1.3) {1}; 
\node[draw=green(munsell), fill=white, text=green(munsell), thick, circle, minimum size=0.01cm, font=\tiny] at (-4.83,-0.05) {1}; 
\node[draw=green(munsell), fill=white, text=green(munsell), thick, circle, minimum size=0.01cm, font=\tiny] at (-4.4,-2.47) {1}; 
\node[draw=green(munsell), fill=white, text=green(munsell), thick, circle, minimum size=0.01cm, font=\tiny] at (-4.4,-4.59) {1}; 
\node[draw=green(munsell), fill=white, text=green(munsell), thick, circle, minimum size=0.01cm, font=\tiny] at (-4.06,-2.86) {1}; 

\end{tikzpicture}
\caption[Count of the contribution of accumulated nesting component]{Count of the contribution of accumulated nesting component, $\mu$, from the example in Figure~\ref{fig:ccExampleInCode}.}
\label{fig:mu_contribution_to_nesting_comp}
\end{figure}

\newpage
\clearpage

\pagenumbering{roman}
\section{Resolution algorithms}
This appendix shows the pseudo-code of all algorithms named in~\ref{subsec:mo-algorithms}.

\begin{algorithm}[!ht]
\DontPrintSemicolon
\SetAlgoLined
\BlankLine
\KwIn{\emph{Number of combinations of weights}}
\KwOut{$PF$}

\For{$i \gets 0$ \KwTo Number of combinations of weights}{
    $(w_1, ..., w_p) \gets$ generate weights for subdivision $i$ \;
    $z \gets \text{solve }\left\{\min \left(\displaystyle\sum_{i=1}^{p} w_i \cdot f_i(x) \right), \text{ subject to } x \in X\right\}$ \;
    $PF = PF \cup \{z\}$ \;
}

\caption{Weighted sum algorithm for $p$ objectives}
\label{alg:weighted-sum-algorithm}
\end{algorithm}

For each iteration, Algorithm~\ref{alg:augmecon} sets a positive value for the $\lambda$ constant, which must be small enough to prevent the algorithm from omitting some of the efficient solutions, and large enough to avoid numerical problems. In general, it is sufficient to take a value of $\lambda$ in $[10^{-6}, 10^{-3}]$. Chicano et al.~\cite{augmecon2016chicano} chose to calculate the value of $\lambda$ at each iteration, taking into account the efficient point obtained previously. Given a multi-objective optimization problem, the \emph{ideal} point is $ID = (id_1, ..., id_n)$, where $\displaystyle id_i = \min_{x \in PF} x_i$, and the \emph{nadir} point as $NA = (na_1, ..., na_n)$, where $na_i = \displaystyle\max_{x \in PF} x_i $. In this work, $\lambda$ is calculated as $\lambda = \dfrac{1}{(f_1(z) - l_1) \cdot 10^{-3}}$, where $l_1 = id_1$ is a lower bound of $\min f_1(x), x \in X$, that is the first component of the ideal point. Variable $s$ is a slack variable.

\begin{algorithm}[!ht]
\DontPrintSemicolon
\SetAlgoLined
\BlankLine
\KwOut{$PF$}

$z \gets \text{solve }\left\{\min f_2(x), \text{ subject to } x \in X\right\}$ \;
$z \gets \text{solve }\left\{\min f_1(x), \text{ subject to } f_2(x) \leq f_2(z), x \in X\right\}$ \;

$PF \gets \{z\}$ // Pareto front \;
$\varepsilon \gets f_1(z) - 1$ \;

\While{$\exists x \in X, f_1(x) \leq \varepsilon$}{
    Estimate a value for $\lambda > 0$ \;
    $z \gets \text{solve }\left\{\min f_2(x) - \lambda \cdot s, \text{ subject to } f_1(x) + s = \varepsilon, x \in X\right\}$ \;
    $PF = PF \cup \{z\}$ \;
    $\varepsilon \gets f_1(x) - 1$ \;
}

\caption{Augmented $\varepsilon$-constraint (AUGMECON)}
\label{alg:augmecon}
\end{algorithm}

\begin{figure}[!ht]
    \centering
    \begin{subfigure}[t]{0.49\textwidth}
        \centering
        \includegraphics[width=\linewidth]{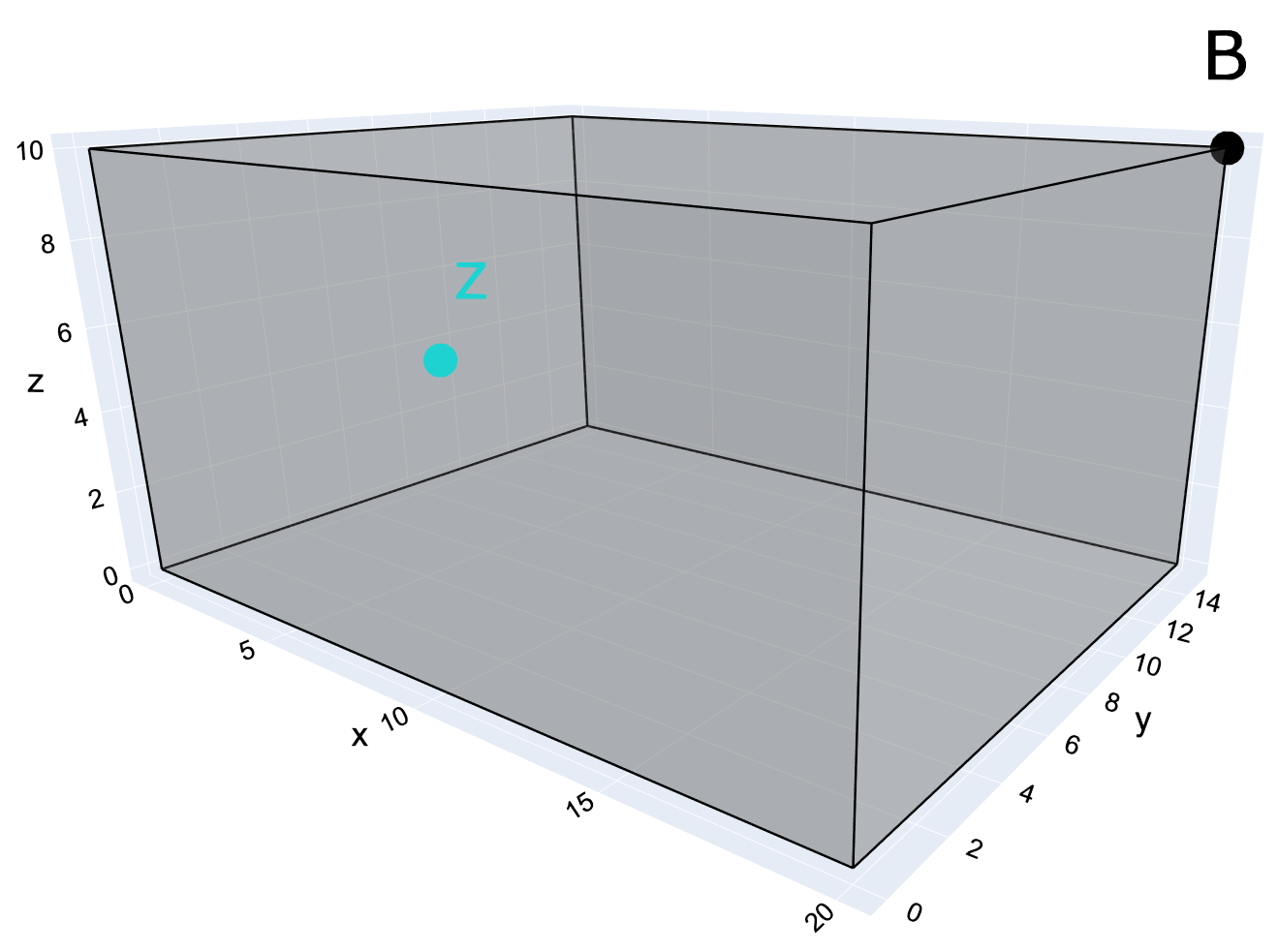}
        \caption{Box $\mathbf{B} = [(0,0,0),(20,15,10)]$, with non-dominated solution $z = (5,5,5)$.}
        \label{subfig:initial_box}
    \end{subfigure}
    \hfill
    \begin{subfigure}[t]{0.49\textwidth}
        \centering
        \includegraphics[width=\linewidth]{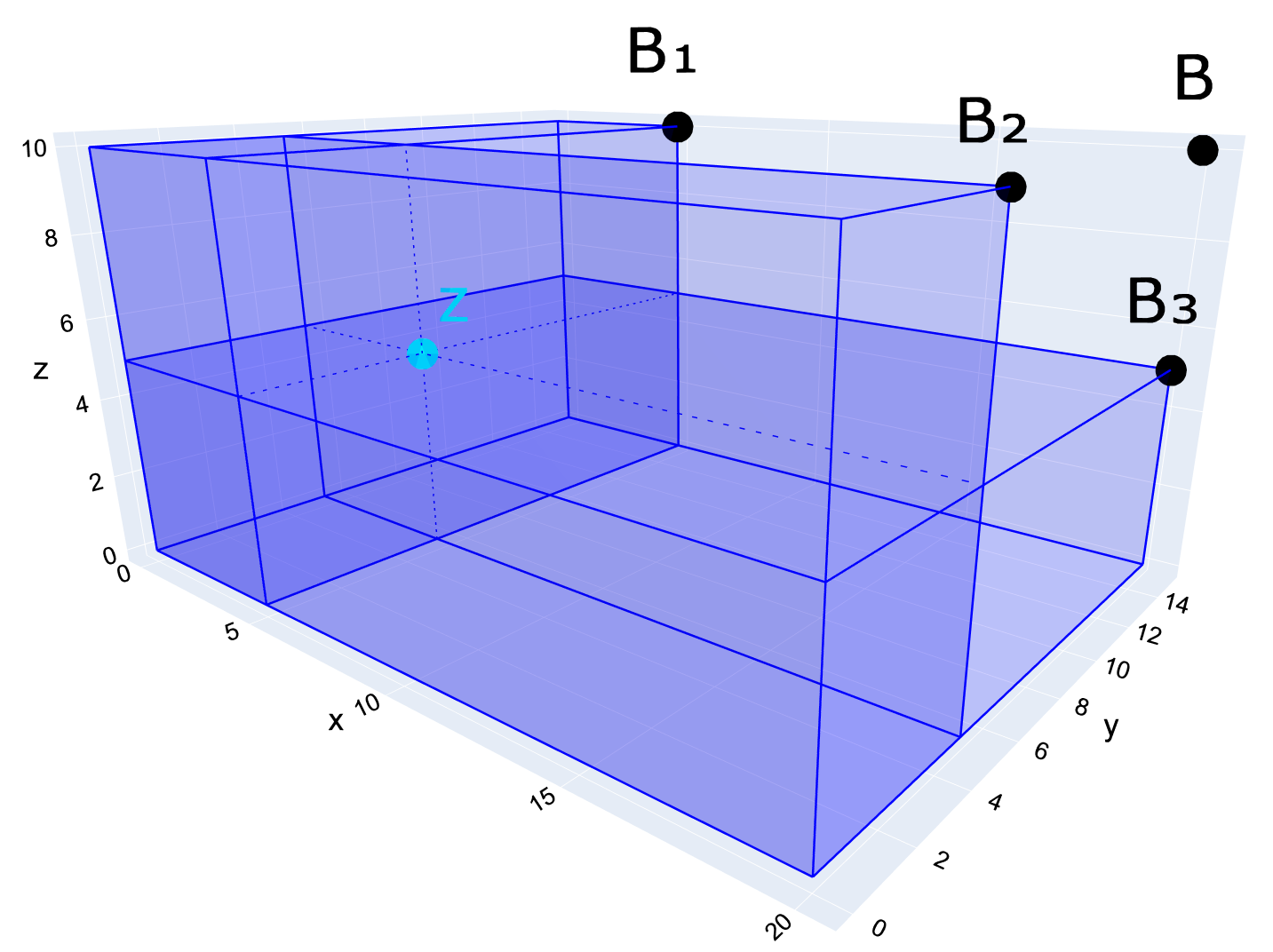}
        \caption{After applying \emph{full-p-split}: $\mathbf{B_1} = [(5,15,10)]$, $\mathbf{B_2} = [(20,5,10)]$, $\mathbf{B_3} = [(20,15,5)]$. The lower bounds of the boxes are all $ID = [(0,0,0)]$.}
        \label{subfig:first_partition}
    \end{subfigure}
\caption{Partitioning a box using \emph{full-p-split}.}
\label{fig:full_p_split_example}
\end{figure}

Algorithm~\ref{alg:hybrid-method} divides the objective space and explores it one section at a time. Each time it finds a new solution within one box, it is divided, and the boxes that contain only dominated solutions are discarded. The result after each iteration is the number of new boxes equal to the objectives for each solution found in the initial box. Once the first box is divided, the new boxes overlap, but they occupy a lower volume than the initial one. When a new solution is found, all the boxes that include that solution are split as well. Figure~\ref{fig:full_p_split_example} shows an example of the first partition of \emph{full p-split}, and it can be found interactively in Zenodo\footnote{https://doi.org/10.5281/zenodo.15682834}. If the algorithm ends, then the complete PF is obtained, including unsupported solutions. This is the great advantage of the algorithm, as it can not only find all solutions for a convex PF, but also for a PF that has unsupported solutions.

To solve the multi-objective problem, Algorithm~\ref{alg:hybrid-method} takes as input the initial box, that is, the whole objective space. One example for an initial box with $NA = (20,15,10)$ is depicted in Figure~\ref{subfig:initial_box}. It finds the complete PF by applying the multi-objective ILP model to each box obtained. An example of the first iteration of \emph{full p-split} is depicted in Figure~\ref{subfig:first_partition}. As a result of the \emph{full p-split}, one box may contain another one, so \emph{lines $4-18$} of Algorithm~\ref{alg:redundancy-elimination} iterate over all pairs of boxes and eliminate those boxes that are contained in larger boxes.

\begin{algorithm}[!ht]
\DontPrintSemicolon
\SetAlgoLined
\KwIn{$\mathcal{U}, z \quad$ \tcp{Set of boxes and solution found}}
\KwOut{$\mathcal{U} \quad$ \tcp{New boxes set}}
\BlankLine

$\mathfrak{R} \gets \{ u \in \mathcal{U} : z < u \}$ \tcp{Boxes that contain $z$}
\For{$j \in 1, ..., p$}{
    $\mathfrak{D}_j \gets \{u \in \mathcal{U} : z_j = u_j \land z_k < u_k, \forall k \neq j \}$ \;
    $ \mathfrak{P}_j \gets \varnothing$ \;
}

\For{$u \in \mathfrak{R}$}{
    \For{$j \in 1, ..., p$}{
        $ \mathfrak{P}_j \gets \mathfrak{P}_j \bigcup \{(u_1, ..., u_{j-1} , z_j, u_{j+1}, ..., u_p) \}$ \;
    }
}

\For{$j \in 1, ..., p$}{
    $ \mathfrak{P}_j \gets \{ z' \in \mathfrak{P}_j  : z' \not\leq u', \forall u' \in \mathfrak{P}_j \cup \mathfrak{D}_j - \{z'\} \}$ \tcp{Filter redundant boxes}
}

$\mathcal{U} \gets (\mathcal{U} - \mathfrak{R}) \bigcup \Bigl( \bigcup_{j=1}^{p} \mathfrak{P}_j \Bigr)$

\caption{RE (Redundancy elimination)}
\label{alg:redundancy-elimination}
\end{algorithm}

\begin{algorithm}[!ht]
\DontPrintSemicolon
\SetAlgoLined
\KwIn{$B_0$}
\KwOut{$PF$}
\BlankLine

$PF \gets \varnothing$ \;
$\mathcal{U} \gets [B_0]$ \;

$P(u) \equiv \displaystyle \left\{\min \left( \sum_{i=1}^{p} f_i(x)\right), \text{ subject to } f(x) \leq u, x \in \mathcal{X}\right\}$ \;

\While{$\bigl( \mathcal{U} \neq \varnothing \bigr)$}{
    $B \gets$ $\mathcal{U}_{first}$ \;
    \eIf{$\bigl( P(B)$ has solution $\bigr)$} {
    
        $x^* \gets$ Optimal solution of $P(B)$  \;
        $PF \gets PF \cup \left\{ f(x^*) \right\}$ \;
        $\mathcal{U} \gets$ FCB$\bigl(\mathcal{U}, f(x^*)\bigr)$ \;
    }{
    $\mathcal{U} \gets \mathcal{U} - {B}$
    }
    
}
\caption{Hybrid Method using \emph{full p-split}}
\label{alg:hybrid-method}
\end{algorithm}

\newpage
\clearpage

\pagenumbering{roman}
\section{Open-source projects additional information}
\label{ch:ApendiceC_tablas_open-source}

This appendix shows all the tables referred to in Section~\ref{sec:results}, which, for reasons of space, are more conveniently consulted here.

Table~\ref{tab:index-methods} displays the index assigned to each method, the project to which it belongs, the class within the project where it can be found, and its name. Due to space constraints, each class is simplified, and only the beginning and the end of the route are shown, connected by ellipses. Tables~\ref{tab:hipervolumen-open-source-dos-objs} and~\ref{tab:hipervolumen-open-source-tres-objs} are described in Sub-subsections~\ref{subsubsec:results_open_source_two_objs} and~\ref{subsubsec:results_open_source_three_objs}, respectively.

If the solution that best balances LOC in Figure~\ref{fig:routeAPacketTo_general} is chosen, i.e., the second solution (\textbf{s2}), the corresponding extractions are the ones in Figure~\ref{fig:refactoring_solution_multi-objective}. These figures are referred to in Sub-subsection~\ref{subsubsec:results_open_source_three_objs}.

\begin{landscape}

\begin{longtable}[!ht]{clll}
\caption{Methods selected from the nine open-source projects studied.} 
\label{tab:index-methods}
\\
\hline
\multicolumn{1}{l}{\textbf{ID}} & \multicolumn{1}{l}{\textbf{project}} & \multicolumn{1}{l}{\textbf{class}} & \multicolumn{1}{l}{\textbf{method}} \\ \hline
1 & bytecode-viewer & src.main.java.the.bytecode...ASMResourceUtil.java & renameClassNode \\
2 & bytecode-viewer & src.main.java.the.bytecode...ASMResourceUtil.java & renameFieldNode \\
3 & bytecode-viewer & src.main.java.the.bytecode...ASMResourceUtil.java & renameMethodNode \\
4 & bytecode-viewer & src.main.java.the.bytecode...Boot.java & downloadZipsOnly \\
5 & bytecode-viewer & src.main.java.the.bytecode...BytecodeViewer.java & compile \\
6 & bytecode-viewer & src.main.java.the.bytecode...CommandLineInput.java & parseCommandLine \\
7 & bytecode-viewer & src.main.java.the.bytecode...ProcyonDecompiler.java & doSaveJarDecompiled \\
8 & bytecode-viewer & src.main.java.the.bytecode....ResourceListIconRenderer.java & getTreeCellRendererComponent \\
9 & bytecode-viewer & src.main.java.the.bytecode...ReplaceStrings.java & scanClassNode \\
10 & bytecode-viewer & src.main.java.the...ViewAPKAndroidPermissions.java & execute \\
11 & bytecode-viewer & src.main.java.the.bytecode...ZStringArrayDecrypter.java & execute \\
12 & bytecode-viewer & src.main.java.the.bytecode...impl.APKExport.java & promptForExport \\
13 & bytecode-viewer & src.main.java.the.bytecode...DirectoryResourceImporter.java & open \\
14 & bytecode-viewer & src.main.java.the.bytecode...util.BootCheck.java & failSafeLoadLibraries \\
15 & bytecode-viewer & src.main.java.the.bytecode...util.FileDrop.java & FileDrop \\
16 & bytecode-viewer & src.main.java.the.bytecode...util.JarUtils.java & importArchiveA \\
17 & bytecode-viewer & src.main.java.the.bytecode...util.JarUtils.java & importArchiveB \\
18 & bytecode-viewer & src.main.java.the.bytecode...util.JTextAreaUtils.java & search \\
19 & bytecode-viewer & src.main.java.the.bytecode...util.MethodParser.java & findNearestMethod \\
20 & bytecode-viewer & src.main.java.the.bytecode...util.SecurityMan.java & checkExec \\
\hline
21 & cybercaptor-server & src.main.java.org.fiware...AttackGraph.java & getRelatedTopologyGraph \\
22 & cybercaptor-server & src.main.java.org.fiware...AttackPath.java & getLeavesThatCanBeRemediated \\
23 & cybercaptor-server & src.main.java.org.fiware...AttackPath.java & getRemedationActions \\
24 & cybercaptor-server & src.main.java.org.fiware...AttackPath.java & getRemediationAction \\
25 & cybercaptor-server & src.main.java.org.fiware...AttackPath.java & getRemediationActionForLeaf \\
26 & cybercaptor-server & src.main.java.org.fiware...AttackPath.java & leavesMandatoryForVertex \\
27 & cybercaptor-server & src.main.java.org.fiware...AttackPath.java & printListRemediations \\
28 & cybercaptor-server & src.main.java.org.fiware...AttackPath.java & remediateLeaf \\
29 & cybercaptor-server & src.main.java.org.fiware...MulvalAttackGraph.java & addArcsAndVerticesFromDomElement \\
30 & cybercaptor-server & src.main.java.org.fiware...Vertex.java & getRelatedMachine \\
31 & cybercaptor-server & src.main.java.org.fiware.cybercaptor...AttackPaths.java & exploreAttackPath \\
32 & cybercaptor-server & src.main.java.org.fiware.cybercaptor...Interface.java & mergeTwoInterfaces \\
33 & cybercaptor-server & src.main.java.org.fiware.cybercaptor...Host.java & equals \\
34 & cybercaptor-server & src.main.java.org.fiware.cybercaptor...Host.java & hostThatPreventToSendAPacket \\
35 & cybercaptor-server & src.main.java.org.fiware.cybercaptor...Host.java & routeAPacketTo \\
36 & cybercaptor-server & src.main.java.org.fiware.cybercaptor...Topology.java & clone \\
37 & cybercaptor-server & src.main.java.org.fiware.cybercaptor...Topology.java & mergeTwoHosts \\
38 & cybercaptor-server & src.main.java.org.fiware.cybercaptor...Topology.java & sendAPacketOnARoute \\
39 & cybercaptor-server & src.main.java.org.fiware.cybercaptor...CPE.java & populateCPECVEDatabase \\
\hline
40 & fastjson & src.main.java.com...JSON.java & toJSON \\
41 & fastjson & src.main.java.com...JSONPath.java & buildArraySegement \\
42 & fastjson & src.main.java.com...JSONPath.java & extract(JSONPath-DefaultJSONParser-Context) \\
43 & fastjson & src.main.java.com...JSONPath.java & getArrayItem \\
44 & fastjson & src.main.java.com...JSONPath.java & getPropertyValue \\
45 & fastjson & src.main.java.com...JSONPath.java & paths \\
46 & fastjson & src.main.java.com...JSONPath.java & readSegement \\
47 & fastjson & src.main.java.com...AbstractDateDeserializer.java & deserialze \\
48 & fastjson & src.main.java.com...DefaultFieldDeserializer.java & parseField \\
49 & fastjson & src.main.java.com...parser.deserializer.Jdk8DateCodec.java & parseDateTime \\
50 & fastjson & src.main.java.com...parser.deserializer.Jdk8DateCodec.java & parseLocalDate \\
51 & fastjson & src.main.java.com...parser.deserializer.Jdk8DateCodec.java & parseZonedDateTime \\
52 & fastjson & src.main.java.com...parser.JSONLexerBase.java & integerValue \\
53 & fastjson & src.main.java.com...parser.JSONLexerBase.java & scanFieldStringArray \\
54 & fastjson & src.main.java.com...parser.JSONLexerBase.java & scanNumber \\
55 & fastjson & src.main.java.com...parser.JSONScanner.java & seekObjectToField(long{[}{]}) \\
56 & fastjson & src.main.java.com...serializer.JodaCodec.java & parseDateTime \\
57 & fastjson & src.main.java.com...serializer.PrimitiveArraySerializer.java & write \\
58 & fastjson & src.main.java.com...util.TypeUtils.java & cast(Object-Class-T--ParserConfig) \\
59 & fastjson & src.main.java.com...util.TypeUtils.java & castToDate \\
60 & fastjson & src.main.java.com...util.TypeUtils.java & castToJavaBean \\
\hline
61 & fiware-commons & src.main.java.com...OpenStackKeystoneV2.java & parseEndpoint \\
62 & fiware-commons & src.main.java.com...OpenStackKeystoneV2.java & parseRegionNames \\
63 & fiware-commons & src.main.java.com...OpenStackKeystoneV3.java & parseRegionNames \\
\hline
64 & iotbroker & src.main.java.com...amqp.AmqpClient.java & processAttach \\
65 & iotbroker & src.main.java.com...amqp.net.AMQPDecoder.java & decode \\
66 & iotbroker & src.main.java.com...amqp.TimersMap.java & store \\
67 & iotbroker & src.main.java.com...coap.CoapClient.java & packetReceived \\
68 & iotbroker & src.main.java.com...mqtt.MqttClient.java & processSuback \\
69 & iotbroker & src.main.java.com...mqtt.net.MQDecoder.java & decode \\
70 & iotbroker & src.main.java.com...mqtt.TimersMap.java & store \\
71 & iotbroker & src.main.java.com...mqttsn.MqttsnClient.java & processSuback \\
72 & iotbroker & src.main.java.com...mqttsn.TimersMap.java & store \\
73 & iotbroker & src.main.java.com...ui.LogInPane.java & loginBtnClicked \\
\hline
74 & jedis & src.main.java...jedis.JedisClusterCommand.java & runWithRetries \\
75 & jedis & src.main.java...jedis.JedisPubSub.java & process \\
76 & jedis & src.main.java...jedis.JedisSentinelPool.java & run \\
\hline
77 & jmetal & src.main.java.org.uma.jmetal...DMOPSO.java & fitnessFunction \\
78 & jmetal & src.main.java.org.uma.jmetal...ScalarizationUtils.java & tradeoffUtility \\
79 & jmetal & src.main.java.org.uma.jmetal...GWASFGARanking.java & computeRanking \\
80 & jmetal & src.main.java.org.uma.jmetal...GWASFGARanking.java & rankUnfeasibleSolutions \\
81 & jmetal & src.main.java.org.uma.jmetal...MOCHC45.java & run \\
82 & jmetal & src.main.java.org.uma.jmetal...AbstractMOEAD.java & fitnessFunction \\
83 & jmetal & src.main.java.org.uma.jmetal...ConstraintMOEAD.java & updateNeighborhood \\
84 & jmetal & src.main.java.org.uma.jmetal...MOEADD.java & deleteCrowdRegion1 \\
85 & jmetal & src.main.java.org.uma.jmetal...MOEADD.java & deleteCrowdRegion2 \\
86 & jmetal & src.main.java.org.uma.jmetal...MOEADD.java & deleteRankOne \\
87 & jmetal & src.main.java.org.uma.jmetal...MOEADD.java & matingSelection \\
88 & jmetal & src.main.java...MOEADD.java\_nondominated\_sorting & delete \\
89 & jmetal & src.main.java.org.uma.jmetal...EnvironmentalSelection.java & execute \\
90 & jmetal & src.main.java.org.uma.jmetal...LZ09.java & objective \\
\hline
91 & knowage-core & src.main.java.it.eng.spagobi...AbstractDriverRuntime.java & loadAdmissibleValues \\
92 & knowage-core & src.main.java.it.eng.spagobi...DriversValidationAPI.java & getValidationErrorOnCheck \\
93 & knowage-core & src.main.java.it.eng.spagobi...ExecutionInstance.java & applyViewpoint \\
94 & knowage-core & src.main.java.it.eng.spagobi...ExecutionInstance.java & getValidationErrorOnCheck \\
95 & knowage-core & src.main.java.it.eng.spagobi...SelfServiceDataSetCRUD.java & validateDataset \\
96 & knowage-core & src.main.java.it.eng...MenuListJSONSerializerForREST.java & createUserMenuElement \\
97 & knowage-core & src.main.java.it.eng...DocumentCompositionUtils.java & getExecutionUrl \\
98 & knowage-core & src.main.java.it.eng.spagobi...KpiService.java & checkConflictsWithKpi \\
99 & knowage-core & src.main.java.it.eng...GeoLayerJSONDeserializer.java & deserialize \\
100 & knowage-core & src.main.java.it.eng.spagobi...SchedulerServiceSupplier.java & getChronExpression \\
101 & knowage-core & src.main.java.it.eng...AbstractAssociativityManager.java & getSelections \\
102 & knowage-core & src.main.java.it.eng...DatasetManagementAPI.java & setDataSetParameters(IDataSet-Map-String,String--String) \\
103 & knowage-core & src.main.java...GeoSpatialDimensionDatasetValidator.java & doValidateDataset \\
104 & knowage-core & src.main.java.it.eng.spagobi...MenuUtilities.java & getMenuItems(SourceBean-SourceBean-IEngUserProfile) \\
105 & knowage-core & src.main.java.it.eng.spagobi...MenuUtilities.java & getMenuPath \\
\hline
106 & MOEAFramework & src.org.moeaframework.algorithm.CMAES.java & checkEigenSystem \\
107 & MOEAFramework & src.org.moeaframework.algorithm.CMAES.java & eigendecomposition \\
108 & MOEAFramework & src.org.moeaframework.algorithm.CMAES.java & samplePopulation \\
109 & MOEAFramework & src.org.moeaframework.algorithm.CMAES.java & tql2 \\
110 & MOEAFramework & src.org.moeaframework.algorithm.CMAES.java & updateDistribution \\
111 & MOEAFramework & src.org.moeaframework.algorithm.DBEA.java & updateIdealPointAndIntercepts \\
112 & MOEAFramework & src.org.moeaframework.algorithm.DBEA.java & updatePopulation \\
113 & MOEAFramework & src.org.moeaframework...jmetal.JMetalAlgorithms.java & getAlgorithm \\
114 & MOEAFramework & src.org.moeaframework.algorithm.MOEAD.java & updateSolution \\
115 & MOEAFramework & src.org.moeaframework...MSOPSRankedPopulation.java & update \\
116 & MOEAFramework & src.org.moeaframework.algorithm.NSGAII.java & iterate \\
117 & MOEAFramework & src...ReferencePointNondominatedSortingPopulation.java & calculateIntercepts \\
118 & MOEAFramework & src...ReferencePointNondominatedSortingPopulation.java & lsolve \\
119 & MOEAFramework & src...ReferencePointNondominatedSortingPopulation.java & truncate \\
120 & MOEAFramework & src.org.moeaframework...SPEA2.java & evaluate \\
121 & MOEAFramework & src.org.moeaframework...StandardAlgorithms.java & getAlgorithm \\
\hline
\end{longtable}
\end{landscape}

\begin{longtable}{rccccccc}
\caption[Results obtained for each method]{Results obtained for methods under study in open-source projects for two objectives (\emph{Extractions~(E), CC~(C)}). Each column represents, respectively, the index of each method (see Table~\ref{tab:index-methods}), number of solutions obtained for the method, reference point used for the method to compute the HV, normalized HV (N-HV), and the median and inter-quartile range (\emph{IQR})  for the two objective functions.}
\label{tab:hipervolumen-open-source-dos-objs}
\footnotesize
\\
\hline
\multirow{2}{1cm}{\centering\textbf{ID}} & \multirow{2}{1.5cm}{\textbf{No. sol}} & \textbf{Reference} & \multirow{2}{1.5cm}{\centering\textbf{N-HV}} & \multicolumn{2}{c}{\textbf{Extractions}} & \multicolumn{2}{c}{\textbf{CC}}\\
& & \emph{(E,C)} & & \emph{Median} & \emph{IQR} & \emph{Median} & \emph{IQR}  \\ 
\hline
  1 & 1 &   (5,2) & 1.00 &  4.0 & 0.00 &  1.0 & 0.00 \\
  2 & 2 &   (5,3) & 0.67 &  3.0 & 1.00 &  1.5 & 0.50 \\
  3 & 3 &   (8,4) & 0.53 &  6.0 & 2.00 &  2.0 & 1.00 \\
  4 & 1 &   (3,3) & 1.00 &  2.0 & 0.00 &  2.0 & 0.00 \\
  5 & 2 &   (5,6) & 0.62 &  3.5 & 0.50 &  3.5 & 1.50 \\
  6 & 2 &   (4,5) & 0.60 &  2.5 & 0.50 &  2.0 & 2.00 \\
  7 & 3 &  (6,12) & 0.57 &  4.0 & 1.00 &  8.0 & 3.00 \\
  8 & 2 &   (5,6) & 0.62 &  3.5 & 0.50 &  3.5 & 1.50 \\
  9 & 1 &   (5,2) & 1.00 &  4.0 & 0.00 &  1.0 & 0.00 \\
 10 & 2 &   (4,3) & 0.67 &  2.5 & 0.50 &  1.0 & 1.00 \\
 11 & 2 &   (4,4) & 0.67 &  2.5 & 0.50 &  2.0 & 1.00 \\
 12 & 1 &   (3,2) & 1.00 &  2.0 & 0.00 &  1.0 & 0.00 \\
 13 & 1 &   (3,3) & 1.00 &  2.0 & 0.00 &  2.0 & 0.00 \\
 14 & 1 &   (3,1) & 1.00 &  2.0 & 0.00 &  0.0 & 0.00 \\
 15 & 4 &  (12,6) & 0.59 &  7.0 & 3.25 &  3.5 & 1.50 \\
 16 & 3 &   (8,6) & 0.73 &  3.0 & 2.50 &  2.0 & 2.00 \\
 17 & 3 &   (6,4) & 0.67 &  3.0 & 1.50 &  2.0 & 1.00 \\
 18 & 3 &  (8,11) & 0.53 &  6.0 & 2.00 &  9.0 & 1.00 \\
 19 & 2 &  (4,14) & 0.60 &  2.5 & 0.50 & 11.0 & 2.00 \\
 20 & 3 &   (5,4) & 0.67 &  3.0 & 1.00 &  2.0 & 1.00 \\
 21 & 4 &   (9,9) & 0.60 &  6.0 & 2.50 &  3.5 & 4.00 \\
 22 & 3 &   (6,4) & 0.58 &  4.0 & 1.50 &  2.0 & 1.00 \\
 23 & 3 &  (11,3) & 0.67 &  3.0 & 4.00 &  1.0 & 1.00 \\
 24 & 3 &  (11,3) & 0.67 &  3.0 & 4.00 &  1.0 & 1.00 \\
 25 & 3 &  (17,4) & 0.59 &  8.0 & 6.00 &  2.0 & 1.00 \\
 26 & 1 &   (3,1) & 1.00 &  2.0 & 0.00 &  0.0 & 0.00 \\
 27 & 1 &   (3,1) & 1.00 &  2.0 & 0.00 &  0.0 & 0.00 \\
 28 & 3 &  (11,4) & 0.58 &  6.0 & 3.50 &  2.0 & 1.00 \\
 29 & 2 &   (5,4) & 0.56 &  3.0 & 1.00 &  2.0 & 1.00 \\
 30 & 8 & (15,15) & 0.59 &  9.5 & 3.75 &  7.0 & 7.00 \\
 31 & 1 &  (4,11) & 1.00 &  3.0 & 0.00 & 10.0 & 0.00 \\
 32 & 1 &   (3,1) & 1.00 &  2.0 & 0.00 &  0.0 & 0.00 \\
 33 & 1 &   (3,4) & 1.00 &  2.0 & 0.00 &  3.0 & 0.00 \\
 34 & 2 &   (4,8) & 0.58 &  2.5 & 0.50 &  4.5 & 2.50 \\
 35 & 2 &   (7,3) & 0.60 &  4.0 & 2.00 &  1.5 & 0.50 \\
 36 & 3 &  (10 3) & 0.62 &  4.0 & 3.50 &  1.0 & 1.00 \\
 37 & 2 &   (8,2) & 0.58 &  4.5 & 2.50 &  0.5 & 0.50 \\
 38 & 1 &   (3,2) & 1.00 &  2.0 & 0.00 &  1.0 & 0.00 \\
 39 & 2 &   (4,2) & 0.75 &  2.5 & 0.50 &  0.5 & 0.50 \\
 40 & 1 &   (4,2) & 1.00 &  3.0 & 0.00 &  1.0 & 0.00 \\
 41 & 4 &  (9,11) & 0.63 &  5.5 & 1.75 &  6.5 & 2.25 \\
 42 & 2 & (10,13) & 0.56 &  8.0 & 1.00 & 11.0 & 1.00 \\
 43 & 1 &   (4,2) & 1.00 &  3.0 & 0.00 &  1.0 & 0.00 \\
 44 & 1 &   (7,7) & 1.00 &  6.0 & 0.00 &  6.0 & 0.00 \\
 45 & 2 &   (7,4) & 0.50 &  4.5 & 1.50 &  2.0 & 1.00 \\
 46 & 1 &   (5,4) & 1.00 &  4.0 & 0.00 &  3.0 & 0.00 \\
 47 & 3 & (10,14) & 0.67 &  8.0 & 1.00 & 11.0 & 1.50 \\
 48 & 2 &   (5,5) & 0.75 &  3.5 & 0.50 &  3.5 & 0.50 \\
 49 & 1 &   (5,4) & 1.00 &  4.0 & 0.00 &  3.0 & 0.00 \\
 50 & 4 &   (7,7) & 0.55 &  4.5 & 1.50 &  4.5 & 1.75 \\
 51 & 3 &   (9,7) & 0.70 &  5.0 & 2.00 &  4.0 & 1.50 \\
 52 & 1 &   (4,2) & 1.00 &  3.0 & 0.00 &  1.0 & 0.00 \\
 53 & 1 &  (6,15) & 1.00 &  5.0 & 0.00 & 14.0 & 0.00 \\
 54 & 2 &   (5,5) & 0.62 &  3.5 & 0.50 &  2.5 & 1.50 \\
 55 & 4 & (18,12) & 0.63 & 12.5 & 6.00 &  7.5 & 2.00 \\
 56 & 2 &  (10,5) & 0.58 &  6.5 & 2.50 &  3.5 & 0.50 \\
 57 & 2 &  (10,4) & 0.44 &  6.5 & 2.50 &  2.0 & 1.00 \\
 58 & 2 &   (8,5) & 0.67 &  6.0 & 1.00 &  3.5 & 0.50 \\
 59 & 1 &   (6,7) & 1.00 &  5.0 & 0.00 &  6.0 & 0.00 \\
 60 & 1 &   (4,3) & 1.00 &  3.0 & 0.00 &  2.0 & 0.00 \\
 61 & 1 &  (3,14) & 1.00 &  2.0 & 0.00 & 13.0 & 0.00 \\
 62 & 2 &   (6,3) & 0.62 &  3.5 & 1.50 &  1.5 & 0.50 \\
 63 & 2 &   (6,3) & 0.62 &  3.5 & 1.50 &  1.5 & 0.50 \\
 64 & 2 &   (6,3) & 0.67 &  4.0 & 1.00 &  1.5 & 0.50 \\
 65 & 1 &   (3,2) & 1.00 &  2.0 & 0.00 &  1.0 & 0.00 \\
 66 & 3 &   (6,4) & 0.67 &  3.0 & 1.50 &  2.0 & 1.00 \\
 67 & 2 &   (6,5) & 0.67 &  4.5 & 0.50 &  3.0 & 1.00 \\
 68 & 2 &   (4,6) & 0.60 &  2.5 & 0.50 &  3.0 & 2.00 \\
 69 & 1 &   (3,2) & 1.00 &  2.0 & 0.00 &  1.0 & 0.00 \\
 70 & 1 &   (3,2) & 1.00 &  2.0 & 0.00 &  1.0 & 0.00 \\
 71 & 2 &   (4,3) & 0.75 &  2.5 & 0.50 &  1.5 & 0.50 \\
 72 & 1 &   (3,2) & 1.00 &  2.0 & 0.00 &  1.0 & 0.00 \\
 73 & 2 &   (4,5) & 0.67 &  2.5 & 0.50 &  3.0 & 1.00 \\
 74 & 4 &  (7,12) & 0.56 &  4.5 & 1.50 &  7.5 & 4.00 \\
 75 & 2 &  (15,3) & 0.54 &  8.5 & 5.50 &  1.5 & 0.50 \\
 76 & 2 &   (5,6) & 0.67 &  3.0 & 1.00 &  4.5 & 0.50 \\
 77 & 1 &   (3,2) & 1.00 &  2.0 & 0.00 &  1.0 & 0.00 \\
 78 & 1 &   (3,3) & 1.00 &  2.0 & 0.00 &  2.0 & 0.00 \\
 79 & 1 &   (4,2) & 1.00 &  3.0 & 0.00 &  1.0 & 0.00 \\
 80 & 1 &   (3,3) & 1.00 &  2.0 & 0.00 &  2.0 & 0.00 \\
 81 & 2 &  (11,2) & 0.56 &  6.0 & 4.00 &  0.5 & 0.50 \\
 82 & 2 &   (6,5) & 0.62 &  3.5 & 1.50 &  3.5 & 0.50 \\
 83 & 1 &   (3,3) & 1.00 &  2.0 & 0.00 &  2.0 & 0.00 \\
 84 & 2 &   (4,2) & 0.75 &  2.5 & 0.50 &  0.5 & 0.50 \\
 85 & 2 &   (6,4) & 0.67 &  4.0 & 1.00 &  2.5 & 0.50 \\
 86 & 2 &   (7,4) & 0.62 &  4.5 & 1.50 &  2.5 & 0.50 \\
 87 & 3 &   (7,7) & 0.50 &  5.0 & 1.00 &  5.0 & 2.50 \\
 88 & 2 &  (10,3) & 0.58 &  6.5 & 2.50 &  1.5 & 0.50 \\
 89 & 1 &   (5,3) & 1.00 &  4.0 & 0.00 &  2.0 & 0.00 \\
 90 & 1 &   (4,2) & 1.00 &  3.0 & 0.00 &  1.0 & 0.00 \\
 91 & 1 &   (6,6) & 1.00 &  5.0 & 0.00 &  5.0 & 0.00 \\
 92 & 2 &   (8,4) & 0.62 &  5.5 & 1.50 &  2.5 & 0.50 \\
 93 & 2 &  (5,12) & 0.58 &  3.5 & 0.50 &  8.5 & 2.50 \\
 94 & 2 &   (8,4) & 0.62 &  5.5 & 1.50 &  2.5 & 0.50 \\
 95 & 3 &   (8,9) & 0.58 &  6.0 & 1.00 &  4.0 & 3.50 \\
 96 & 1 &  (7,15) & 1.00 &  6.0 & 0.00 & 14.0 & 0.00 \\
 98 & 3 &  (14,4) & 0.63 &  7.0 & 4.00 &  2.0 & 1.00 \\
 99 & 2 &  (11,3) & 0.57 &  7.0 & 3.00 &  1.5 & 0.50 \\
100 & 3 &  (15,7) & 0.41 & 12.0 & 5.00 &  4.0 & 1.50 \\
101 & 4 & (17,10) & 0.82 &  5.5 & 3.75 &  3.0 & 3.50 \\
102 & 2 &   (6,6) & 0.62 &  4.5 & 0.50 &  3.5 & 1.50 \\
103 & 3 &  (14,5) & 0.68 &  6.0 & 4.50 &  2.0 & 1.50 \\
104 & 2 &  (4,11) & 0.62 &  2.5 & 0.50 &  8.5 & 1.50 \\
105 & 3 &   (6,4) & 0.67 &  3.0 & 1.50 &  2.0 & 1.00 \\
106 & 1 &   (3,3) & 1.00 &  2.0 & 0.00 &  2.0 & 0.00 \\
107 & 2 &   (4,4) & 0.67 &  2.5 & 0.50 &  2.0 & 1.00 \\
108 & 1 &   (4,3) & 1.00 &  3.0 & 0.00 &  2.0 & 0.00 \\
109 & 2 &   (7,3) & 0.62 &  4.5 & 1.50 &  1.5 & 0.50 \\
110 & 2 &   (5,4) & 0.75 &  3.5 & 0.50 &  2.5 & 0.50 \\
111 & 2 &   (7,3) & 0.62 &  4.5 & 1.50 &  1.5 & 0.50 \\
112 & 2 &   (4,3) & 0.67 &  2.5 & 0.50 &  1.0 & 1.00 \\
113 & 2 &   (5,3) & 0.75 &  3.5 & 0.50 &  1.5 & 0.50 \\
114 & 2 &   (4,5) & 0.62 &  2.5 & 0.50 &  2.5 & 1.50 \\
115 & 1 &   (3,2) & 1.00 &  2.0 & 0.00 &  1.0 & 0.00 \\
116 & 3 &   (7,4) & 0.67 &  3.0 & 2.00 &  2.0 & 1.00 \\
117 & 2 &   (4,6) & 0.60 &  2.5 & 0.50 &  3.0 & 2.00 \\
118 & 3 &   (5,4) & 0.67 &  3.0 & 1.00 &  1.0 & 1.50 \\
119 & 2 &  (12,3) & 0.55 &  6.5 & 4.50 &  1.5 & 0.50 \\
120 & 1 &   (3,1) & 1.00 &  2.0 & 0.00 &  0.0 & 0.00 \\
121 & 1 &   (4,2) & 1.00 &  3.0 & 0.00 &  1.0 & 0.00 \\
\hline
\end{longtable}

\begin{longtable}{cccccccccc}
\caption[Results obtained for each method]{Results obtained for methods under study in open-source projects. Each column represents, respectively, the index of each method (see Table~\ref{tab:index-methods}), number of solutions obtained for the method, the reference point used for the method (\emph{Extractions~(E), CC~(C), LOC~(L)}) to compute the HV, normalized HV (N-HV), and the median and inter-quartile range (\emph{IQR})  for the three objective functions.}
\label{tab:hipervolumen-open-source-tres-objs}
\footnotesize
\\
\hline
\multirow{2}{1cm}{\centering\textbf{ID}} & \multirow{2}{1.5cm}{\textbf{No. sol}} & \textbf{Reference} & \multirow{2}{1cm}{\centering\textbf{N-HV}} & \multicolumn{2}{c}{\textbf{Extractions}} & \multicolumn{2}{c}{\textbf{CC}} & \multicolumn{2}{c}{\textbf{LOC}} \\
& & \emph{(E,C,L)} & & \emph{Median} & \emph{IQR} & \emph{Median} & \emph{IQR} & \emph{Median} & \emph{IQR} \\ 
\hline
  1 & 19 & (20,11,11) & 0.86 &  9.0 & 7.50 &  3.0 & 1.50 &  5.0 &  3.00 \\
  2 &  2 &    (5,3,3) & 0.58 &  3.0 & 1.00 &  1.5 & 0.50 &  1.5 &  0.50 \\
  3 & 13 & (12,10,15) & 0.79 &  7.0 & 4.00 &  3.0 & 2.00 &  4.0 &  3.00 \\
  4 &  5 &   (8,8,24) & 0.83 &  4.0 & 4.00 &  2.0 & 1.00 & 13.0 & 11.00 \\
  5 & 13 & (12,10,26) & 0.73 &  5.0 & 3.00 &  5.0 & 2.00 & 11.0 &  5.00 \\
  6 &  6 &   (5,8,23) & 0.49 &  3.0 & 1.50 &  3.5 & 1.75 &  6.5 & 10.75 \\
  7 & 10 &  (8,14,26) & 0.59 &  5.0 & 1.75 &  8.0 & 3.50 & 10.5 &  9.00 \\
  8 & 11 &  (11,8,31) & 0.66 &  5.0 & 4.00 &  4.0 & 3.00 &  9.0 &  6.00 \\
  9 & 12 &  (16,8,20) & 0.55 &  9.0 & 6.25 &  3.5 & 2.25 & 11.0 &  6.75 \\
 10 &  8 &  (8,10,11) & 0.61 &  3.5 & 2.50 &  2.0 & 2.00 &  3.5 &  4.50 \\
 11 & 15 &  (10,8,20) & 0.66 &  5.0 & 3.00 &  3.0 & 2.00 &  8.0 &  5.00 \\
 12 &  1 &    (3,2,2) & 1.00 &  2.0 & 0.00 &  1.0 & 0.00 &  1.0 &  0.00 \\
 13 &  1 &    (3,3,3) & 1.00 &  2.0 & 0.00 &  2.0 & 0.00 &  2.0 &  0.00 \\
 14 &  1 &    (3,1,1) & 1.00 &  2.0 & 0.00 &  0.0 & 0.00 &  0.0 &  0.00 \\
 15 &  9 &  (13,8,42) & 0.50 &  8.0 & 2.00 &  4.0 & 2.00 & 36.0 &  4.00 \\
 16 &  8 &   (9,9,16) & 0.54 &  4.0 & 3.50 &  3.5 & 3.00 &  6.5 &  6.00 \\
 17 &  6 &   (7,6,15) & 0.67 &  3.0 & 2.25 &  2.5 & 1.00 &  4.0 &  4.00 \\
 18 &  5 &  (9,11,33) & 0.35 &  7.0 & 1.00 &  9.0 & 1.00 & 30.0 &  2.00 \\
 19 &  2 &  (4,14,14) & 0.40 &  2.5 & 0.50 & 11.0 & 2.00 & 12.5 &  0.50 \\
 20 & 20 & (13,12,32) & 0.68 &  6.5 & 6.25 &  5.0 & 3.25 & 10.0 &  6.75 \\
 21 & 12 &   (9,9,16) & 0.40 &  7.0 & 1.25 &  5.0 & 3.25 &  7.0 &  7.25 \\
 22 &  7 &   (7,12,9) & 0.72 &  3.0 & 2.00 &  3.0 & 1.50 &  6.0 &  5.00 \\
 23 & 11 &  (18,9,24) & 0.85 &  9.0 & 7.50 &  1.0 & 1.50 &  4.0 &  8.00 \\
 24 & 11 &  (18,5,24) & 0.74 &  9.0 & 7.50 &  1.0 & 1.50 &  4.0 &  7.00 \\
 25 & 19 & (18,12,29) & 0.58 &  8.0 & 4.50 &  4.0 & 4.00 & 17.0 &  7.50 \\
 26 &  1 &    (3,1,1) & 1.00 &  2.0 & 0.00 &  0.0 & 0.00 &  0.0 &  0.00 \\
 27 &  1 &    (3,1,2) & 1.00 &  2.0 & 0.00 &  0.0 & 0.00 &  1.0 &  0.00 \\
 28 & 12 & (12,11,19) & 0.51 &  6.0 & 3.25 &  4.0 & 3.50 & 12.0 &  3.50 \\
 29 &  6 &   (11,5,9) & 0.75 &  4.5 & 4.75 &  1.5 & 1.75 &  2.5 &  2.50 \\
 30 & 10 & (15,15,47) & 0.44 & 10.5 & 4.50 &  5.0 & 5.50 & 37.0 &  7.50 \\
 31 &  2 &  (5,11,14) & 0.57 &  3.5 & 0.50 & 10.0 & 0.00 & 10.0 &  3.00 \\
 32 &  1 &    (3,1,2) & 1.00 &  2.0 & 0.00 &  0.0 & 0.00 &  1.0 &  0.00 \\
 33 &  1 &    (3,4,3) & 1.00 &  2.0 & 0.00 &  3.0 & 0.00 &  2.0 &  0.00 \\
 34 &  2 &   (4,8,15) & 0.51 &  2.0 & 0.50 &  4.5 & 2.50 &  8.0 &  6.00 \\
 35 &  5 &    (9,4,7) & 0.62 &  4.0 & 3.00 &  2.0 & 1.00 &  3.0 &  2.00 \\
 36 & 12 &  (10,6,18) & 0.69 &  4.5 & 3.50 &  2.0 & 1.25 &  4.5 &  6.50 \\
 37 &  2 &    (8,2,7) & 0.51 &  4.5 & 2.50 &  0.5 & 0.50 &  3.0 &  3.00 \\
 38 &  1 &    (3,2,3) & 1.00 &  2.0 & 0.00 &  1.0 & 0.00 &  2.0 &  0.00 \\
 39 &  2 &    (4,2,6) & 0.54 &  2.5 & 0.50 &  0.5 & 0.50 &  2.5 &  2.50 \\
 40 & 20 & (15,12,27) & 0.73 &  6.5 & 6.00 &  4.0 & 3.25 &  7.5 &  3.75 \\
 41 & 11 & (14,11,19) & 0.68 &  8.0 & 4.00 &  6.0 & 2.00 &  9.0 &  7.00 \\
 42 &  4 & (10,14,48) & 0.49 &  7.5 & 1.25 & 12.5 & 1.50 & 34.0 &  6.75 \\
 43 &  9 &  (15,8,10) & 0.55 &  6.0 & 6.00 &  3.0 & 2.00 &  6.0 &  3.00 \\
 44 &  2 &   (8,7,20) & 0.75 &  6.5 & 0.50 &  6.0 & 0.00 & 18.5 &  0.50 \\
 45 & 16 & (15,10,20) & 0.64 &  7.0 & 5.25 &  3.5 & 3.00 &  7.5 &  6.00 \\
 46 &  4 &  (9,10,14) & 0.42 &  6.0 & 2.50 &  4.0 & 3.00 & 10.0 &  2.75 \\
 47 &  3 & (10,14,26) & 0.67 &  8.0 & 1.00 & 11.0 & 1.50 & 25.0 &  0.00 \\
 48 &  2 &   (5,5,10) & 0.53 &  3.5 & 0.50 &  3.5 & 0.50 &  5.5 &  3.50 \\
 49 &  7 & (10,11,12) & 0.35 &  6.0 & 2.50 &  8.0 & 3.50 &  9.0 &  1.50 \\
 50 &  5 &   (7,7,13) & 0.32 &  5.0 & 2.00 &  4.0 & 2.00 &  9.0 &  1.00 \\
 51 &  5 &  (10,7,10) & 0.52 &  8.0 & 3.00 &  4.0 & 2.00 &  6.0 &  2.00 \\
 52 & 17 &  (9,11,40) & 0.49 &  5.0 & 2.00 &  5.0 & 4.00 & 20.0 & 15.00 \\
 53 &  1 &  (6,15,43) & 1.00 &  5.0 & 0.00 & 14.0 & 0.00 & 42.0 &  0.00 \\
 54 &  5 &   (5,8,17) & 0.61 &  3.0 & 1.00 &  4.0 & 3.00 &  7.0 &  3.00 \\
 55 &  9 & (18,13,31) & 0.64 & 10.0 & 2.00 &  9.0 & 3.00 & 21.0 &  5.00 \\
 56 & 14 & (13,12,14) & 0.59 &  9.0 & 3.50 &  4.5 & 2.00 &  8.0 &  3.75 \\
 57 & 17 &  (23,7,15) & 0.70 & 10.0 & 4.00 &  2.0 & 3.00 &  6.0 &  4.00 \\
 58 &  4 &   (8,6,20) & 0.45 &  6.0 & 0.50 &  4.0 & 0.50 &  9.5 &  6.25 \\
 59 &  2 &   (6,8,11) & 0.67 &  5.0 & 0.00 &  6.5 & 0.50 &  9.0 &  1.00 \\
 60 &  2 &    (7,6,8) & 0.53 &  4.5 & 1.50 &  3.5 & 1.50 &  6.5 &  0.50 \\
 61 &  2 &  (4,15,33) & 0.62 &  2.5 & 0.50 & 13.5 & 0.50 & 31.5 &  0.50 \\
 62 &  2 &    (6,3,3) & 0.56 &  3.5 & 1.50 &  1.5 & 0.50 &  1.5 &  0.50 \\
 63 &  2 &    (6,3,4) & 0.53 &  3.5 & 1.50 &  1.5 & 0.50 &  1.5 &  1.50 \\
 64 &  5 &   (7,7,11) & 0.55 &  4.0 & 2.00 &  3.0 & 2.00 &  4.0 &  1.00 \\
 65 &  1 &   (3,2,15) & 1.00 &  2.0 & 0.00 &  1.0 & 0.00 & 14.0 &  0.00 \\
 66 &  4 &   (6,6,15) & 0.69 &  2.5 & 1.50 &  2.5 & 1.75 &  3.5 &  4.00 \\
 67 &  5 &  (10,7,18) & 0.46 &  5.0 & 5.00 &  4.0 & 1.00 & 12.0 &  4.00 \\
 68 &  4 &    (5,6,7) & 0.38 &  3.5 & 1.25 &  2.5 & 1.75 &  2.5 &  2.25 \\
 69 &  1 &   (3,2,15) & 1.00 &  2.0 & 0.00 &  1.0 & 0.00 & 14.0 &  0.00 \\
 70 &  5 &   (7,5,10) & 0.48 &  4.0 & 2.00 &  2.0 & 1.00 &  6.0 &  3.00 \\
 71 &  6 &   (7,7,17) & 0.56 &  3.5 & 1.75 &  2.5 & 1.75 &  9.0 &  4.75 \\
 72 &  5 &    (7,5,9) & 0.51 &  4.0 & 2.00 &  2.0 & 1.00 &  5.0 &  3.00 \\
 73 &  3 &   (4,9,16) & 0.55 &  2.0 & 0.50 &  4.0 & 3.00 & 10.0 &  3.00 \\
 74 &  9 &  (8,14,41) & 0.57 &  5.0 & 2.00 &  9.0 & 5.00 & 27.0 &  7.00 \\
 75 & 12 &   (16,8,8) & 0.65 &  9.0 & 4.25 &  2.5 & 1.25 &  3.0 &  1.50 \\
 76 &  2 &   (5,6,12) & 0.52 &  3.0 & 1.00 &  4.5 & 0.50 &  6.5 &  4.50 \\
 77 &  8 &   (9,8,12) & 0.66 &  4.5 & 3.50 &  2.0 & 2.25 &  4.5 &  2.25 \\
 78 &  1 &    (3,3,1) & 1.00 &  2.0 & 0.00 &  2.0 & 0.00 &  0.0 &  0.00 \\
 79 &  5 &   (17,9,4) & 0.55 &  8.0 & 5.00 &  5.0 & 3.00 &  2.0 &  1.00 \\
 80 &  1 &   (3,3,16) & 1.00 &  2.0 & 0.00 &  2.0 & 0.00 & 15.0 &  0.00 \\
 81 &  5 &   (11,4,6) & 0.61 &  3.0 & 2.00 &  1.0 & 0.00 &  2.0 &  1.00 \\
 82 &  5 &  (7,14,19) & 0.76 &  3.0 & 3.00 &  4.0 & 1.00 & 10.0 &  9.00 \\
 83 &  4 &    (6,9,7) & 0.50 &  2.5 & 1.50 &  4.5 & 5.25 &  3.5 &  1.75 \\
 84 &  2 &   (4,2,22) & 0.51 &  2.5 & 0.50 &  0.5 & 0.50 & 11.0 & 10.00 \\
 85 & 17 &  (9,11,42) & 0.66 &  6.0 & 2.00 &  4.0 & 3.00 & 11.0 &  8.00 \\
 86 & 13 &  (14,9,27) & 0.74 &  6.0 & 3.00 &  3.0 & 1.00 &  8.0 &  8.00 \\
 87 & 20 & (15,11,16) & 0.61 &  8.0 & 5.00 &  5.0 & 3.00 &  5.5 &  6.50 \\
 88 & 20 &  (16,8,18) & 0.70 &  7.0 & 4.50 &  3.0 & 2.50 &  6.0 &  5.50 \\
 89 & 15 &  (17,9,12) & 0.69 &  9.0 & 6.50 &  3.0 & 2.50 &  5.0 &  3.00 \\
 90 &  1 &    (4,2,2) & 1.00 &  3.0 & 0.00 &  1.0 & 0.00 &  1.0 &  0.00 \\
 91 &  5 &  (8,15,35) & 0.32 &  6.0 & 1.00 &  7.0 & 2.00 & 29.0 &  5.00 \\
 92 &  5 &   (12,9,6) & 0.54 &  8.0 & 2.00 &  4.0 & 2.00 &  3.0 &  1.00 \\
 93 &  2 &  (5,12,15) & 0.54 &  3.5 & 0.50 &  8.5 & 2.50 & 13.5 &  0.50 \\
 94 &  6 &   (12,9,7) & 0.60 &  8.5 & 1.75 &  4.5 & 3.25 &  3.0 &  0.75 \\
 95 & 12 & (15,11,49) & 0.70 &  7.0 & 2.25 &  4.5 & 5.25 & 34.0 & 12.25 \\
 96 &  3 &  (9,15,41) & 0.67 &  7.0 & 1.00 & 14.0 & 0.00 & 25.0 &  8.00 \\
 97 &  7 &  (9,14,70) & 0.39 &  7.0 & 1.00 & 10.0 & 4.50 & 43.0 & 32.00 \\
 98 & 11 &   (17,9,5) & 0.64 & 10.0 & 6.00 &  3.0 & 3.00 &  2.0 &  1.50 \\
 99 &  7 &  (15,7,11) & 0.74 &  5.0 & 4.00 &  2.0 & 1.00 &  7.0 &  2.50 \\
100 &  9 & (16,11,40) & 0.52 &  6.0 & 8.00 &  6.0 & 3.00 & 24.0 &  6.00 \\
101 & 11 & (17,10,10) & 0.69 &  7.0 & 5.00 &  3.0 & 3.00 &  4.0 &  3.50 \\
102 &  8 &   (12,9,7) & 0.69 &  5.0 & 2.50 &  5.0 & 2.25 &  4.0 &  2.25 \\
103 & 17 & (15,12,22) & 0.80 &  6.0 & 4.00 &  4.0 & 4.00 &  8.0 &  3.00 \\
104 &  6 &  (6,13,34) & 0.54 &  3.5 & 1.00 &  9.0 & 3.50 & 13.0 & 17.00 \\
105 &  4 &    (7,4,8) & 0.50 &  4.0 & 2.50 &  2.0 & 0.50 &  3.5 &  3.75 \\
106 &  7 &  (6,15,18) & 0.34 &  3.0 & 1.00 &  8.0 & 8.00 &  9.0 &  6.50 \\
107 &  3 &   (4,6,13) & 0.61 &  2.0 & 0.50 &  3.0 & 2.00 &  2.0 &  5.50 \\
108 &  9 & (10,12,19) & 0.73 &  4.0 & 2.00 &  4.0 & 3.00 &  6.0 &  3.00 \\
109 & 18 & (11,11,32) & 0.72 &  4.0 & 2.75 &  3.5 & 4.75 & 11.5 & 12.75 \\
110 & 20 &  (23,6,36) & 0.78 &  9.5 & 8.25 &  3.0 & 2.00 &  9.5 &  8.50 \\
111 & 10 &   (18,6,9) & 0.67 &  6.5 & 7.25 &  2.0 & 0.75 &  3.5 &  3.50 \\
112 &  7 &  (11,7,18) & 0.66 &  5.0 & 4.00 &  5.0 & 3.00 &  3.0 &  1.00 \\
113 & 12 &  (20 8,12) & 0.70 &  5.5 & 7.00 &  2.5 & 3.50 &  3.5 &  2.25 \\
114 &  3 &    (4,5,6) & 0.42 &  3.0 & 0.50 &  2.0 & 1.50 &  3.0 &  2.00 \\
115 &  3 &    (4,5,8) & 0.52 &  2.0 & 0.50 &  3.0 & 1.50 &  1.0 &  3.50 \\
116 &  5 &   (9,4,17) & 0.55 &  5.0 & 3.00 &  2.0 & 0.00 &  7.0 &  6.00 \\
117 &  8 &  (10,8,14) & 0.73 &  4.0 & 3.50 &  3.0 & 3.25 &  5.0 &  5.50 \\
118 & 11 &   (22,4,8) & 0.67 &  8.0 & 7.00 &  1.0 & 0.50 &  4.0 &  4.00 \\
119 & 18 & (15,12,16) & 0.67 &  8.5 & 6.75 &  3.5 & 2.00 &  6.0 &  5.00 \\
120 &  8 &   (18,8,5) & 0.60 &  5.5 & 6.00 &  2.0 & 2.50 &  2.0 &  2.00 \\
121 &  5 &   (11,3,9) & 0.74 &  5.0 & 3.00 &  1.0 & 0.00 &  5.0 &  3.00 \\
\hline
\end{longtable}

\begin{landscape}
\begin{figure}[!ht]
\centering

\vspace{-0.7cm}
\begin{tikzpicture}
\node[inner sep=0pt] (code) {
\begin{minipage}{0.77\textwidth}

\vspace{0.5cm}
\begin{lstlisting}[numbers=left,xleftmargin=0cm,xrightmargin=0cm,frame=single,framexleftmargin=0em,escapechar=ñ,stepnumber=1,
    showstringspaces=false,
    tabsize=1,
    breaklines=true,
    breakatwhitespace=false,
    basicstyle=\tiny,
    language=Java,
    mathescape=true]
public void routeAPacketTo(IPAddress ip, int ttl, List<Host> usedHosts) throws Exception {
    usedHosts.add(this);
    if (ttl == 0) {//Problem in routing
        throw new Exception("Routing problem, TTL is null : packet to " 
                            + ip + " deleted on host " + this.getName());
    }
    if (!hasIP(ip)) { //Packet not arrived
        Host nextHost = null;
        List<Host> directlyAccessibleHosts = getDirectlyAccessibleHosts();
        for (Host directlyAccessibleHost : directlyAccessibleHosts) {
            if (directlyAccessibleHost.hasIP(ip)) // If the packet is for a neighbour,
                                                  // we send it to him
                nextHost = directlyAccessibleHost;
        }
        if (nextHost != null) {
            nextHost.routeAPacketTo(ip, ttl - 1, usedHosts);
        } else {//We have to look in the routing table
            List<Host> directlyAccessible = getDirectlyAccessibleHosts();
            IPAddress nextIP = this.getRoutingTable().getNextHop(ip);
            boolean nextHostFound = false;
            for (Host aDirectlyAccessible : directlyAccessible) {
                if (aDirectlyAccessible.hasIP(nextIP)) { //Search the nextHop host object
                    aDirectlyAccessible.routeAPacketTo(ip, ttl - 1, usedHosts);
                    nextHostFound = true;
                }
            }
            if (!nextHostFound) { //Routing problem
                throw new Exception("Routing problem, there is no route corresponding 
                                     to the packet or the destination host is on the internet");
            }
        }
    }
}
\end{lstlisting}
\vspace{-0.5cm}
\caption*{(a) Original method.}
\label{fig:original-code-solution-example}
\end{minipage}
\hfill
\hspace{1cm}
\begin{minipage}{0.7\textwidth}
\vspace{-1cm}
\begin{lstlisting}[numbers=left,xleftmargin=0cm,xrightmargin=0cm,frame=single,framexleftmargin=0em,escapechar=ñ,stepnumber=1,
    showstringspaces=false,
    tabsize=1,
    breaklines=true,
    breakatwhitespace=false,
    basicstyle=\tiny,
    language=Java,
    mathescape=true]{}
public void routeAPacketTo(IPAddress ip, int ttl, List<Host> usedHosts) throws Exception {
    usedHosts.add(this);
    if (ttl == 0) {//Problem in routing
        throw new Exception("Routing problem, TTL is null : packet to " 
                            + ip + " deleted on host " + this.getName());
    }
    if (!hasIP(ip)) { //Packet not arrived
        Host nextHost = null;
        List<Host> directlyAccessibleHosts = getDirectlyAccessibleHosts();
        forwardToDirectHostOrRoutingTable(ip, ttl, usedHosts, nextHost, directlyAccessibleHosts);
    }
}

private void forwardToDirectHostOrRoutingTable(IPAddress ip, int ttl, List<Host> usedHosts, List<Host> directlyAccessibleHosts) {
    for (Host directlyAccessibleHost : directlyAccessibleHosts) {
        if (directlyAccessibleHost.hasIP(ip)) // If the packet is for a neighbour,
                                              // we send it to him
            nextHost = directlyAccessibleHost;
    }
    if (nextHost != null) {
        nextHost.routeAPacketTo(ip, ttl - 1, usedHosts);
    } else {//We have to look in the routing table
        List<Host> directlyAccessible = getDirectlyAccessibleHosts();
        forwardViaRoutingTable(ip, ttl, usedHosts, directlyAccessible);
    }
}

private void forwardViaRoutingTable(IPAddress ip, int ttl, List<Host> usedHosts, List<Host> directlyAccessible) throws Exception {
    IPAddress nextIP = this.getRoutingTable().getNextHop(ip);
    boolean nextHostFound = false;
    for (Host aDirectlyAccessible : directlyAccessible) {
        if (aDirectlyAccessible.hasIP(nextIP)) { //Search the nextHop host object
            aDirectlyAccessible.routeAPacketTo(ip, ttl - 1, usedHosts);
            nextHostFound = true;
        }
    }
    if (!nextHostFound) { //Routing problem
        throw new Exception("Routing problem, there is no route corresponding 
                             to the packet or the destination host is on the internet");
    }
}
\end{lstlisting}
\vspace{-0.5cm}
\caption*{(b) Refactored version.}
\end{minipage}
};

\draw[amber, ultra thick] (-11.5,-5.3) rectangle (-0.1,2.9);
\draw[airforceblue, ultra thick] (-11,-5.24) rectangle (-0.17,-0.38);

\draw[amber, ultra thick, fill=amber, fill opacity=0.05] (1.75,4.8) rectangle (10.4,5.55);
\draw[airforceblue, ultra thick, fill=airforceblue, fill opacity=0.05] (2,-0.72) rectangle (10.4,-0.4);

\draw[amber, ultra thick, fill=amber, fill opacity=0.05] (0.45,-1.5) rectangle (12.5,3.7);
\draw[airforceblue, ultra thick, fill=airforceblue, fill opacity=0.05] (0.45,-7.4) rectangle (12.5,-1.85);

\node[draw=amber, fill=white, text=amber, ultra thick, circle, minimum size=0.05cm] at (11,5.2) {1};
\node[draw=airforceblue, fill=white, text=airforceblue, ultra thick, circle, minimum size=0.05cm] at (11,-0.55) {2};

\node[draw=amber, fill=white, text=amber, ultra thick, circle, minimum size=0.05cm] at (12.5,3.7) {1};
\node[draw=airforceblue, fill=white, text=airforceblue, ultra thick, circle, minimum size=0.05cm] at (12.5,-2) {2};

\draw[->, ultra thick, bend left=35] (-2,2.9) to (1.75,5.2);
\draw[->, ultra thick, bend left=24] (-2.5,-0.38) to (2,-0.55);

\end{tikzpicture}
\vspace{-1.1cm}
\caption[Extract method applied to solution \textbf{s2} in \texttt{routeAPacketTo} method]{Extract method refactoring applied to solution \textbf{s2} in method \texttt{routeAPacketTo} of the running example in Figure~\ref{fig:ccExampleInCode}, using the hybrid method algorithm to solve the multi-objective ILP problem.}
\label{fig:refactoring_solution_multi-objective}
\end{figure}
\end{landscape}

\clearpage
\newpage
\bibliographystyle{splncs04}
\bibliography{references}

\end{document}